\documentclass[11pt,reqno,notitlepage]{article}
\usepackage{setspace}
\usepackage{amssymb}
\usepackage{amsthm}
\usepackage{graphicx}
\usepackage{amsmath}
\usepackage{fancyhdr}
\usepackage{datetime}
\usepackage[toc]{appendix}
\usepackage[usenames,dvipsnames]{color}
\usepackage{tikz}
\usepackage{multirow}
\usepackage{color}
\usepackage{colortbl}
\usepackage{mathrsfs}
\usepackage[normalem]{ulem}
\usepackage{graphicx, xcolor}
\usepackage{hyperref}
\usepackage{enumerate}

\usepackage{pdflscape}

\usepackage{setspace}
\usepackage{graphicx, xcolor}												
\usepackage[mathscr]{eucal}												
\usepackage{subcaption}
\usepackage{color}
\usepackage{epstopdf}

\allowdisplaybreaks

\makeatletter
\newcommand{\labitem}[2]{%
\def\@itemlabel{\textbf{#1}}
\item
\def\@currentlabel{#1}\label{#2}}

\renewcommand\paragraph{\@startsection{paragraph}{4}{\z@}%
                                      {\parskip}
                                      {0em}%
                                      {\noindent\normalfont\normalsize\bfseries}}
\makeatother

\usetikzlibrary{arrows,decorations.pathreplacing}
\usepackage[top=1in, bottom=1in, left=1in, right=1in]{geometry}
\usepackage{mathtools}									%
\mathtoolsset{showonlyrefs=true}							%

\newtheorem{theorem}{Theorem}
\newtheorem{lemma}[theorem]{Lemma}								%
\newtheorem{proposition}[theorem]{Proposition}	
\newtheorem{corollary}[theorem]{Corollary}	
\newtheorem{assumption}[theorem]{Assumption}	
\newtheorem{definition}[theorem]{Definition}
\newtheorem{remark}{Remark}

\numberwithin{equation}{section}	
\numberwithin{theorem}{section}
\renewcommand{\vec}[1]{\mathbf{#1}}

\def\R{\mathbb{R}}
\def\X{\mathcal{X}}
\def\Y{\mathcal{Y}}

\def\C{\mathcal{C}}

\def\eps{\varepsilon}

\def \Y {\mathcal{Y}}
\def \Imp {I}

\def\d{\partial}

\newcommand{\e}[1]{\operatorname{e}^{#1}}

\newcommand{\dom}{\operatorname{dom}}

\renewcommand{\vec}[1]{\mathbf{#1}}

\newtheorem{example}[theorem]{Example}

\long\def\symbolfootnote[#1]#2{\begingroup\def\thefootnote{\fnsymbol{footnote}}\footnote[#1]{#2}\endgroup}

\begin{document}

\title{Axioms for Automated Market Makers: A Mathematical Framework in FinTech and Decentralized Finance} 

\author{
Maxim Bichuch
\thanks{
Department of Mathematics,
SUNY at Buffalo
Buffalo, NY 14260. 
{\tt mbichuch@buffalo.edu}. Work  is partially supported by NSF grant DMS-1736414. }
\and  Zachary Feinstein
\thanks{
School of Business,
Stevens Institute of Technology,
Hoboken, NJ 07030, USA,
{\tt  zfeinste@stevens.edu}. }
}
\date{\today}
\maketitle

\begin{abstract}
Within this work we consider an axiomatic framework for Automated Market Makers (AMMs). AMMs are smart contracts that set prices for swaps on a pool of assets. By imposing reasonable axioms on the underlying utility function, we are able to characterize the properties of the swap size of the assets and of the resulting pricing oracle. In providing these general axioms, we define a novel measure of price impacts that can be used to quantify those costs between different AMM constructions. We have analyzed many existing AMMs and shown that the vast majority of them satisfy our axioms. We have also considered the question of fees and divergence loss. In doing so, we have proposed a new fee structure so as to make the AMM indifferent to transaction splitting. Finally, we have proposed a novel AMM that has nice analytical properties and provides a large range over which there is no divergence loss. 
\end{abstract}


\section{Introduction}\label{sec:intro}

Decentralized Finance (DeFi) is a new paradigm for finance which replaces traditional intermediaries with innovative financial technologies based on blockchain.
DeFi companies are providing services in areas such as lending and borrowing, insurance underwriting, and trading \emph{without} working with the traditional financial intermediaries.  Within this work, we focus on the use of DeFi for constructing markets to trade financial instruments; specifically, Automated Market Makers (AMMs) are a decentralized approach for creating financial markets. The key idea is to create a (liquidity) pool of assets against which a trader can transact; a swap is executed on the pool at a price determined by the AMM according to an invariance function.  A key benefit of DeFi and AMMs is that any individual can invest in the pool so that he or she can share in the benefits from providing these key services in the financial system.  

The organization of this paper is as follows.  The motivation for studying DeFi and, more specifically, AMMs is provided within Section~\ref{sec:intro-motivation}.  A review of the literature on AMMs is provided within Section~\ref{sec:intro-lit}.  The introduction is then concluded by an overview of the primary contributions and results of this work in Section~\ref{sec:intro-contributions}.  In Section~\ref{sec:design}, the desirable economic and financial properties for markets made by AMMs are provided. The mathematical construction of a generic AMM is presented within Section~\ref{sec:amm}.  With that discussion, the properties of the realized swap value and the pricing oracle derived from the AMM are provided and compared with those desirable properties in Section~\ref{sec:design}.  In Section~\ref{sec:examples}, these axioms and properties are then validated against AMMs that exist in practice as well as a new mathematical structure that can be used for generating new AMMs.  Finally, in Section~\ref{sec:fees}, we present a discussion on how the pool can collect fees along with the possible risks that are involved with becoming a liquidity provider (i.e., the divergence or impermanent loss). We summarize and conclude in Section \ref{sec:conclude}. The proofs of all results are provided within Appendix~\ref{sec:proofs}.

\subsection{Motivation}\label{sec:intro-motivation}
At the most recent peak for cryptocurrency valuation on Nov 8, 2021, the total market capitalization of various DeFi projects was nearly \$180B.\footnote{\url{https://tradingview.com/markets/cryptocurrencies/global-charts/}} Since then, and especially after the onset of the so-called ``crypto winter'' in May 2022, the risk of various DeFi projects has come into full view of the public.  To highlight just a few DeFi failures: the demise of the algorithmic stablecoin TerraUSD -- which was triggered by a run on the Terra Luna coin -- wiped out billions of dollars in wealth in a single week; separately, the DeFi lending platform Celsius Networks, which had approximately \$12B in assets under management prior to the crypto winter, filed for bankruptcy in July 2022. With the onset of the crypto winter and the crash in cryptocurrencies, the total value locked in to pools at Decentralized Exchanges has dropped precipitously as well with, e.g., Curve falling from a market capitalization of almost \$3B in early January 2022 to just over \$300M in June 2022.\footnote{\url{https://coinmarketcap.com/currencies/curve-dao-token/}}  With the irrational exuberance subsiding, now is the perfect time to explore the viability and risks associated with DeFi projects. Specifically, for the purposes of this work, we wish to quantify the risks, and understand the pitfalls, of AMMs while highlighting the benefits for investors and liquidity providers.

One beneficial aspect of AMMs, and a defining property of them, is the decentralization. While much has been written on the decentralization of transaction verification through blockchain, AMMs also allow for decentralization in rewards. This is because anybody can become a liquidity provider in, and therefore a shareholder of, the pool.  Due to this democratization of market making, AMMs lower the barriers for listing new securities (or tokens) considerably.

Of course, with any type of investment also comes risk; as such, understanding the downsides for liquidity providers (i.e., investors in a pool) is vital. We tackle this problem by formalizing the axioms that AMMs operate under. For example, we prove that the typical constructions of AMMs in practice (without fees) always have a divergence loss -- a detrimental result for liquidity providers.
In fact \cite{xu2021sok} states ``this ``loss'' only disappears when the current proportions of the pool assets equal exactly those at liquidity provision, which is rarely the case.'' 
As such, fees are necessary to provide a cushion for the investors against this loss and, thereby, to encourage investment in pools; however, not every fee structure works well. For example, a pool can charge fees on any of its quoted assets either before or after the verification of a trade. While any combination of fees can intuitively work, some fee structures can lead to, e.g., the pool charging smaller fees for bulk trades.  As far as we are aware, the implication of fees on optimal investor behavior has not previously been studied. 

Due to these (and other) conceptual benefits and risks for AMMs, in this work we postulate (intuitive) axioms for AMMs to follow and deduce their implications for traders and liquidity providers. 
Providing such a mathematical foundation for AMMs permits an exploration of the fundamental properties of these DeFi products; this includes both investigating real-world AMMs and proposing new AMMs that satisfy the appropriate axioms.  As touched on above, studying these AMMs mathematically allows us to consider the implications of different fee structures and, importantly, propose a new fee structure for AMMs which is ambivalent to trade execution (i.e., between trading in bulk or splitting a transaction).

\subsection{Literature review}\label{sec:intro-lit}

AMMs for cryptocurrencies and other digital assets have existed since, at least, 2018 with the launch of Uniswap.  As summarized by the defining whitepapers~\cite{uniswapv1,uniswapv2}, this initial AMM follows a constant product rule for swapping assets.  Briefly, and explained more in depth within Section~\ref{sec:examples-real} and Appendix~\ref{sec:uniswapv2}, if the pool holds two types of assets then the product of the reserves of those assets must be the same before and after any swap is realized. We refer the interested reader to, e.g.,~\cite{lehar2021decentralized} for a thorough comparison of the constant product market with a limit order book market.
This notion of keeping the value of a function (e.g., the product) of asset reserves invariant to swaps was later generalized into the idea of \emph{constant function market making}.

The basic construction of constant function market makers was formalized in~\cite{angeris2020improved}.  These structures were studied for use with making foreign exchange markets for digital assets in~\cite{lipton2021automated}.
Typically for simplicity, and as is taken within this work, these pools are presented for markets with two assets only.  In~\cite{angeris2021constant}, trading in multi-asset pools was considered.  Alternatively, trading against multiple pools was presented within~\cite{engel2021composing}.
\cite{bartoletti2021theory} provides a generalized structure for the interactions that can occur between investors and the pool.
We refer the interested reader to~\cite{xu2021sok} for a summary of terminology and structures that are currently used in practice in this field. \cite{angeris2020does} studies specific widely-used AMM structures to investigate the implications that this parameterized AMM has on price stability to determine scenarios in which different structures may be most appropriate. 

The profitability of AMMs in an economic framework with different investor classes was studied in~\cite{capponi2021adoption}.
The risks, and the appropriate hedging of those risks, for specific AMM structures have been studied in numerous works; we highlight the study of the divergence loss of Uniswap V2 \cite{aigner2021uniswap} and Uniswap V3 \cite{deng2022static}.  (These constant product market makers were studied in a number of other works as well, e.g.,~\cite{clark2020replicating,cartea2022decentralised}.)
The costs of being a liquidity provider were further analyzed within~\cite{milionis2022automated}.
Conversely, \cite{angeris2021replicating} proposes a method to construct AMMs which replicate the payoff of a financial derivative allowing for the study of derivative pricing to inform AMM valuation.

In a separate context, AMMs for prediction markets were first proposed in~\cite{hanson2007logarithmic}.  Such a structure is fundamentally different from the constant function market makers considered herein insofar as a prediction market includes a terminal time at which bets are realized. As summarized above, constant function market making has no terminal time but rather is focused merely on the spot market between two (or more) assets.

Though most prior works on AMMs have focused on specific market structures such as the constant product market maker of Uniswap V2, select literature has considered the generalized AMM construction from a mathematical perspective. 
For instance, \cite{capponi2021adoption} introduces specific sufficient conditions for the constant function construction for the results provided in that work. 
Similarly to this work, \cite{ferreira2022credible,schlegel2022axioms} explicitly propose axioms for AMMs as well though the former only loosely imposes conditions and the latter focuses on the relation to prediction markets.
Additionally, \cite{angeris2023geometry} provides geometric axioms for AMMs in a similar vein to those taken herein.
We refer the interested reader to Appendix~\ref{sec:comparison} for a detailed overview of the axioms assumed in these other works and how they compare with the results presented herein.

\subsection{Primary contributions}\label{sec:intro-contributions}
In light of the aforementioned growth and contraction of DeFi, and keeping the specific motivations provided within Section~\ref{sec:intro-motivation} in mind, the primary contributions of this paper are as follows.
\begin{itemize}
\item We construct an \textbf{axiomatic definition for AMMs} and characterize the economic implications of these various axioms on the swap amount and pricing oracle within Section~\ref{sec:amm}.  As highlighted in Section~\ref{sec:intro-lit}, prior studies of AMMs have imposed strict requirements on the structure of the AMM for mathematical simplicity. As discussed in Section~\ref{sec:examples-real}, some of these strict mathematical structures are \emph{not} satisfied by many widely used AMMs, whereas the axioms proposed herein are satisfied.
\item Beyond providing axioms and properties for AMMs, we consider (as far as the authors are aware) a novel \textbf{fee structure} for an AMM in Section~\ref{sec:fees}.  This framework, e.g., imposes a logical indifference to trade execution which is lost by other constructions presented within the literature.  Along with the construction of fees, we provide a mathematical description and consideration of the divergence loss -- also called impermanent loss -- which defines the risk a liquidity provider assumes by pooling with the AMM.
\item Of particular use for practitioners, we mathematically \textbf{characterize widely-used real-world AMMs} in Section~\ref{sec:examples} and expanded upon in Appendix~\ref{sec:current-amm}.  By characterizing these AMMs, we are able to demonstrate which axioms are satisfied and violated by these structures.  We then propose some new AMM constructions which extend real-world AMM constructions in novel ways satisfying the desired mathematical and financial properties discussed within this work. 
\end{itemize}

\section{Market design}\label{sec:design}
When creating a new financial market, there are certain properties that are desirable for the efficient use of liquidity. As the purpose of this work is to study markets made by AMMs, we will first review the financial properties we deem desirable for these novel markets.  Specifically, we consider two types of properties: those that are are desirable for any market structure (e.g., for traditional limit-order books) and those that are specific to AMMs. 

First, as discussed in, e.g.,~\cite{angeris2020improved,li2013axiomatic,adams2023costs}, any traditional market (i.e., based on a limit-order book) should satisfy the following properties:
\paragraph{(No Arbitrage)}\label{NA}\hangindent=\parindent~ Buying and, immediately, selling (resp.\ selling and, immediately, purchasing) assets results in no more value at the end than at the beginning of the transaction. As a consequence, e.g., no amount of assets are recovered when no cash is paid (resp.\ no cash is paid when 0 assets are sold).
\paragraph{(Path Independent)}\label{SPI}\hangindent=\parindent~ Neglecting fees, the number of assets purchased in a single large trade is equal to the sum total from placing multiple small trades in rapid succession (that sum to the large transaction) (resp.\ the total cash recovered from a large liquidation is bounded by that which is recovered from small transactions).\footnote{This property was independently defined in \cite{frongillo2023axiomatic} as \emph{PathIndependence}.}\par
\paragraph{(No Wasted Liquidity)}\label{SNW}\hangindent=\parindent~ The entire market liquidity can be purchased for a, potentially infinite, cost (resp.\ recovered from an infinite-sized sale).\footnote{For a traditional market, this corresponds to a sufficiently wealthy trader's ability to purchase the entire book.}
\paragraph{(Positive Marginal Return)}\label{SSI}\hangindent=\parindent~ More assets are purchased with a larger expenditure of cash (resp.\ more cash is recovered when more assets are sold).\footnote{This property was independently defined in \cite{frongillo2023axiomatic} as \emph{NoDominatedTrades}.}
\paragraph{(Decreasing Marginal Return)}\label{SCC}\hangindent=\parindent~ Market prices are nondecreasing in the number of units purchased (resp.\ market prices are nonincreasing in the number of units sold). 

In order to consider the properties specific for AMMs, we need to introduce a few notions of decentralized exchanges which will be more formally defined below. AMM pools function by holding reserves of liquidity in all of the assets that it trades. One of these assets, typically, acts as the num\'eraire (``cash'') against which all prices are quoted. The pool uses these reserves in order to execute swaps desired by external traders. The pool quotes prices on these assets in such a way so as to be consistent with the swaps that it permits, i.e., via the cost of a marginal swap.
With these ideas, an AMM should satisfy the following properties so that the designed market is usable and follows economic logic:
\paragraph{(Infinite Liquidity)}\label{SIL}\hangindent=\parindent~ The market can act as a counterparty for any trade, i.e., assets are always available for purchase or sale.\footnote{This property was independently defined in \cite{frongillo2023axiomatic} as \emph{BoundedReserves}. We also wish to note that this is similar to \emph{Liquidation} as defined therein except that that property assumes that the market maker can pay out infinite amounts of any asset for a high enough price.} 
\paragraph{(Monotone in Liquidity)}\label{ML}\hangindent=\parindent~ As the liquidity of a single asset increases, while that of all others remains unchanged, its price drops. Conversely, as the liquidity of only the num\'eraire asset increases, the price of all other assets grow. 
\paragraph{(Pooling Increases Liquidity)}\label{PL}\hangindent=\parindent~ Larger asset reserves in the pool, starting from the same initial price, lead to lower execution costs for buying and a higher recovery value for selling assets.

In the following section, we will formally introduce the mathematical structure utilized by AMMs and introduce a number of axioms on that framework in order to guarantee the aforementioned properties. The relation between these axioms and desired properties is summarized within Table~\ref{table:properties} of Appendix~\ref{sec:summary} (with axioms defined in Definition~\ref{defn:amm}). A simple review of this summary table provides a clear indication that some properties are unnecessary to guarantee any of these desirable market properties and, thus, may not be needed in practice.

\section{Constant function market maker}\label{sec:amm}
Constant function market makers are the most prevalent AMMs that exist in practice; for instance, the most prominent AMM -- Uniswap -- is a constant function market maker.  Due to this prominence, we will often equate the concepts of AMMs and constant function market makers within this work.
Fundamentally, a constant function market maker is a multivariate utility function which codifies the value placed on a portfolio by the liquidity providers.  This utility function has the dual task of providing liquidity to the market via swaps and acting as a pricing oracle to quote spot prices on the assets. The construction and axioms of such a utility function are provided within Section~\ref{sec:utility}. For this reason, within this work, we will equate the AMM smart contract with its underlying utility function.
For simplicity, we will assume the pool of asset reserves only permits swaps between two assets $A$ and $B$, although similar axioms can be developed for multi-asset swaps as well.
Simply put, and as expressed within e.g.~\cite{angeris2020improved}, a trader exchanging quantity $x$ of asset $A$ with a constant function market maker $u$ receives quantity $\Y(x)$ of asset $B$ such that $u(a+x,b-\Y(x)) = u(a,b)$ where the market maker has initial reserves $(a,b)$ in the two assets; this is explored within Section~\ref{sec:swap}.  The pricing oracle provides exactly the swap amounts for a marginally small trade, i.e., $\Y'(0)$; this is explored in depth within Section~\ref{sec:price}. 
Throughout these discussions, we relate the provided axioms of AMM utility functions to the properties of a well-designed market as provided in Section~\ref{sec:design}.
Recall that the proofs for all results are provided within Appendix~\ref{sec:proofs}.

\subsection{Axioms}\label{sec:utility}

As expressed in the introduction of this section, an AMM is a multivariate (herein always taken to be bivariate) utility function which codifies the value placed on a portfolio by the liquidity providers.  This utility function defines the price impacts of swaps and acts as a pricing oracle to quote spot prices on the assets.  Within the following definition, we encode the axioms on these AMMs we impose at various points within this work. 
To correspond more closely to the typical utility theory (see, e.g.,~\cite{von1947theory}), we note that the functions provided for AMMs in practice (see Section~\ref{sec:examples-real} and Appendix~\ref{sec:current-amm}) are often characterized as the exponential of the formulations given herein. We will clarify this point in Example~\ref{ex:uniswap} below. By formulating AMMs in this way, we are able to directly interpret the provided axioms in the same way as taken in utility theory.
To simplify notation, throughout we use subscripts to denote partial derivatives; in line with the nomenclature for our two assets, subscript $A$ (respectively $B$) denotes the partial derivative w.r.t.\ the first (second) input.
\begin{definition}\label{defn:amm}
An AMM is a utility function  $u: \R^2_+ \to \R \cup \{-\infty\}$ that may satisfy the following properties:
\begin{enumerate}
\labitem{(UfB)}{normalized} \textbf{\underline{Unbounded from below}:} $u(x,0) = u(0,y) = -\infty$ for every $x,y \geq 0$ and $u(z) > -\infty$ for every $z \in \R^2_{++}$.
\labitem{(UfA)}{inf} \textbf{\underline{Unbounded from above}:} $\lim_{\bar x \to \infty} u(\bar x,y) = \lim_{\bar y \to \infty} u(x,\bar y) = \infty$ for every $x,y > 0$.
\labitem{(SM)}{strict_monotonic} \textbf{\underline{Strictly monotonic}:} $u(z) > u(\bar z)$ if $z - \bar z \in \R^2_+ \backslash \{0\}$ for $z,\bar z \in \R^2_{++}$.
\labitem{(C)}{cont} \textbf{\underline{Continuous}:} $u$ is continuous.
\labitem{(QC)}{convex} \textbf{\underline{Quasiconcave}:} $u$ is quasiconcave.
\labitem{(SI)}{scale} \textbf{\underline{Scale invariant}:} if $u(z) \geq u(\bar z)$ for $z,\bar z \in \R^2_+$ then $u(tz) \geq u(t\bar z)$ for any $t \geq 0$.
\labitem{(I+)}{inada} \textbf{\underline{Inada+}:} $u$ is differentiable with: 
    \begin{itemize}
    \item $\lim_{\bar x \to \infty} u_A(\bar x,y) = \lim_{\bar y \to \infty} u_B(x,\bar y) = 0$, 
    \item $\lim_{\bar x \to \infty} u_B(\bar x,y) \in (0,\infty)$, $\lim_{\bar y \to \infty} u_A(x,\bar y) \in (0,\infty)$, 
    \item $\lim_{\bar x \to 0} u_A(\bar x,y) = \lim_{\bar y \to 0} u_B(x,\bar y) = \infty$,
    \item $\lim_{\bar x \to 0} u_B(\bar x,y) < \infty$, $\lim_{\bar y \to 0} u_A(x,\bar y) < \infty$
    \end{itemize}
    for every $x,y > 0$.
\labitem{(SC)}{deriv} \textbf{\underline{Single crossing}:} $u$  is twice continuously differentiable with $u_A(z),u_B(z) > 0$, $u_B(z)u_{AA}(z) \leq u_A(z)u_{AB}(z)$, and $u_A(z)u_{BB}(z) \leq u_B(z)u_{AB}(z)$ for every $z \in \R^2_{++}$. 
\end{enumerate}
\end{definition}

\begin{example}\label{ex:uniswap}
The most commonly reported AMM is the constant product market maker with function $F(x,y) = xy$ which is utilized by, e.g., Uniswap V2. As noted above, to correspond more closely with the literature on utility functions, herein we formalize Uniswap V2 as the logarithmic utility $u(x,y) := \log(F(x,y)) = \log(x) + \log(y)$ instead.  We expand on this AMM in Section~\ref{sec:examples-real} and Appendix~\ref{sec:uniswapv2}. Notably, and as can be trivially verified, the logarithmic utility function satisfies all axioms proposed within Definition~\ref{defn:amm}.
\end{example}

\begin{remark}\label{rem:deriv}
We wish to note that some of the axioms of Definition~\ref{defn:amm} can imply others.  
For instance, \ref{deriv} implies \ref{strict_monotonic} and \ref{cont} and, additionally, \ref{convex} if at least one of the defining inequalities is strict; this condition is further implied by~\ref{strict_monotonic} and a monotone differences property ($u_{AA},u_{BB} \leq 0$ and $u_{AB} \geq 0$) that is explicitly given in Assumption 1 of \cite{capponi2021adoption} for AMMs.
Within our discussion of some real-world AMMs (see Section~\ref{sec:examples-real}), we note that this monotone difference condition is not satisfied in many cases.  However, the relevant mathematical properties from monotone differences of an AMM -- as shown in~\cite{capponi2021adoption} -- appear to be satisfied under the weaker \ref{deriv} condition proposed herein. 
A full description of the axioms used within~\cite{capponi2021adoption} (as well as other works) and their relation to those provided within Definition~\ref{defn:amm} is provided in Appendix~\ref{sec:comparison}.
\end{remark}

\begin{remark}
We refer to condition~\ref{inada} as the Inada+ condition as the usual Inada conditions (i.e., $\lim_{\bar x \to 0} u_A(\bar x,y) = \lim_{\bar y \to 0} u_B(x,\bar y) = \infty$ and $\lim_{\bar x \to \infty} u_A(\bar x,y) = \lim_{\bar y \to \infty} u_B(x,\bar y) = 0$ for any $x,y > 0$) are satisfied; in addition to these properties,~\ref{inada} also provides the strict monotonicity of $u(x,y)$ in $y$  as $x \to \infty$ (and vice versa for monotonicity of $x$ as $y \to \infty$) as well as behavior of the derivatives at $0$. 
\end{remark}

\begin{remark}
Theorem~\ref{thm:monotone-reserve}\eqref{thm:mr-monotone} provides the interpretation of~\ref{deriv} as providing a single crossing condition for both $f^A((x,y),\delta) := u(a+\delta+x,b-y)$ and $f^B((x,y),\delta) := u(a+x,b+\delta-y)$ for any $a,b > 0$. Specifically, for any feasible $(x',y') \geq (x'',y'')$ and $\delta' \geq \delta''$,
\begin{align*}
f^A((x',y'),\delta') \geq f^A((x'',y''),\delta') \; \Rightarrow \; f^A((x',y'),\delta'') \geq f^A((x'',y''),\delta'') \\
f^B((x',y'),\delta'') \geq f^B((x'',y''),\delta'') \; \Rightarrow \; f^B((x',y'),\delta') \geq f^B((x'',y''),\delta'). 
\end{align*}
Similarly, the results can be shown for subtracting $x$ and adding $y$.
\end{remark}

The first task of an AMM pool is to act as a liquidity provider to facilitate swaps between the two assets. The size of these swaps is provided by the functions $\Y,\X$.
\begin{definition}\label{defn:Y}
Let $u: \R^2_+ \to \R \cup \{-\infty\}$ be an AMM.  Let $(a,b) \in \R^2_+$ be the size of the reserves of asset $A$ and $B$ within the pool respectively.  The swap values $\Y: \R^3_+ \to \R_+$ and $\X: \R^3_+ \to \R_+$ are defined for any $x,y \geq 0$ as:
\begin{align*}
\Y(x;a,b) &:= \sup\{y \in [0,b] \; | \; u(a+x,b-y) \geq u(a,b)\},\\
\X(y;a,b) &:= \sup\{x \in [0,a] \; | \; u(a-x,b+y) \geq u(a,b)\}.
\end{align*}
Where clear, we will drop the explicit dependence of $\Y,\X$ on the pool sizes $(a,b) \in \R^2_+$.
\end{definition}
\begin{remark}
Though Definition~\ref{defn:Y} differs slightly from the typical form of a constant function market maker (see, e.g.,~\cite{angeris2020improved}) in that the ``constant function'' can be an inequality here.  Provided the equality is attained (see, e.g., the conditions in Lemma~\ref{lemma:amm-liq}\eqref{thm:amm-util-2} below), i.e., $u(a+x,b-\Y(x)) = u(a,b)$, we can recover a notion of the indifference price for the quantity $x$ (see, e.g.,~\cite{carmona2008indifference} but for AMMs). As discussed above, in this way the axioms of Definition~\ref{defn:amm} can be interpreted as in standard utility theory (see, e.g.,~\cite{von1947theory}).
\end{remark}

\begin{example}\label{ex:uniswap-Y}
Consider again the Uniswap V2 logarithmic utility function introduced within Example~\ref{ex:uniswap}. The resulting swap functions $\Y,\X: \R^3_+ \to \R_+$ have analytical forms
$\Y(x;a,b) = \frac{bx}{a+x}$ 
and $\X(y;a,b) = \frac{ay}{b+y}$, for any $x,y \geq 0$ and $a,b > 0$.
\end{example}

The second task of an AMM, encoded by the utility function $u: \R^2_+ \to \R \cup \{-\infty\}$, is to act as a pricing oracle $P: \R^2_{++} \to \R_{++}$.  Throughout this work we will consider the price $P$ of asset $A$ denominated in units of asset $B$.  For $B$ in terms of $A$, the reciprocal $1/P$ is taken instead; further considerations of the change of num\'eraire and a bid-ask spread are presented within Sections~\ref{sec:price} and~\ref{sec:fees}.  For the purposes of the definition of the pricing oracle, we will assume that the swap function $\Y$ is differentiable at $(0;a,b)$; we will use the above axioms on the utility $u$ to guarantee this derivative exists within Section~\ref{sec:swap} below.
\begin{definition}\label{defn:P}
The pricing oracle $P: \R^2_{++} \to \R_{++}$ provides the marginal units of asset $B$ obtained from selling a marginal number of units of asset $A$, i.e., $P(a,b) := \Y'(0;a,b)$ for any $a,b > 0$. 
\end{definition}

\begin{example}\label{ex:uniswap-P}
Consider again the Uniswap V2 logarithmic utility function introduced within Example~\ref{ex:uniswap}. The resulting pricing oracle $P: \R^2_{++} \to \R_{++}$ is provided by the ratio of the reserves in the pool, i.e., $P(a,b) = b/a$ for any $a,b > 0$.
\end{example}

\subsection{Swaps}\label{sec:swap}
Herein we wish to study the swap values $\Y,\X$ when transferring between the assets in a pool using an AMM.  To reduce redundancy, throughout this section, we will focus on the properties of the swap of units of $A$ for $B$ only, i.e., properties of $\Y$ as given in Definition~\ref{defn:Y}.  By symmetry of the AMM, comparable results can be provided for the swap of units of $B$ for $A$ (i.e., $\X$).

First, we will consider whether the AMM and the current pool reserves can satisfy the constant function framework, i.e., $u(a,b) = u(a+x,b-\Y(x))$. As highlighted above, this constant function framework is how the swap amounts $\Y$ are normally defined (see, e.g.,~\cite{angeris2020improved}) even though a solution to the constant function need not exist nor be unique for general utility functions $u$. 
For instance, we highlight mStable (details provided in Appendix~\ref{sec:mstable}) for which no feasible solution exists for a trade of size $x > b$.
Notably, this constant function structure follows trivially when the AMM satisfies \nameref{SIL} so that the pool never depletes its reserves. As provided in the following lemma, an AMM satisfying \ref{normalized} and \ref{cont} is guaranteed to generate a market with \nameref{SIL}.
\begin{lemma}\label{lemma:amm-liq}
Consider an AMM $u: \R^2_+ \to \R \cup \{-\infty\}$.  Fix the initial pool $a,b > 0$ and swap amount $x \geq 0$.
\begin{enumerate}
\item\label{thm:amm-util} If~\ref{cont} then $u(a+x,b-\Y(x)) \geq u(a,b)$, i.e., market utility never drops.
\item\label{thm:amm-util-2} If~\ref{cont} and $\Y(x) \neq b$ then $u(a+x,b-\Y(x)) = u(a,b)$, i.e., the constant function market maker structure is satisfied.
\item\label{thm:amm-liquidity} \textbf{\nameref{SIL}:} If~\ref{normalized} and~\ref{cont} then $\Y(x) < b$.
\end{enumerate}
\end{lemma}
Having determined how the properties of the AMM utility relates to those of the swapped amount $\Y$, we want to consider fundamental properties of financial markets. Within the subsequent lemma, we consider sufficient conditions for the resulting market to satisfy~\nameref{SSI}, i.e., the price of assets never drops below 0. Furthermore, in related results, we also find that a simple no-arbitrage condition ($\Y(0) = 0$) so that no assets are recovered if the trader pays nothing. Finally, augmenting also Lemma~\ref{lemma:amm-liq}\eqref{thm:amm-liquidity}, sufficient conditions are provided so that the pool has access to all of its asset reserves if a sufficiently large trade is undertaken, i.e.,~\nameref{SNW}.
\begin{lemma}\label{lemma:amm-monotone}
Consider an AMM $u: \R^2_+ \to \R \cup \{-\infty\}$.  Fix the initial pool $a,b > 0$ and swap amount $x \geq 0$.
\begin{enumerate}
\item\label{thm:amm-zero} If~\ref{strict_monotonic} then $\Y(0) = 0$, i.e., no assets are recovered when no payment is made.
\item\label{thm:amm-monotone} \textbf{\nameref{SSI}:} If~\ref{strict_monotonic} then $\Y$ is nondecreasing in $x$. If, additionally,~\ref{normalized} and~\ref{cont} then $\Y$ is strictly increasing in $x$. 
\item\label{thm:amm-limit-x} \textbf{\nameref{SNW}:} If~\ref{inf},~\ref{strict_monotonic}, and~\ref{cont} then $\lim_{x \to \infty} \Y(x) = b$. 
\end{enumerate}
\end{lemma}
In addition to~\nameref{SSI}, financial markets should have~\nameref{SCC}, i.e., increasing costs as more assets are purchased. This notion can be mathematically encoded by concavity of $\Y$. Sufficient conditions for such a result are provided in the subsequent lemma. 
In addition, by applying~\ref{scale}, a relation between the pool sizes and trade sizes can be provided; specifically, the recovered assets grow in proportion to, jointly, the trade size and asset reserves in the pool. This simplifying structure matches what occurs on, e.g., a limit-order book when all trades in the book are scaled proportionally.
\begin{lemma}\label{lemma:amm-concave}
Consider an AMM $u: \R^2_+ \to \R \cup \{-\infty\}$.  Fix the initial pool $a,b > 0$ and swap amount $x \geq 0$.
\begin{enumerate}
\item\label{thm:amm-usc} If~\ref{cont} then $\Y$ is upper semicontinuous.
\item\label{thm:amm-concave} \textbf{\nameref{SCC}:} If~\ref{cont} and~\ref{convex} or~\ref{deriv} then $\Y$ is concave, continuous and a.e.\ differentiable.
\item\label{thm:amm-ph} If~\ref{cont} and~\ref{scale} then $\Y$ is positive homogeneous in $(x;a,b)$, i.e., $\Y(tx;ta,tb) = t\Y(x;a,b)$ for any $t > 0$. 
\item\label{thm:amm-subadditive} If~\ref{strict_monotonic},~\ref{cont}, and~\ref{convex} or~\ref{deriv} then $\Y$ is subadditive, i.e., the number of assets purchased in a single large trade is no larger than sum total from placing multiple small trades (that sum to the large transaction) when all such trades are assessed at the current market snapshot.
\end{enumerate}
\end{lemma}
\begin{remark}\label{rem:subadditive}
Subadditivity of $\Y$ (Lemma~\ref{lemma:amm-concave}\eqref{thm:amm-subadditive}) differs from~\nameref{SPI} in that the former property assumes that, for all trades, the market conditions (i.e., pool reserves) are frozen at the current state; in the latter property, the market conditions evolve as swaps are actualized.
\end{remark}
\begin{remark}
Assume~\ref{cont} and~\ref{convex} or~\ref{deriv}. 
It follows from Lemma~\ref{lemma:amm-concave}\eqref{thm:amm-concave} that $\Y$ and $\X$ are concave. Therefore, sub-differentials of $\Y$ and $\X$ are always well defined, and will a.s.\ coincide with the derivative. This allows us to give the following definition.
\end{remark}

\begin{definition}
Denote $\Y'(x;a,b) := \lim\limits_{\eps\searrow 0} \frac{\Y(x+\eps;a,b)-\Y(x;a,b)}{\eps}$, and similarly for $\X'(x;a,b)$ if they exist. 
If the AMM is differentiable (with nonzero partial derivatives) then $\Y'(x;a,b) = \frac{u_A(a+x,b-\Y(x;a,b))}{u_B(a+x,b-\Y(x;a,b))}$ by implicit differentiation (similarly for $\X'(x;a,b)$).
Furthermore, assuming differentiability of the AMM, denote $\Y_a(x;a,b) = \lim\limits_{\eps \searrow 0} \frac{\Y(x;a+\eps,b)-\Y(x;a,b)}{\eps}$, and similarly for $\Y_b(x;a,b),\X_a(x;a,b),\X_b(x;a,b)$, which can all be formulated via implicit differentiation as well.
\end{definition}

We conclude our discussion of the properties of swaps without fees by considering a simple strategy to optimally execute transactions. In particular, we consider how a trader may want to split transactions, either into smaller trades or as a round-trip transaction (i.e., potential arbitrage opportunities). As described by \nameref{SPI} and \nameref{NA}, without fees,\footnote{Fees are introduced and studied in Section~\ref{sec:fees} and Appendix~\ref{sec:fee-price-details} below.} a trader should be indifferent in how their swap is executed. Formally, \nameref{SPI} can be viewed as a type of no arbitrage argument for trades going in the same direction whereas \nameref{NA} considers trades that reverse directions.
\begin{lemma}\label{lemma:arbitrage}
Consider an AMM $u: \R^2_+ \to \R \cup \{-\infty\}$.  Fix the initial pool $a,b > 0$ and swap amounts $x_1,x_2 \geq 0$.
\begin{enumerate}
\item\label{thm:amm-split} \textbf{\nameref{SPI}:} If~\ref{normalized},~\ref{strict_monotonic}, and~\ref{cont} then $\Y(x_1+x_2;a,b) = \Y(x_1;a,b) + \Y(x_2;a+x_1,b-\Y(x_1;a,b))$.
\item\label{thm:amm-arbitrage} \textbf{\nameref{NA}:} If~\ref{strict_monotonic} and~\ref{cont} then $\X(\Y(x;a,b);a+x,b-\Y(x;a,b)) \leq x$ and $\Y(\X(y;a,b);a-\X(y;a,b),b+y) \leq y$ for any $x,y \geq 0$ and $a,b > 0$.
    If, additionally,~\ref{normalized} then these inequalities hold as equalities for every $x,y \geq 0$ and $a,b > 0$.
\end{enumerate}
\end{lemma}

\begin{remark}
Lemma~\ref{lemma:arbitrage}\eqref{thm:amm-split} implies an investor is indifferent to trade execution, i.e., \nameref{SPI}.  However, as discussed in Remark~\ref{rem:fee-structure} in Appendix~\ref{sec:fee-price-details}, naively assessing fees to transactions can result in this property no longer holding in general.
As discussed in Remark~\ref{rem:subadditive}, Lemma~\ref{lemma:amm-concave}\eqref{thm:amm-subadditive} implies splitting a transaction \emph{with pool recovery} receives a higher value.
\end{remark}

Within the next theorem, we want to study the implications of changing the pool sizes $a,b$.  Specifically, how altering the pool composition changes the value of a swap and the limiting behavior as pool sizes shrink to 0 or grow to infinity.
\begin{theorem}\label{thm:monotone-reserve}
Assume~\ref{normalized} and~\ref{deriv}. 
\begin{enumerate}
\item\label{thm:mr-monotone} \textbf{\nameref{ML}:} $\Y_a(x;a,b) \leq 0$ and $\Y_b(x;a,b) \in [0,1)$ for any $x \geq 0$ and $a,b > 0$. 
\item\label{thm:mr-limit-0} $\lim_{\bar a \searrow 0} \Y(x;\bar a,b) = b$ and $\lim_{\bar b \searrow 0} \Y(x;a,\bar b) = 0$ for every $a,b,x > 0$ with the latter limit converging uniformly in $a,x$, i.e., as asset reserves in $A$ drop to 0 the value of asset $A$ grows to $\infty$ whereas the value of asset $A$ decreases to $0$ when reserves of $B$ drop to 0.
\item\label{thm:mr-limit-inf} If additionally~\ref{convex} and~\ref{inada} then $\lim_{\bar a \nearrow \infty} \Y(x;\bar a,b) = 0$ and $\lim_{\bar b \nearrow \infty} \Y(x;a,\bar b) = \infty$ for every $a,b,x > 0$, i.e., as asset reserves in $A$ grow to $\infty$ the value of asset $A$ drops to $0$ whereas the value of asset $A$ grows to $\infty$ when reserves of $B$ grow to $\infty$.
\end{enumerate}
\end{theorem}

\subsection{Pooling and the pricing oracle}\label{sec:price}
Recall from Definition~\ref{defn:P} that the pricing oracle $P: \R^2_{++} \to \R_{++}$ provides the marginal units of asset $B$ obtained from selling a marginal number of units of asset $A$, i.e., $P(a,b) := \Y'(0;a,b)$ for any $a,b > 0$.
\begin{remark}\label{rem:no-spread}
Following Lemma~\ref{lemma:arbitrage}, if~\ref{strict_monotonic} and the AMM $u$ is differentiable then there does \emph{not} exist a bid-ask spread in the pricing oracle for strictly positive reserves, i.e., $P(a,b) = \Y'(0;a,b) = \frac{1}{\X'(0;a,b)}$ for any $a,b>0$.  
\end{remark}

In the following proposition, we consider simple properties on the pricing oracle. Specifically, though already considered above for the swap function in Section~\ref{sec:swap}, we consider ways in which \nameref{ML} and \nameref{SIL} can be represented with the pricing oracle. Further details of \nameref{SIL} are provided in Remark~\ref{rem:price-sil}.
\begin{proposition}\label{prop:price}
Assume~\ref{deriv}.  
\begin{enumerate}
\item\label{prop:price-monotone} \textbf{\nameref{ML}:} The pricing oracle $P$ is differentiable with $P_A \leq 0$ and $P_B \geq 0$.
\item\label{prop:price-b} If, additionally,~\ref{normalized},~\ref{convex}, and~\ref{inada} then for any $a > 0$, $b \mapsto P(a,b)$ is nondecreasing and surjective on $\R_{++}$.
\item\label{prop:price-a} If, additionally,~\ref{normalized},~\ref{convex}, and~\ref{inada} then for any $b > 0$, $a \mapsto P(a,b)$ is nonincreasing and surjective on $\R_{++}$.
\item\label{prop:price-si} If, additionally,~\ref{scale} then $P$ is scale invariant, i.e., $P(ta,tb) = P(a,b)$ for any reserves $a,b > 0$ with scaling $t > 0$.
\end{enumerate}
\end{proposition}
\begin{remark}\label{rem:price-sil}
\nameref{SIL} implies that that swaps are able to fluctuate prices within $\operatorname{range}P \subseteq \R_{++}$ with $\inf\operatorname{range}P = 0$ and $\sup\operatorname{range}P = \infty$. In contrast, Proposition~\ref{prop:price}\eqref{prop:price-b} and \eqref{prop:price-a} imply it is possible to trace the entire price curve $\R_{++}$ by manipulating the reserves in each asset independently.
\end{remark}

As opposed to more traditional market makers, an AMM allows investors to join the liquidity pool and capture a fraction of the profits (see Section~\ref{sec:fees} and Appendix~\ref{sec:fee-price-details} below for a discussion of fee structures).
As presented in, e.g., \cite{angeris2021constant}, pooling should be accomplished so that the price is unaffected by the injection of additional liquidity  into the market.  That is, adding $(\alpha,\beta) \in \R^2_+ \backslash \{0\}$ to a pool of size $(a,b) \in \R^2_{++}$ so that $P(a+\alpha,b+\beta) = P(a,b)$.  
Within the following theorem, we consider a sufficient condition on the pricing oracle so that increasing the size of the reserves actually increases market liquidity \nameref{PL}, i.e., pooling $(\alpha,\beta) \in \R^2_+ \backslash \{0\}$ results in
\begin{align}
\label{eq:pool-Y} \Y(x;a+\alpha,b+\beta) &\geq \Y(x;a,b), 
\quad
\X(y;a+\alpha,b+\beta) \geq \X(y;a,b), \quad \forall x,y \geq 0.
\end{align}
\begin{theorem}\label{thm:pooling} 
\textbf{\nameref{PL}:} Assume~\ref{deriv}. Pooling (i.e., injecting $(\alpha,\beta) \in \R^2_+ \backslash \{0\}$ to the initial pool $(a,b) \in \R^2_{++}$ such that $P(a+\alpha,b+\beta) = P(a,b)$) increases market liquidity as defined in~\eqref{eq:pool-Y} if either: 
(1) the AMM satisfies \ref{scale}; or
(2) the utility function $u: \R^2_+ \to \R \cup \{-\infty\}$ is thrice continuously differentiable, \ref{normalized}, \ref{convex}, \ref{inada}, and
\begin{equation}\label{eq:P-cond} 
P_B(z)P_{AA}(z) - (P(z) P_B(z) + P_A(z))P_{AB}(z) + P(z) P_A(z) P_{BB}(z) \geq 0,  \quad \forall z \in \R^2_{++}.
\end{equation}
\end{theorem}

\begin{remark}\label{rem:pooling}
Maintaining a constant price is a necessary condition for increased liquidity as $P(a,b) = \Y'(0;a,b) = \frac{1}{\X'(0;a,b)}$.  Under~\ref{scale}, due to Proposition~\ref{prop:price}\eqref{prop:price-si}, this reduces to the simpler ``proportional to current reserves'' rule-set that is often assumed instead (see, e.g., \cite{capponi2021adoption}).  
\end{remark}

\begin{remark}\label{rem:P-cond}
As far as the authors are aware, there is no meaningful financial condition that implies \eqref{eq:P-cond}. This is merely a mathematical argument on the pricing oracle $P$ that provides sufficient conditions for the desired increasing liquidity property of~\eqref{eq:pool-Y}.
As highlighted in the proof of Theorem~\ref{thm:pooling} in Appendix~\ref{sec:proofs-pooling}, \eqref{eq:P-cond} arises so as to guarantee that the number of units $\alpha \mapsto \beta(\alpha)$ of $B$ being pooled is \emph{more} sensitive to the number of units $\alpha$ of $A$ being pooled before any swap $(x,\Y(x))$ is actualized on the pool (with decreasing sensitivity in the size of the swap $x \geq 0$).
As highlighted in Section~\ref{sec:examples} below, all AMMs studied within this work satisfy this condition, though as demonstrated in Section~\ref{sec:examples-sum} this curvature condition is independent of the other axioms on AMMs.

Inspired by Proposition~\ref{prop:sdamm} below, it can be shown that \eqref{eq:P-cond} follows from \ref{deriv} along with $u_A,u_B$ log-convex (with $u$ thrice differentiable) and $u_{AB} \geq 0$; we wish to recall from Remark~\ref{rem:deriv} that Assumption 1 of \cite{capponi2021adoption} imposes the second derivative condition $u_{AB} \geq 0$ though this stronger condition is not satisfied for every AMM utilized in practice as highlighted in Section~\ref{sec:examples-real}.
\end{remark}

\begin{example}\label{ex:uniswap-pooling}
Consider again the Uniswap V2 logarithmic utility function introduced within Example~\ref{ex:uniswap}. As this utility is such that $u_A(a,b) = 1/a,~u_B(a,b) = 1/b$ are log-convex and $u_{AB} = 0$ is non-negative, as provided in Remark~\ref{rem:P-cond}, Uniswap V2 satisfies \eqref{eq:P-cond}. We wish to note that the product formulation of this AMM is such that $F_A(a,b) = b$ and $F_B(a,b) = a$ are \emph{not} log-convex; even though \eqref{eq:P-cond} is invariant to monotonic transformations of the utility function, the use of the logarithmic transformation can assist in verifying these essential properties. 
\end{example}

We conclude this section on pooling and the impacts of liquidity by quantifying a measure of price impacts. We consider this in three parts: first, we determine the effect that pooling has on price impacts from executing a swap; second, we define a (approximating) price impact oracle that only depends on the current pool reserves; finally, we find conditions so that this price impact oracle can be used to bound the actualized price impacts from any swap. In studying this measure we are able to compare AMM constructions as well as measure the level of liquidity impacts.
\begin{corollary}\label{cor:impacts}
Define the price impacts for swapping assets $A$ for $B$ and vice versa, respectively, as:
$\Imp_\Y(x;a,b) = P(a,b)x - \Y(x;a,b),~\Imp_\X(y;a,b) = y/P(a,b) - \X(y;a,b),~(a,b)\in \R^2_{++},~x,y\ge0.$
Under the conditions of Theorem \ref{thm:pooling}, 
pooling decreases the price impact from swapping. In other words, for any $x,y \geq 0$, pooling additional liquidity $(\alpha,\beta) \in \R^2_+ \backslash \{0\}$ to the initial pool $(a,b) \in \R^2_{++}$ such that $P(a+\alpha,b+\beta) = P(a,b)$ decreases the price impacts
\begin{align}
\Imp_\Y(x;a,b) \ge \Imp_\Y(x;a+\alpha,b+\beta),~\Imp_\X(x;a,b) \ge \Imp_\X(x;a+\alpha,b+\beta).
\label{eq:impct1}
\end{align}
\end{corollary}
Corollary \ref{cor:impacts} introduce the new notations $\Imp_\X, \Imp_\Y$ of price impact from swapping that quantifies this quantity. The key result shows that this impact decreases as liquidity is added to the pool. This is of course a very desirable conclusion---recall \nameref{PL} from the market design list---as this implies that the more liquid the market is, the closer to linear the price is even for large transactions. Note that the linear part $P(a,b)x$ (and $y/P(a,b)$ respectively) is invariant to additional liquidity.

\begin{figure}[h!]
\centering
\begin{subfigure}[t]{0.3\textwidth}
\centering
\includegraphics[width=\textwidth]{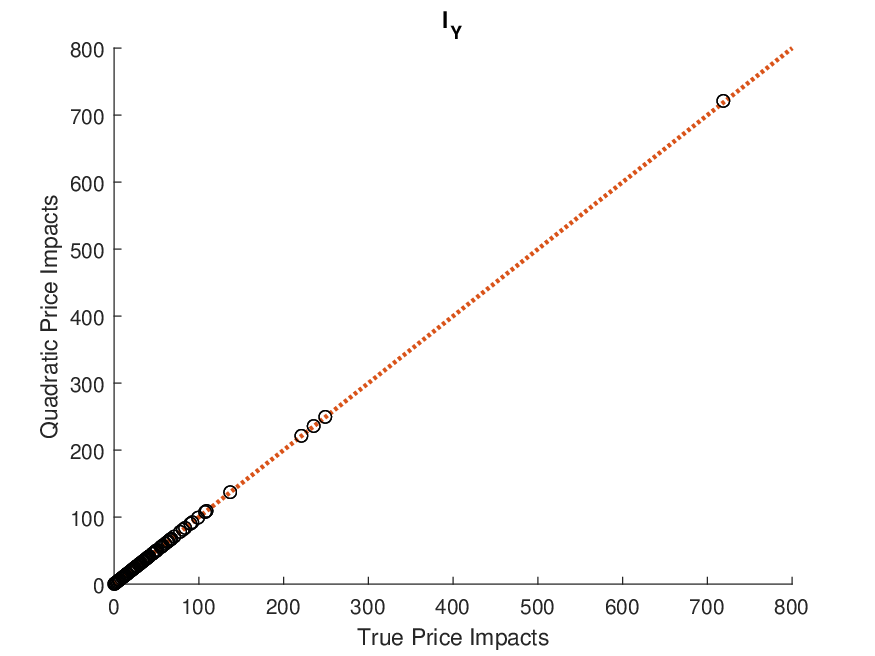}
\caption{Graph of quadratic error term $\Imp(a,b)x^2$ vs.\ the true price impact $\Imp_\Y$.}
\label{fig:Impct-IY}
\end{subfigure}
~
\begin{subfigure}[t]{0.3\textwidth}
\centering
\includegraphics[width=\textwidth]{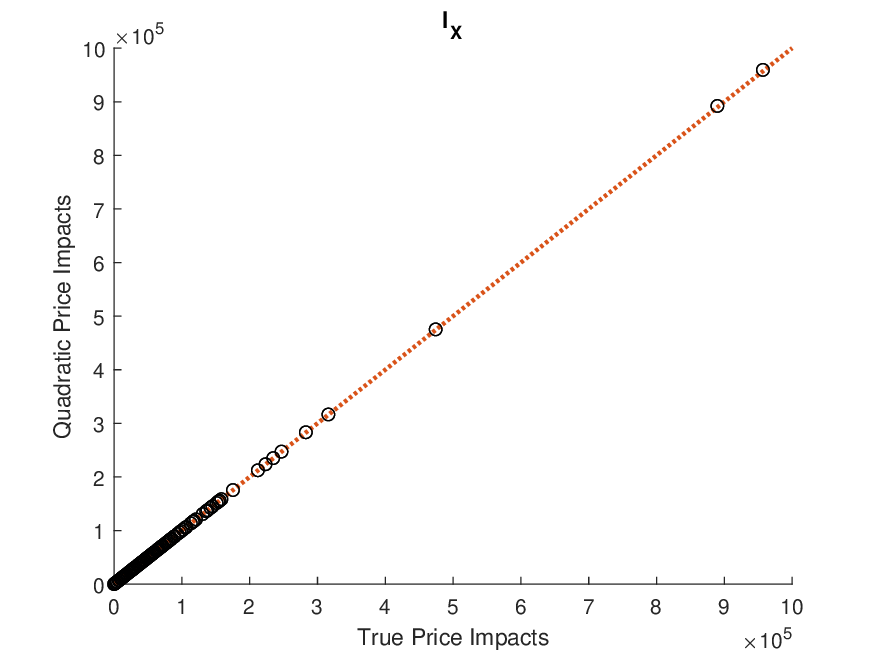}
\caption{Graph of quadratic error term $\frac{\Imp(a,b)}{P(a,b)^3}y^2$ vs.\ the true price impact $\Imp_\X$.}
\label{fig:Impct-IX}
\end{subfigure}
~
\begin{subfigure}[t]{0.3\textwidth}
\centering
\includegraphics[width=\textwidth]{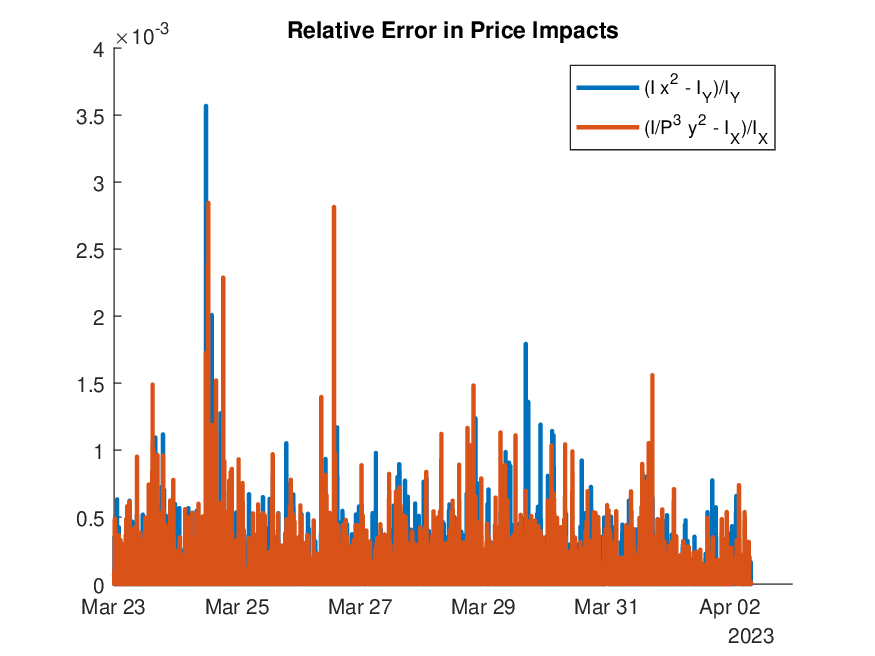}
\caption{Graph of the relative error
 $\frac{\Imp(a,b)x^2 - \Imp_\Y(x;a,b)}{\Imp_\Y(x;a,b)}$ and  $\frac{\frac{\Imp(a,b)}{P(a,b)^3}y^2 - \Imp_\Y(x;a,b)}{\Imp_\Y(x;a,b)}$ between March 23 and April 02, 2023.}
\label{fig:Impct-err}
\end{subfigure}
\caption{Comparison of the approximation of the price impact and the true price impact using Uniswap data between March 23 and April 02, 2023 from a USDC/WETH pool on the Polygon blockchain (smart contract 0x45dda9cb7c25131df268515131f647d726f50608).}
\label{fig:Impct}
\end{figure}

Though $\Imp_\X,\Imp_\Y$ capture the full price impacts from swapping, these functions explicitly depend on the size of the swap. We now wish to introduce a novel price impact oracle which is independent to the swap size $x,y$.
\begin{definition}\label{defn:impacts}
The price impact oracle $\Imp: \R^2_{++} \to \R$ provides (a multiplier of) the marginal change in the pricing oracle from selling a marginal number of units of asset $A$, i.e., $\Imp(z) := -\frac{1}{2}\Y''(0;z) = \frac{1}{2}[P(z) P_B(z) - P_A(z)]$ for any $z \in \R^2_{++}$. 
\end{definition}
\begin{proposition}\label{prop:quadratic-approx}
Consider a thrice continuously differentiable AMM $u: \R^2_+ \to \R \cup \{-\infty\}$.
Fix the pool reserves $(a,b) \in \R^2_{++}$, then the price impacts from swapping can be approximated via the price impact oracle for any $x,y \geq 0$ as: 
\begin{align}
\Imp_\Y(x;a,b) &= \Imp(a,b)x^2 + O\left(\left(\frac{x}{a}\right)^3\right), \quad
\Imp_\X(y;a,b) = \frac{\Imp(a,b)}{(P(a,b))^3}y^2 + O\left(\left(\frac{y}{b}\right)^3\right).
\end{align}
Furthermore, pooling decreases the price impact oracle (i.e., $\Imp(a+\alpha,b+\beta) \leq \Imp(a,b)$ when pooling ($\alpha,\beta) \in \R^2_+ \backslash \{0\}$ into the pool with reserves $(a,b)$) if either of the conditions of Theorem \ref{thm:pooling} are satisfied. Moreover, if \ref{deriv} and \ref{scale} then the price impact oracle is positive homogeneous of degree $-1$ (i.e., $\Imp(ta,tb) = t^{-1} \Imp(a,b)$ for any $t > 0$).
\end{proposition}
Proposition \ref{prop:quadratic-approx} formally presents how the price impact oracle approximates the actualized price impacts, i.e., as a \emph{quadratic} approximation. If trade sizes are sufficiently large then the identified error term will become dominant. However, empirical analysis of a Uniswap pool finds that trade sizes remain sufficiently small (in relation to the pool size) so that the price impacts can be accurately captured by the quadratic term in practice.
Specifically, Figures \ref{fig:Impct-IY} and \ref{fig:Impct-IX} show that, empirically on data obtained from a Uniswap pool, the price impacts $\Imp_\Y(x;a,b),\Imp_\X(y;a,b)$ are (almost) entirely captured by the quadratic terms $\Imp(a,b)x^2,~\frac{\Imp(a,b)}{P(a,b)^3} y^2$, the coefficients of which are independent of the trade size and only depend on the current market liquidity. 
Furthermore, Figure \ref{fig:Impct-err} shows that even the relative errors for these quadratic approximations are empirically very small. Thus we conclude that $\Imp(a,b)$ is a highly effective empirical measure for price impact.

Finally, to further quantify the accuracy of the price impact oracle, we further investigate its use as an explicit bound on the price impacts in the following lemma.
\begin{corollary}\label{cor:quadratic-approx}
Consider a thrice continuously differentiable AMM $u: \R^2_+ \to \R \cup \{-\infty\}$.
Assume \ref{deriv} holds, and fix the pool $(a,b) \in \R^2_{++}$, 
then:
\begin{enumerate}

\item The pricing oracle bounds the swap amount, i.e., $\Imp_\Y(x;a,b),\Imp_\X(y;a,b) \ge 0$ for any $x,y \geq 0$. Additionally, the price impact oracle is nonnegative, i.e., $\Imp(a,b) \ge 0$.
\item If, additionally, for every $z \in \R^2_{++},$ 
\begin{equation}\label{eq:P-cond-quad}
\Psi(z) := P_{AA}(z) - 2 P(z) P_{AB}(z) + P(z)^2 P_{BB}(z) + (P(z) P_B(z) - P_A(z)) P_B(z) \geq 0, 
\end{equation}
then 
$\Imp_\Y(x;a,b) \le  \Imp(a,b) x^2$ for any $x \geq 0$. 

\item If, additionally, for every $z \in \R^2_{++},$ 
\begin{equation}\label{eq:P-cond-quad-X}
-P(z)\Psi(z) + 3 (P(z) P_B(z) - P_A(z))^2 \geq 0, 
\end{equation}
then 
$\Imp_\X(y;a,b) \leq \frac{\Imp(a,b)}{P(a,b)^3} y^2$ for any $y \geq 0$. 
\end{enumerate}
\end{corollary}

\begin{remark}
Condition~\eqref{eq:P-cond-quad} guarantees that, with a fixed amount of liquidity (i.e., along the utility indifference curve), that the change in prices is convex as a trader swaps $A$ for $B$.
Condition~\eqref{eq:P-cond-quad-X} guarantees the same result when swapping $B$ for $A$. As highlighted in Remark~\ref{rem:riskaverse}, in the special case of a decomposable utility function $u(z) = U(z_1) + U(z_2)$, these conditions are related to the absolute risk aversion of the univariate utility function $U$.
\end{remark}

\begin{remark}
\cite{angeris2020does} introduced a definition for $\mu$-price stability so that no swap $(x,\Y(x))$ moves the price by more than $\mu x$ with implicit dependence on the pool reserves $(a,b) \in \R^2_{++}$. Any AMM satisfying~\eqref{eq:P-cond-quad} is, therefore, $\Imp(a,b)$-price stable. 
Similarly, when considering swaps $(\X(y),y)$, the pricing oracle is $\frac{\Imp(a,b)}{P(a,b)^3}$-price stable if \eqref{eq:P-cond-quad-X} holds.
\end{remark}

As the above results show, the approximation of $\Imp_\Y(x;a,b)$ via $\Imp(a,b)x^2$ (respectively $\Imp_\X(y;a,b)$ via $\Imp(a,b)/(P(a,b)^3)y^2$)  is not only accurate empirically, but also is an upper bound on the price impact encountered by the trader and  is entirely captured by the quadratic term $\Imp(a,b)$. For instance, if this term were to equal $0$ then $\Imp_\Y(x;a,b) = 0$ via the results of Corollary~\ref{cor:quadratic-approx}, i.e., no price impacts would be experienced by the trader.  Additionally, the quantity $\Imp(a,b)$ is independent of the trade sizes $x,y$, which makes the approximation simpler and even more desirable. Therefore, we will use $\Imp(a,b)$ as a measure of price impact.

\begin{example}
Consider again the Uniswap V2 logarithmic utility function introduced within Example~\ref{ex:uniswap}. As the pricing oracle for this AMM is given by $P(a,b) = b/a$, and the price impact oracle $\Imp(a,b)=\frac{b}{a^2}.$ We can directly check conditions \eqref{eq:P-cond-quad} and \eqref{eq:P-cond-quad-X}. In particular, we find that 
$
0 \leq \frac{6z_2}{z_1^3} = \Psi(z)  
\quad \text{ and } \quad 0 \leq \frac{6z_2^2}{z_1^4} = -P(z)\Psi(z) + 3 (P(z) P_B(z) - P_A(z))^2 
$
for any $z \in \R^2_{++}$.
Therefore, following the results of Corollary \ref{cor:quadratic-approx}, $\Imp_\Y(x;a,b) \in [0,\frac{bx^2}{a^2}]$ and $\Imp_\X(y;a,b) \in [0,\frac{ay^2}{b^2} ]$ for any $x,y \geq 0$ and $(a,b) \in \R^2_{++}$. 
\end{example}

\section{Examples of AMMs}\label{sec:examples}
Within this section we wish to explore a number of different AMMs. Within Section~\ref{sec:examples-real}, we will summarize a number of AMMs that exist in practice; the detailed mathematical representations of these AMMs are provided within Appendix~\ref{sec:current-amm}. All such results have been empirically validated through reference to the applicable white papers and documentation. 
A summary of the axioms satisfied by these real-world AMMs is provided within Table~\ref{table:amm}.  
Within Section~\ref{sec:examples-sum}, we construct a novel generalized AMM structure; we use that structure to provide a new concentrated liquidity swap structure which has strong theoretical stability properties.  We wish to remind the reader that many of the AMMs presented below, and detailed in Appendix~\ref{sec:current-amm}, are often characterized as the exponential of the utility functions provided herein; as the definition of the swap values $\X,\Y$ are based on relative utilities, this exponentiation does not impact the implementation, only the description, of the AMM.

\subsection{AMMs in practice}\label{sec:examples-real}
\begin{table}
\centering
\resizebox{\textwidth}{!}{
\begin{tabular}{|c||c||c|c|c|c|c|c|c|c|c|c|c|}
\hline
{\bf AMM} & {\bf Details} & \ref{normalized} & \ref{inf} & \ref{strict_monotonic} & \ref{cont} & \ref{convex} & \ref{scale} & \ref{inada} & \ref{deriv} & \eqref{eq:P-cond} & \eqref{eq:P-cond-quad} & \eqref{eq:P-cond-quad-X}\\ \hline\hline
\emph{Uniswap V2} & \ref{sec:uniswapv2} & X & X & X & X & X & X & X & X & X & X & X \\ \hline
\emph{Balancer} & \ref{sec:balancer} & X & X & X & X & X & X & X & X & X & X & X \\ \hline
\emph{Uniswap V3} & \ref{sec:uniswapv3} & & X & X & X & X & & & X & X & X & X \\ \hline
\emph{mStable} & \ref{sec:mstable} & & X & X & X & X & X & & X & X & X & X \\ \hline
\emph{StableSwap} & \ref{sec:stableswap} & & X & X & X & X & & & X & X & X & X \\ \hline
\emph{L.StableSwap} & \ref{sec:l.stableswap} & X & X & X & X & X & X & X & X & X &  &  \\ \hline
\emph{Curve} & \ref{sec:curve} & X & X & X & X & X$^*$ & X & X & X$^*$ & X$^*$ &  &  \\ \hline
\emph{Dodo} & \ref{sec:dodo} & X & X & X & X & X & X & X & X & \cellcolor{gray} & \cellcolor{gray} & \cellcolor{gray} \\ \hline
\emph{SDAMM} & \ref{sec:examples-sum} & X$^\dagger$ & X$^\dagger$ & X$^\dagger$ & X$^\dagger$ & X$^\dagger$ &  & X$^\dagger$ & X$^\dagger$ & X$^\dagger$ & X$^\dagger$ & X$^\dagger$ \\ \hline
\end{tabular}
}
\caption{Summary of popular AMMs and the axioms satisfied. 
$\dagger$: Under some conditions specified in Proposition \ref{prop:sdamm}.\\
$*$: Axiom verified numerically.}
\label{table:amm}
\end{table}

Herein we summarize a number of AMMs that exist in practice.  The details for all of these AMMs are provided within Appendix~\ref{sec:current-amm}.  We, also, refer the reader to Table~\ref{table:amm} which summarizes the findings about all example AMMs considered within this work.

Fundamentally, most current AMMs are built based on two structures:
\begin{itemize}
\item \emph{Uniswap V2}: As presented in the running example throughout Section \ref{sec:amm} (beginning with Example \ref{ex:uniswap}) and Appendix~\ref{sec:uniswapv2}, Uniswap V2 is the AMM with logarithmic utility function, i.e., $u(x,y) = \log(x) + \log(y)$.  Notably, this simple structure satisfies all axioms presented within this work.  However, Uniswap V2 is often criticized as it is subject to high price impacts as it tends to save a lot of liquidity solely for the tail of the price distribution (so as to guarantee infinite liquidity).
\item \emph{mStable}: As presented in Appendix~\ref{sec:mstable}, mStable is an AMM that has no price impacts at all, i.e., $u(x,y) = \log(x+y)$.  In order to guarantee the constant price of mStable, this AMM fails to satisfy~\ref{normalized} and~\ref{inada}.
\end{itemize}
By combining the notions of these AMMs in different ways, a liquidity provider is able to concentrate liquidity, i.e., lower the price impacts from trades, when the pool is ``balanced'' (i.e., when the asset pools are of comparable size).
\emph{StableSwap} ($u(x,y) = \log(C(x+y) + xy)$ for parameter $C > 0$; see Appendix~\ref{sec:stableswap}) considers a linear combination of Uniswap V2 and mStable \emph{within} the logarithm.  By doing so, there is less price impact than Uniswap V2 but at the expense of infinite liquidity and positive homogeneity (i.e.,~\ref{normalized},~\ref{scale} and~\ref{inada} are not satisfied). Such a construction is especially valuable when pairing stablecoins (i.e., assets which are constructed so as to keep a stable price against a reference instrument).
\emph{Curve} is an extension of StableSwap that satisfies~\ref{normalized} and~\ref{scale} at the expense of \eqref{eq:P-cond-quad}, \eqref{eq:P-cond-quad-X} and is only implicitly defined ($u(x,y) = \log(D(x,y))$ for $D(x,y)^3 + 4(C-1)xyD(x,y) - 4C(x+y)xy = 0$ with parameter $C \geq 1$; see Appendix~\ref{sec:curve}).  To maintain the infinite liquidity and concentrate liquidity around a price of 1, the price impacts in the tails (i.e., when one of the pools is nearly exhausted) become exceedingly high.  
Within Appendix~\ref{sec:l.stableswap}, we introduce a new AMM structure, as the linear combination of UniSwap V2 and mStable, which we call \emph{L.StableSwap} ($u(x,y) = C\log(x+y) + \log(x) + \log(y)$ for parameter $C > 0$).  As with Uniswap V2 and Curve, all fundamental axioms are satisfied for this new construction and it is subject to low price impacts near pool balance ($a \approx b$ for reserves $a,b > 0$). However, in contrast to Uniswap V2 (but similar to Curve), \eqref{eq:P-cond-quad} and \eqref{eq:P-cond-quad-X} are not satisfied for this construction as it does not exhibit price stability in the edges of its reserves.  Notably, in contrast to Curve, L.StableSwap is constructed with a simple analytical utility function. Both Curve and L.StableSwap, due to their low price impacts\footnote{Verified empirically for Curve.}, are able to provide additional liquidity near the theoretical constant price for stablecoin pairs, but still capable of providing some limited liquidity in the tails of the price distribution. Due to the uniform distribution of liquidity in Uniswap V2, that AMM can serve well for asset pairs with high volatility whereas these more stable AMMs are more appropriate for pairs with low volatility.

\begin{remark}\label{rem:p=1}
In each of the aforementioned AMMs, we have described them so that $P(t,t) = 1$ for any $t > 0$. Especially when considering pairs of stablecoins, a different balanced price may be desirable. Consider the desired balanced price $p > 0$, then setting $\bar{u}^p(x,y) := u(px,y)$ satisfies all of the same axioms as $u$ with pricing oracle $\bar{P}^p(x,y) = p P(px,y)$; in particular, $\bar{P}^p(t,pt) = p$ by construction for any $t > 0$. Notably, in this case, the balance occurs not when the pool of asset $A$ and $B$ are equal but rather when their ratio coincides with the desired price $p$.
\end{remark}

Other AMMs are built from the structure of Uniswap V2 directly.  \emph{Balancer} ($u(x,y) = w\log(x) + (1-w)\log(y)$ for parameters $w \in (0,1)$; see Appendix~\ref{sec:balancer}) is merely a weighted version of Uniswap V2 so as to scale the asset pools when pricing.  \emph{Uniswap V3} ($u(x,y) = \log(\alpha + x) + \log(\beta + y)$ for parameters $\alpha,\beta > 0$; see Appendix~\ref{sec:uniswapv3}) introduces ``virtual reserves'' so as to concentrate liquidity and decrease price impacts.  However, the concentrated liquidity introduced in Uniswap V3 comes at the expense of~\ref{normalized},~\ref{scale}, and~\ref{inada} so that, much like StableSwap, it only has finite liquidity and does not scale linearly.\footnote{As detailed in \cite{uniswapv3} and provided in Appendix~\ref{sec:uniswapv3}, the virtual liquidity $\alpha,\beta$ for Uniswap V3 are dynamic in practice; in fact, they are constructed in such a way that Uniswap V3 satisfies \ref{scale}.} 

The final AMM which we consider that is used in practice is \emph{Dodo} ($u(x,y) = \log(P\alpha(x,y) + \beta(x,y))$ for external price $P > 0$; see Appendix~\ref{sec:dodo}).  Dodo is fundamentally different from all other AMMs considered within this work as it uses an \emph{exogenous} pricing oracle (such as a centralized exchange) and does not provide its own pricing oracle.  Because there is no endogenous pricing oracle, Dodo permits pooling and withdrawing in any combination of assets (though with the possibility of withdrawal fees so as to guarantee the withdrawal is possible).  The utility for Dodo is constructed in such a way to attempt to match the external price. 

\begin{remark}
As noted within Remark~\ref{rem:deriv},~\cite{capponi2021adoption} assumes AMMs satisfy a monotone differences property ($u_{AA},u_{BB} \leq 0$ and $u_{AB} \geq 0$).  However, the utility functions for mStable, StableSwap, Curve, and Dodo do \emph{not} satisfy this stronger property.
\end{remark}

Select AMMs are compared visually in Appendix~\ref{sec:visualization}.

\subsection{Symmetric Decomposable AMM}\label{sec:examples-sum}
Herein we propose a new class of AMMs which we name \emph{Symmetric Decomposable AMMs} (SDAMMs) that provides a simple analytical structure for the utility function.  This is in contrast to the majority of currently existing AMMs which, by and large, are combinations of Uniswap V2 and mStable.
Specifically, define the SDAMM utility function as \[u(x,y) = U(x) + U(y)\] for any $x,y \geq 0$ with the univariate utility function $U: \R_+ \to \R \cup \{-\infty\}$.
This can be seen as a generalization of Uniswap V2 since taking $U(z) := \log(z)$ for any $z \geq 0$ exactly replicates this well-known AMM.

The following proposition relates the properties of the univariate utility function $U$ to the axioms of the associated SDAMM.  
\begin{proposition}\label{prop:sdamm}
Let $U: \R_+ \to \R \cup \{-\infty\}$ be a thrice differentiable univariate utility function and let $u$ be the associated SDAMM.  Immediately, $u$ satisfies \ref{cont}.
\begin{enumerate}
\item\label{prop:sdamm-1} If $U(0) = -\infty$ and $U(z) > -\infty$ for any $z > 0$ then \ref{normalized}.
\item If $\lim_{z \to \infty} U(z) = \infty$ then \ref{inf}. 
\item If $U$ is strictly increasing then \ref{strict_monotonic}.
\item If $U$ is concave then \ref{convex}. If, additionally, $U$ is strictly increasing then \ref{deriv}.
\item If $U$ is strictly increasing and both $\lim_{z \to \infty} U'(z) = 0$ and $\lim_{z \to 0} U'(z) = \infty$ then \ref{inada}.
\item\label{prop:sdamm-6} If $U$ is strictly increasing, concave, and $U'$ is log-convex then \eqref{eq:P-cond}.
\item\label{prop:sdamm-quad} If $U$ is strictly increasing, concave, and $3U''(z)^2 \geq U'(z) U'''(z) \geq 0$ for any $z > 0$ then \eqref{eq:P-cond-quad} and \eqref{eq:P-cond-quad-X}.
\end{enumerate}
\end{proposition}
Proposition~\ref{prop:sdamm} provides simple conditions on $U: \R_+ \to \R \cup \{-\infty\}$ to guarantee the various axioms proposed within this work.  As these properties are easily satisfied (see, e.g., Uniswap V2 and Example~\ref{ex:sinh}), we highlight in Table~\ref{table:amm} that SDAMM satisfies these axioms.  
\begin{remark}\label{rem:riskaverse}
Proposition~\ref{prop:sdamm}\eqref{prop:sdamm-quad} provides a financial interpretation to \eqref{eq:P-cond-quad} and \eqref{eq:P-cond-quad-X}. Specifically, $3U''(z)^2 \geq U'(z) U'''(z)$ for every $z > 0$ if and only if the absolute risk aversion of $U$ does not decrease too rapidly. In fact, if $U$ has nondecreasing absolute risk aversion then this condition is automatically satisfied. It is for these reasons that we view conditions \eqref{eq:P-cond-quad} and \eqref{eq:P-cond-quad-X} as properties on the absolute risk aversion of AMMs. 
Furthermore, to complete the discussion of Proposition~\ref{prop:sdamm}\eqref{prop:sdamm-quad}, the additional condition $U'(z) U'''(z) \geq 0$ for every $z > 0$ is equivalent to the convexity of $U'$ due to the monotonicity of $U$.
\end{remark}

With this utility function, the pricing oracle is defined as $P(x,y) = \frac{U'(x)}{U'(y)}$ for $x,y \geq 0$.  In contrast to many of the aforementioned AMMs, pooling for SDAMM can be much more complex as it need not satisfy \ref{scale}.\footnote{As proven in~\cite[Theorem 8]{skiadas2016scale}, this decomposable structure for SDAMM satisfies~\ref{scale} if and only if $U$ is a power utility function.}  In fact, to accomplish pooling in such a case, we recommend a scheme in which the liquidity provider supplies the assets in any ratio which the pool then swaps appropriately to maintain a constant price.  
For instance, in Example~\ref{ex:sinh} below, we consider a specific function $U$ that satisfies 
Properties~\eqref{prop:sdamm-1}-\eqref{prop:sdamm-6} of Proposition~\ref{prop:sdamm} but is not scale invariant~\ref{scale} nor does it satisfy \eqref{eq:P-cond-quad} and \eqref{eq:P-cond-quad-X}; this proposed new AMM structure concentrates liquidity around the balanced reserve price of 1. 
Within Section~\ref{sec:divergence} below, we investigate some important implications of dropping the scale invariance~\ref{scale} axiom.

\begin{example}\label{ex:sinh}
Let $U(z) = \log(\sinh(Cz^q))$ with $C > 0$ and $q \in (0,1]$ for any $z \geq 0$.  This satisfies all the properties considered within Proposition~\ref{prop:sdamm} except \eqref{prop:sdamm-quad} (nor does it satisfy \ref{scale})  if $q < 1$; however, if $q = 1$, then \ref{inada} is no longer satisfied as well.  We use the hyperbolic sine function because it limits to the exponential function as the pool size grows to infinity.  As such this AMM limits to mStable when the pool size is large enough (and as $q \to 1$).  In this way, this AMM pool can have extremely stable prices at $1$, even more than Curve, and have infinite liquidity.  In order to gain this price stability, an (extremely) unbalanced pool has extremely high price impacts in order to remain liquid. Much like Curve and L.StableSwap, these stable AMMs fail to satisfy \eqref{eq:P-cond-quad} and \eqref{eq:P-cond-quad-X} in order to still provide liquidity throughout the price curve. 
\end{example}

\section{Fee structures and divergence loss}\label{sec:fees}
Thus far, we have only discussed AMMs without any explicit fees collected by the liquidity providers.
Within this section we propose a novel structure for assessing fees on a swap.\footnote{We directly compare this construction to the prior methods of assessing fees discussed within the literature (.e.g,~\cite{angeris2020improved,lipton2021automated}) in Remark~\ref{rem:fee-structure} in the appendix.} In doing so, we explore the costs that liquidity providers may incur by investing in the pool. In particular, we focus primarily on the so-called ``impermanent loss'' or divergence loss \cite{xu2021sok}.
We refer the interested reader to~\cite{lehar2021decentralized} for a discussion of the tradeoff between fees and adverse selection for the liquidity providers.
As before, we will primarily focus on the properties of $\Y$ as the results are symmetric to $\X$.

\begin{assumption}\label{ass:fees}
Throughout the remainder of this paper, we will consider AMMs satisfying \ref{normalized}, \ref{convex}, \ref{inada}, and \ref{deriv}. 
\end{assumption}

\subsection{Assessing fees on the marginal price}\label{sec:fee-price}

Herein we, first, wish to propose a novel structure for assessing fees on a swap.  Specifically, we propose that a fee $\gamma \in [0,1]$ is assessed to each marginal unit being bought.  This is accomplished by assessing the fees directly on the price as given by the pricing oracle $P$.  This interpretation is consistent with a typical financial interpretation of a bid-ask spread (see, e.g., \cite{demsetz1968cost}).
\begin{definition}\label{defn:fees}
Let $u: \R^2_+ \to \R \cup \{-\infty\}$ be an AMM with associated pricing oracle $P: \R^2_{++} \to \R_{++}$.  Let $a,b > 0$ be the size of the reserves of asset $A$ and $B$ within the pool respectively.  Let $\gamma \in [0,1]$ be the fee level assessed on a transaction.  
The swap functions with fees $\Y_\gamma: \R^3_+ \to \R_+$ and $\X_\gamma: \R^3_+ \to \R_+$ are defined for any $x,y \geq 0$ as: 
\begin{align}
\Y_{\gamma}(x;a,b) &:= (1-\gamma)\int_0^x P(a+z , b-\Y_{\gamma}(z;a,b))dz \label{eq:Y-gamma},\\
\X_{\gamma}(y;a,b) &:= (1-\gamma)\int_0^y \frac{1}{P(a-\X_{\gamma}(z;a,b) , b+z)}  dz.\label{eq:X-gamma}
\end{align}
Where clear, we will drop the explicit dependence of $\Y_{\gamma},\X_{\gamma}$ on the pool reserves $(a,b) \in \R^2_{++}$.
\end{definition}

The intuition behind the constructions \eqref{eq:Y-gamma} and \eqref{eq:X-gamma} is that the AMM should be indifferent to the size of transactions; a sequence of unidirectional small trades should have the same result for the pool as a single large trade (assuming nothing happens in between the transactions). Lemma~\ref{lemma:exist} proves the existence and uniqueness of these swap functions $\Y_\gamma,\X_\gamma$ for any $\gamma \in [0,1]$.

\begin{remark}\label{rem:zero-fee}
By construction of the pricing oracle of an AMM satisfying Assumption~\ref{ass:fees}, if the fees are zero ($\gamma = 0$), we recover $\Y_0 \equiv \Y$.  That is, the pool recovers the swap amounts $\Y$ as presented in the prior sections when no fees are assessed.  In addition, as proven in Proposition~\ref{prop:amm-fee-bound}, the incurred fees are monotonic in $\gamma$, i.e., $0 \leq \Y_{\gamma_2}(x;a,b) < \Y_{\gamma_1}(x;a,b) \leq \Y(x;a,b)$ for any $0 \leq \gamma_1 < \gamma_2 \leq 1$. Further properties of this fee structure are provided in Appendix~\ref{sec:fee-price-details}; we wish to highlight that this fee structure is designed to satisfy the properties introduced in Section~\ref{sec:design}.
\end{remark}

\subsection{Divergence loss}\label{sec:divergence}
We will conclude our discussion of AMMs by considering the profits and losses observed by liquidity providers. The greater the gains (or smaller the losses), the more incentive there is for users to deposit liquidity in order to capture a portion of the fees. The most commonly reported metric for the costs to liquidity providers is called the divergence loss or impermanent loss (see, e.g., \cite{aigner2021uniswap,deng2022static}) which measures the opportunity cost of investing in the pool. Notably, the divergence loss differs from the accounting profits and losses for the liquidity provider.\footnote{Another widely cited metric is the loss-versus-rebalancing proposed within \cite{milionis2022automated}.}

The divergence loss measures the mark-to-market loss from repurchasing the original pooled portfolio after withdrawing the liquidity position (accounting for any asset rebalancing and fee collection), i.e., the difference between the value of a buy-and-hold strategy and of a liquidity position in the pool.  Though conceptually this is a dynamic measure due to the relation between fee collection and price volatility, the divergence loss is traditionally simplified so that it measures the gains or losses to the liquidity position after only a single swap is undertaken. That is, first the liquidity provider invests the funds, second the price moves because of a trade, and finally the liquidity is withdrawn to provide the value of the pooled position; the divergence loss then is the difference between the value of the original liquidity position (under the updated price) and this withdrawn position.  
Formally, the divergence loss is defined directly below.

\begin{definition}\label{defn:divergence}
Consider a pool with reserves $(a,b) \in \R^2_{++}$ and fee level $\gamma \in [0,1]$.  Let $(\alpha,\beta) \in \R^2_+ \backslash \{0\}$ such that $P(a+\alpha,b+\beta) = P(a,b)$.
The divergence loss $\Delta: (0,\infty) \to \R$ 
depending implicitly on $(a,b,\alpha,\beta)$ is the difference between the mark-to-market value of $(\alpha,\beta)$ being held and being pooled, i.e.,
\begin{equation*}
\Delta(p) := \begin{cases} \left[p\alpha + \beta\right] - \frac{P(a,b)\alpha + \beta}{P(a,b)[a+\alpha] + [b+\beta]}\left[p(a+\alpha+x_p)+(b+\beta-\Y_{\gamma}(x_p;a+\alpha,b+\beta)\right] &\text{if } p \leq P(a,b) \\ 
    \left[p\alpha + \beta\right] - \frac{P(a,b)\alpha + \beta}{P(a,b)[a+\alpha] + [b+\beta]}\left[p(a+\alpha-\X_{\gamma}(y_p;a+\alpha,b+\beta))+(b+\beta+y_p)\right] &\text{if } p > P(a,b) \end{cases}
\end{equation*}
where $x_p > 0$ is such that $p = P(a+\alpha+x_p,b+\beta-\Y_{\gamma}(x_p;a+\alpha,b+\beta))$ and $y_p > 0$ is such that $p = P(a+\alpha-\X_{\gamma}(y_p;a+\alpha,b+\beta),b+\beta+y_p)$.
\end{definition}
\begin{remark}
Note that the domain of $\Delta$ is $(0,\infty)$ under Assumption~\ref{ass:fees}. In other words, for any $p < P(a,b)$ there exists some $x_p > 0$ such that $p = P(a+\alpha+x_p,b+\beta-\Y_{\gamma}(x_p;a+\alpha,b+\beta))$, and similarly for any $p > P(a,b)$ there exists some $y_p > 0$ such that $p = P(a+\alpha-\X_{\gamma}(y_p;a+\alpha,b+\beta),b+\beta+y_p)$.
To see this, recall that the pricing oracle $P$ is differentiable with $P_A \leq 0$ and $P_B \geq 0$. As $\Y_{\gamma}$ is strictly increasing, we find that $P(a+\alpha+x,b+\beta-\Y_{\gamma}(x;a+\alpha,b+\beta)) < P(a,b)$ for any $x > 0$.  Furthermore, $0\le \lim\limits_{x\to\infty} P(a+\alpha+x,b+\beta-\Y_{\gamma}(x;a+\alpha,b+\beta)) \le \lim\limits_{x\to\infty} P(a+\alpha+x,b+\beta) =0$.   Therefore, together with continuity, this guarantees the existence of such $x_p$.
\end{remark}
Often it is convenient to consider the divergence loss as a function of the swap amounts rather than prices.  That is, $\bar\Delta: \R \to \R$ defined as:
\begin{equation*}
\bar\Delta(z):= \begin{cases} \Delta(P(a+\alpha+z,b+\beta-\Y_{\gamma}(z;a+\alpha,b+\beta))) &\text{if } z \geq 0, \\
    \Delta(P(a+\alpha-\X_{\gamma}(-z;a+\alpha,b+\beta),b+\beta-z)) &\text{if } z < 0, \end{cases}
\end{equation*}
so as to remove the explicit dependence of $\Delta$ on the mappings $x_p,y_p$.

\begin{remark}\label{rem:pnl}
Recall that the divergence loss $\Delta(p)$ is a form of opportunity cost from investing the portfolio $(\alpha,\beta)$ into the pool. It can be seen as the difference between the \emph{accounting} profits and losses from the buy-and-hold strategy and the liquidity providing position. Specifically, these accounting profits $\Pi_H(p),\Pi_L(p)$, respectively, are provided by: $\Pi_H(p) := (p\alpha+\beta) - (P(a,b)\alpha+\beta) = (p-P(a,b))\alpha$
and, using Definition~\ref{defn:divergence}, $ \Pi_L(p)= \Pi_H(p) - \Delta(p)$ .
Further details and discussion of the accounting profits and losses are provided in Appendix~\ref{sec:pnl}. In particular, we highlight that both the liquidity provider and arbitrageur can profit simultaneously as the liquidity provider is long in asset $A$ whereas the arbitrageur is direction agnostic. That is, if the arbitrageur takes advantage of stale prices in such a way so as to increase the value of asset $A$, then a liquidity provider will also make an accounting profit. On net, the profits of all liquidity providers and arbitrageurs replicate the buy-and-hold strategy. In other words, the liquidity providers and arbitragers are not locked in a zero sum game. Rather they split the profits (or losses) of the buy-and-hold position between them. Thus both can have accounting profits if the price has increased.
\end{remark}

\begin{figure}
\centering
\begin{subfigure}[t]{0.4\textwidth}
\includegraphics[width=\textwidth]{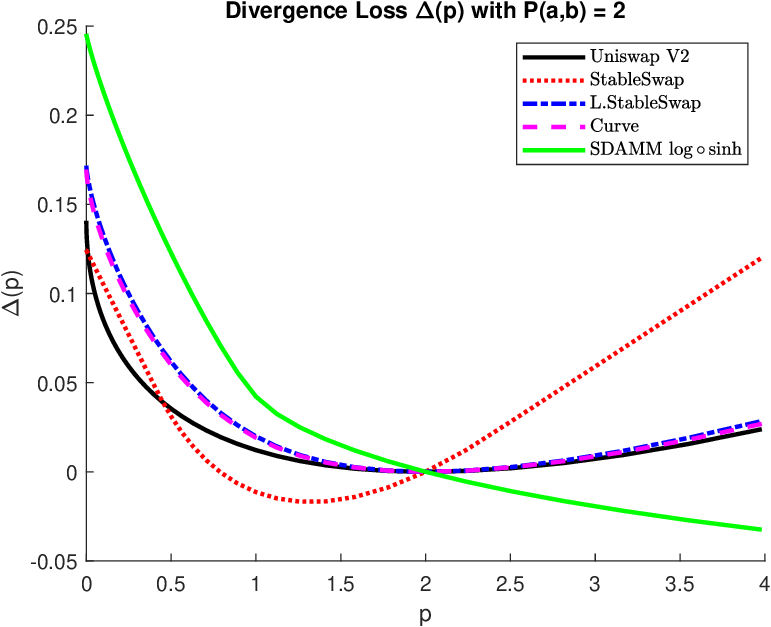}
\caption{The divergence loss $\Delta(p)$ for Uniswap V2, StableSwap, L.StableSwap, Curve, and the hyperbolic sine SDAMM.}
\label{fig:comparison-delta}
\end{subfigure}
~~
\begin{subfigure}[t]{0.4\textwidth}
\includegraphics[width=\textwidth]{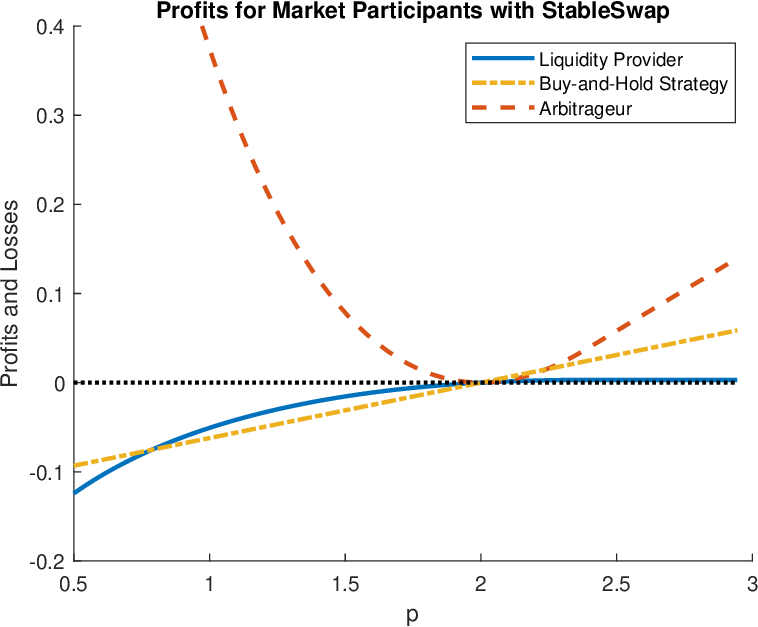}
\caption{Accounting profits and losses for different strategies with StableSwap.} 
\label{fig:pnl}
\end{subfigure}
\caption{Pools all considered with $u(a,b) = 1$, $P(a,b) = 2$, and $\frac{P(a,b)\alpha+\beta}{P(a,b)(a+\alpha)+(b+\beta)} = \frac{1}{11}$. Note that StableSwap is only able to support prices in $[0.4195,2.3837]$.}
\label{fig:divergence}
\end{figure}

As noted above, the divergence loss is important as it quantifies an opportunity cost for the liquidity provider in pooling his or her assets. As stated in~\cite{xu2021sok}, the divergence loss $\Delta(p)$ of well-studied AMMs is strictly positive so long as $p \neq P(a,b)$ when neglecting the fees. As we will see in Lemma~\ref{lemma:divergence} and Figure~\ref{fig:comparison-delta}, this result is a direct consequence of \ref{scale}. This creates a tradeoff between the easy implementation of the proportional rule for pooling assets as noted in Remark~\ref{rem:pooling} and losses for the liquidity providers.

\begin{lemma}\label{lemma:divergence}
Consider an AMM satisfying Assumption~\ref{ass:fees} and \ref{scale}.
The divergence loss can be simplified as:
\begin{equation*}
\Delta(p) = \begin{cases} \frac{\delta}{1+\delta}[\Y_{\gamma}(x_p;(1+\delta)a,(1+\delta)b) - px_p] &\text{if } p \leq P(a,b), \\
    \frac{\delta}{1+\delta}[p\X_{\gamma}(y_p;(1+\delta)a,(1+\delta)b) - y_p] &\text{if } p > P(a,b) ,\end{cases}
\end{equation*}
where, implicitly, $\alpha := \delta a$ and $\beta := \delta b$ for some $\delta > 0$. 
Furthermore, the sign of the divergence loss can be characterized w.r.t.\ the fees:
\begin{enumerate}
\item if $\gamma = 0$ then $\Delta(p) \geq 0$ with strict inequality if $p \neq P(a,b)$;
\item\label{lemma:divergence-gamma} if $\gamma \in (0,1]$ then there exist $p_* < P(a,b) < p^*$ such that $\Delta(p) < 0$ for $p \in (p_*,p^*)\backslash\{p\}$. 
\end{enumerate}
\end{lemma}

We now want to consider the divergence loss for Uniswap V2 with fees (Example~\ref{ex:uniswap}) in order to formalize the results of Lemma~\ref{lemma:divergence} for this well-known AMM.  
Second, we numerically study multiple different AMMs to consider the divergence loss when \ref{scale} is not satisfied to highlight the value in \emph{dropping} this axiom that is widely assumed in practice (see Table~\ref{table:amm}) and within the literature (see, e.g., \cite{capponi2021adoption,schlegel2022axioms}).
\begin{example}\label{ex:uniswapv2-divergence}
Consider the Uniswap V2 AMM with fees $\gamma \in (0,1)$, in other words, $\Y_{\gamma}(x;a,b) = b\left(1 - \left(\frac{a}{a+x}\right)^{1-\gamma}\right)$ and $\X_{\gamma}(y;a,b) = a\left(1 - \left(\frac{b}{b+y}\right)^{1-\gamma}\right)$ for any transactions $x,y \geq 0$.\footnote{For details, see Example~\ref{ex:uniswapv2}.}
As Uniswap V2 satisfies~\ref{scale}, the results of Lemma~\ref{lemma:divergence} hold.  Furthermore, as noted within Remark~\ref{rem:pooling} and used in Lemma~\ref{lemma:divergence}, we take the liquidity injection to be $\alpha = \delta a$ and $\beta = \delta b$ for some $\delta > 0$.
We wish to consider the form of the divergence loss (characterized as $\bar\Delta$) and the threshold prices $p_*,p^*$ which construct the interval of divergence gains for the liquidity provider.
Let $z \in \R$ then
\begin{equation*}
\bar\Delta(z) = \begin{cases} \delta b \left(1 - 2\left(\frac{(1+\delta)a}{(1+\delta)a + z}\right)^{1-\gamma} + \left(\frac{(1+\delta)a}{(1+\delta)a+z}\right)^{2-\gamma}\right) &\text{if } z \geq 0 \\
    \delta b \left(1 - 2\left(\frac{(1+\delta)b - z}{(1+\delta)b}\right) + \left(\frac{(1+\delta)b - z}{(1+\delta)b}\right)^{2-\gamma}\right) &\text{if } z < 0. \end{cases}
\end{equation*}
We will now direct our attention to determining the threshold prices $p_*,p^*$.  
Consider, first, $p_* \leq P(a,b) = b/a$ which involves studying $\bar\Delta(z)$ for $z \geq 0$.  It can be determined that $\bar\Delta(z) < 0$ for $z \geq 0$ if and only if $z \in (0 , x_* := \frac{(1+\delta)(1-t_*)a}{t_*})$ where $t_* \in (0,1-\gamma)$ solves $1 - 2t^{1-\gamma} + t^{2-\gamma} = 0$. 
The lower bound $p_*$ for divergence gains can be determined by finding the price assuming $x_*$ assets were transacted, i.e.,
\[p_* := P((1+\delta)a+x_*,(1+\delta)b-\Y_{\gamma}(x_*;(1+\delta)a,(1+\delta)b) = \frac{b}{a}t_*^{2-\gamma} < \frac{b}{a} = P(a,b).\]
Following similar arguments, we determine that the divergence loss is negative for $z < 0$ if and only if $z \in (-y^* := -(1+\delta)b(t^*-1) , 0)$ where $t^* \in (1+\gamma)^{\frac{1}{1-\gamma}} + (0 , -\frac{1-2(1+\gamma)^{\frac{1}{1-\gamma}} + (1+\gamma)^{\frac{2-\gamma}{1-\gamma}}}{\gamma(1-\gamma)})$ solves $1 - 2t + t^{2-\gamma} = 0$, i.e., the upper bound $p^*$ for divergence gains is
\[p^* := P((1+\delta)a-\X_{\gamma}(y^*;(1+\delta)a,(1+\delta)b),(1+\delta)b+y^*) = \frac{b}{a}(t^*)^{2-\gamma} > \frac{b}{a} = P(a,b).\]
\end{example}

\begin{remark}\label{rem:divergence}
Though only provided as a sufficient condition for divergence gain, we hypothesize that Lemma~\ref{lemma:divergence}\eqref{lemma:divergence-gamma} can be strengthened insofar as we conjecture that $\Delta(p) > 0$ for $p \not\in [p_*,p^*]$ under~\ref{scale} with fees $\gamma \in (0,1)$.  We highlight in Example~\ref{ex:uniswapv2-divergence}, that this stronger conjecture is satisfied in that special case.  
To be proven for a generic AMM, this requires that $\Y_{\gamma}(x;a,b) - P(a+x,b-\Y_{\gamma}(x;a,b))x \geq 0$ for $x > x^*$ (wlog suppressing the need for $\delta \geq 0$ in this expression).
However, this expression is not necessarily monotonic nor convex in $x$ which makes such a proof highly non-trivial and beyond the scope of this work. 
\end{remark}

In Lemma~\ref{lemma:divergence}, we have seen how \ref{scale} guarantees the divergence loss when $\gamma = 0$ fees are assessed. In the following corollary, we generalize this result to provide necessary and sufficient conditions for the existence of divergence gain without fees. In fact, except in degenerate cases, \ref{scale} is necessary and sufficient for guaranteed divergence loss.
\begin{corollary}\label{cor:divergence}
Consider an AMM satisfying Assumption~\ref{ass:fees} with $\gamma = 0$. Consider the pool reserves $(a,b) \in \R^2_{++}$ and let $(\alpha,\beta) \in \R^2_+ \backslash \{0\}$ such that $P(a+\alpha,b+\beta) = P(a,b)$. There exists a price $p \neq P(a,b)$ such that $\Delta(p) < 0$ if and only if $\alpha/\beta \neq a/b$.
\end{corollary}
\begin{remark}\label{rem:divergence-scale}
Corollary~\ref{cor:divergence} replicates the $\gamma = 0$ fee case of Lemma~\ref{lemma:divergence} as pooling under \ref{scale} implies $\alpha/\beta = a/b$ as discussed in Remark~\ref{rem:pooling}. In contrast, if \ref{scale} does \emph{not} hold then, except in degenerate cases, $\alpha/\beta \neq a/b$.
This result contrasts with the approach taken in, e.g., \cite{milionis2022automated} in which the divergence loss is considered on the entire pool reserves rather than for the individual LPs. Under \ref{scale}, it is functionally equivalent to consider the entire pool or a single LP position; however, without that property, the value of providing liquidity depends on the original pool reserves at the time of the liquidity provision. 
\end{remark}

Finally, we want to demonstrate how the divergence loss differs based on the AMM construction used. In Figure~\ref{fig:comparison-delta}, we compare the divergence loss $\Delta(p)$ for the same AMMs as considered in Figure~\ref{fig:comparison} (i.e., Uniswap V2, StableSwap, L.StableSwap, Curve, and the hyperbolic sine SDAMM) without fees, i.e., $\gamma = 0$. Notably, both StableSwap and the hyperbolic sine SDAMM are able to capture divergence gains in a portion of the price range, whereas the other AMMs considered all suffer divergence losses throughout. As provided in Table~\ref{table:amm}, neither StableSwap nor the hyperbolic sine SDAMM satisfy \ref{scale} whereas the rest of the considered AMMs are scale invariant. 

Note that the divergence loss is an opportunity cost. Even though AMMs without \ref{scale} can have divergence gains, these do not lead to accounting losses for the arbitrageurs. In Figure~\ref{fig:pnl}, we consider the profits and losses of different investment strategies when considering StableSwap. Specifically, we plot the profits and losses for liquidity providers (holding 1/11th of the pool), buy-and-hold investors, and arbitrageurs. By doing this, we are able to highlight how the divergence loss is the difference between the profits of two different trading strategies so that it is possible for, e.g., simultaneous divergence gains and accounting losses and vice versa.
In addition, we wish to note that both the liquidity providers and arbitrageurs can experience accounting profits at the same time. In fact, as noted in Remark~\ref{rem:pnl}, the collection of all liquidity providers and arbitrageurs split the profits and loses of the buy-and-hold strategy exactly.

\begin{remark}
As discussed above, though \ref{scale} leads to simplifying mathematical properties (see Lemma~\ref{lemma:amm-concave}\eqref{thm:amm-ph} and Proposition~\ref{prop:price}\eqref{prop:price-si}), it introduces significant risks to liquidity providers (see Lemma~\ref{lemma:divergence}). This is especially notable in comparison to AMMs that are utilized in practice that are, frequently, satisfying~\ref{scale} as highlighted within Table~\ref{table:amm}. Furthermore, when reviewing Table~\ref{table:properties}, we note that \ref{scale} is not necessary for any of the fundamental properties for financial markets.
\end{remark}

\section{Conclusion}\label{sec:conclude}
Within this work we have considered an axiomatic framework for AMMs. By imposing reasonable axioms on the underlying utility function, we are able to characterize the properties of the swap size of the assets and of the resulting pricing oracle. In addition, we have introduced a novel price impact oracle which quantifies these costs for traders. We have analyzed many existing AMMs and shown that the vast majority of them satisfy our axioms. Finally, we have also considered the question of fees and divergence loss. In doing so, we have proposed a new fee structure so as to make the AMM pool indifferent to trade execution. Finally, we have proposed a novel AMM that, while it does not satisfy all of our axioms, has nice analytical properties and provides a large range over which there is no divergence loss.

We wish to provide a few extensions for this work.  
First, and importantly, a rigorous study of the conjecture provided within Remark~\ref{rem:divergence} would greatly enhance our understanding of the divergence loss and the impacts that scale invariance~\ref{scale} has on those costs. 
Second, within this work and throughout the literature on constant function market makers, a single utility function $u: \R^2_+ \to \R \cup \{-\infty\}$ is always considered; herein we propose considering the generalized AMM with incomplete preference relation $\succeq$ on $\R^2_+$ so that, e.g., $\Y(x;a,b) = \sup\{y \in [0,b] \; | \; (a+x,b-y) \succeq (a,b)\}$.  In this way the fees can be endogenized within the preference relation itself.
Related to the first two extensions, we propose further studies on new AMM constructions so as to minimize divergence loss (and maximize revenue for liquidity providers) much as we undertook with the SDAMM example within this work. 
Finally, within this paper, we only considered the traditional style AMMs, i.e., for swap markets.  In order to grow the decentralized finance offerings, a rigorous study of AMMs for derivatives and other complex financial securities needs to be undertaken.

\bibliographystyle{plain}
\small{\bibliography{bibtex2}}

\newpage
\setcounter{page}{1}
\appendix
\begin{landscape}
\section{Summary Table of Main Results}\label{sec:summary}
\begin{table}[h!]
\centering
\begin{tabular}{|c||l||c|c|c|c|c|c|c|c|c|c|c|}
\hline
{\bf Property} & \multicolumn{1}{c||}{\bf Details} & \ref{normalized} & \ref{inf} & \ref{strict_monotonic} & \ref{cont} & \ref{convex} & \ref{scale} & \ref{inada} & \ref{deriv} & \eqref{eq:P-cond} & \eqref{eq:P-cond-quad} & \eqref{eq:P-cond-quad-X} \\ \hline\hline
\nameref{NA} & L.~\ref{lemma:arbitrage}\eqref{thm:amm-arbitrage} &  &  & X & X &  &  &  &  &  &  &  \\ \hline
\nameref{SPI} & L.~\ref{lemma:arbitrage}\eqref{thm:amm-split} & X &  & X & X &  &  &  &  &  &  &  \\ \hline
\nameref{SNW} & L.~\ref{lemma:amm-monotone}\eqref{thm:amm-limit-x} &  & X & X & X &  &  & &  &  &  &  \\ \hline
\nameref{SSI} & L.~\ref{lemma:amm-monotone}\eqref{thm:amm-monotone} & X &  & X & X &  & &  &  &  &  &  \\ \hline
\nameref{SCC} & L.~\ref{lemma:amm-concave}\eqref{thm:amm-concave} &  &  &  & (1) & (1) &  &  & (2) &  &  &  \\ \hline
\nameref{SIL} & L.~\ref{lemma:amm-liq}\eqref{thm:amm-liquidity} & X &  &  & X &  &  &  &  &  &  &  \\ \hline
\nameref{ML} & \multicolumn{1}{p{1cm}||}{T.~\ref{thm:monotone-reserve}\eqref{thm:mr-monotone} \newline P.~\ref{prop:price}\eqref{prop:price-monotone}}  & X &  &  &  &  &  &  & X &  &  &  \\ \hline
\nameref{PL} & T.~\ref{thm:pooling} & (2) &  &  &  & (2) & (1) & (2) & X & (2) &  &  \\ \hline
\end{tabular}
\caption{Summary of main results and necessary axioms. (1) and (2) denote alternative conditions where either all axioms in (1) must be satisfied or axiom (2) is satisfied.}
\label{table:properties}
\end{table}
\end{landscape}

\section{Details of Existing Automated Market Makers}\label{sec:current-amm}

\subsection{Uniswap V2}\label{sec:uniswapv2}
Though we refer to this example as Uniswap V2, the same mathematical structure is also utilized within many other AMMs in practice such as SushiSwap, DefiSwap, and Quickswap.  Structurally, Uniswap V2 is the logarithmic utility function, i.e., \[u(x,y) = \log(x) + \log(y)\] for $x,y \geq 0$. 
With this utility function, the pricing oracle and price impact oracle are defined as \[P(x,y) = \frac{y}{x},~\Imp(x,y) = \frac{y}{x^2},~x,y \geq 0.\]  This AMM permits pooling at the ratio of the current reserves.
As highlighted within Table~\ref{table:amm}, this AMM satisfies every axiom proposed within this work.

\subsection{Balancer}\label{sec:balancer}
Though we refer to this example as Balancer, the same mathematical structure is also utilized within many other AMMs in practice such as Bancor and Loopring.  Structurally, Balancer is a weighted version of Uniswap V2, i.e., \[u(x,y) = w \log(x) + (1-w) \log(y)\] with $w \in (0,1)$ for $x,y \geq 0$. 
With this utility function, the pricing oracle and price impact oracle are defined as \[P(x,y) = \frac{wy}{(1-w)x}, ~\Imp(x,y) = \frac{w y}{2(1-w)^2 x^2},~ x,y \geq 0.\] 
Notably, this AMM always achieves a lower price impact than Uniswap V2 for $w\in(0,\frac12)$. It also permits pooling at the ratio of the current reserves.
As highlighted within Table~\ref{table:amm}, this AMM satisfies every axiom proposed within this work.

\subsection{Uniswap V3}\label{sec:uniswapv3}
Though we refer to this example as Uniswap V3, the same mathematical structure is also utilized within many other AMMs in practice such as KyberSwap and MooniSwap.  The generic structure of all of these AMMs is the same as Uniswap V2 but with ``virtual reserves'' to provide concentrated liquidity, i.e., \[u(x,y) = \log(\alpha+x) + \log(\beta+y)\] with $\alpha,\beta > 0$ for $x,y \geq 0$.  Each of these real-world AMMs select $\alpha,\beta$ in different ways which may be dynamic in time or based on the implemented trades.  For instance, Uniswap V3 implements this AMM in such a way that $\alpha,\beta$ are adjusted dynamically so that the AMM maintains a fixed maximum and minimum quoted price.  
With this utility function, the pricing oracle and price impact oracle are defined as \[P(x,y) = \frac{\beta+y}{\alpha+x},~\Imp(x,y)=\frac{\beta +y}{(\alpha +x)^2},~x,y \geq 0.\]  
This AMM  permits pooling at the ratio of the current reserves \emph{inclusive} of the virtual reserves $\alpha,\beta$.
As mentioned, these virtual reserves concentrate the liquidity to reduce price impacts from a transaction when the virtual reserves $(\alpha,\beta)$ are close to the true reserves $(a,b)$, but this comes at the cost of unbounded from below. 
Additionally, as constructed here with static $\alpha,\beta$, Uniswap V3 fails to be scale invariant.
As highlighted in Table~\ref{table:amm}, Uniswap V3 does \emph{not} satisfy~\ref{normalized},~\ref{scale}, or~\ref{inada}. 

In practice this is undertaken with functional forms for $\alpha,\beta$ so that the liquidity is concentrated within constant upper $P^U$ and lower $P^L$ prices without regard to the amount of physical liquidity provided. Within the actual Uniswap V3 this is formulated via:
\begin{align*}
\alpha(x,y) &= \frac{\sqrt{P^L P^U}x + y + \sqrt{(\sqrt{P^L P^U}x + y)^2 + 4\sqrt{P^U}(\sqrt{P^U}-\sqrt{P^L})xy}}{2\sqrt{P^U}(\sqrt{P^U}-\sqrt{P^L})} \\
\beta(x,y) &= \frac{\sqrt{P^L}\left[\sqrt{P^L P^U}x + y + \sqrt{(\sqrt{P^L P^U}x + y)^2 + 4\sqrt{P^U}(\sqrt{P^U}-\sqrt{P^L})xy}\right]}{2(\sqrt{P^U}-\sqrt{P^L})} 
\end{align*}
Notably, following this functional form (as $\alpha,\beta$ are positive homogeneous), \ref{scale} is now recovered.

\subsection{mStable}\label{sec:mstable}
mStable is an AMM constructed to have no price impacts from trading.  This is accomplished through the mathematical structure \[u(x,y) = \log(x+y)\] for $x,y \geq 0$.  
This AMM comes with the constant pricing oracle and zero price impacts \[P(x,y) = 1,~\Imp(x,y)=0,~x,y \geq 0.\]  Pooling for mStable can, in theory, be accomplished with any combination of assets; traditionally, pooling is done either at the current ratio of assets or so that the pooled assets are in equal proportion.
As with Uniswap V3, the ability to reduce price impacts (in this case to 0) comes at the expense of unbounded from below~\ref{normalized}; in fact, these zero price impacts also cause the AMM to lose~\ref{inada}.  Furthermore, though~\ref{deriv} is satisfied, it is only satisfied with an equality (and thus does not guarantee quasiconcavity by itself).

\subsection{Stable swaps}
\subsubsection{StableSwap}\label{sec:stableswap} 
Though we refer to this example as StableSwap, the same mathematical structure is also utilized with many other AMMs in practice such as Saber and Saddle. Much like Uniswap V3, StableSwap aims to concentrate liquidity towards the ``balanced'' pool.  This is accomplished by taking a linear combination of (the exponentials of) Uniswap V2 and mStable, i.e., \[u(x,y) = \log(C(x+y) + xy)\] with $C > 0$ for $x,y \geq 0$.  
With this utility function, the pricing oracle and price impact oracle are defined as \[P(x,y) = \frac{C+y}{C+x},~\Imp(x,y) = \frac{C+y}{(C+x)^2},~x,y \geq 0.\]  This AMM permits pooling at the ratio of the current reserves inclusive of the parameter $C$.
As with Uniswap V3, the ability to reduce price impacts for a (potentially wide) neighborhood around the balanced pool comes at the expense of unbounded from below~\ref{normalized} and is neither scale invariant~\ref{scale} nor satisfies \ref{inada}. 

\subsubsection{Liquid StableSwap [L.StableSwap]}\label{sec:l.stableswap}
As far as we are aware, this construction has never been implemented before as an AMM in practice.  Rather than taking the linear combination of exponentials of Uniswap V2 and mStable, here we take the linear combination directly, i.e., \[u(x,y) = C\log(x+y) + \log(x) + \log(y)\] with $C > 0$ for $x,y \geq 0$.  This concentrates liquidity when the reserves of the pool are not too far out of balance, but exacerbates price impacts once the reserves of the pool become too skewed towards one asset.  
With this utility function, the pricing oracle and price impact oracle are defined as \[P(x,y) = \frac{y[(C+1)x+y]}{x[x+(C+1)y]},~\Imp(x,y)=\frac{(C+2) y (x+y) \left(C \left(x^2+y^2\right)+(x+y)^2\right)}{2 x^2 (C y+x+y)^3} ,~x,y \geq 0.\]  
By taking this structure all fundamental axioms proposed within this work are satisfied in comparison to Uniswap V3, mStable and StableSwap. However, in order to have both a stable price near the balanced pool but provide liquidity throughout the price curve, L.StableSwap loses~\eqref{eq:P-cond-quad} and \eqref{eq:P-cond-quad-X}. Indeed, near the balanced pool, the price impacts are lower than those of Uniswap V2.

\subsection{Curve}\label{sec:curve}
Curve is a popular AMM that, much like our newly proposed Liquid StableSwap, extends StableSwap in such a way so as to guarantee infinite liquidity through satisfying~\ref{normalized}. 
For the construction of Curve, let $D(x,y)$ denote the total number of coins when the reserves $(x,y) \in \R^2_{++}$ are traded into balance from a StableSwap AMM, i.e., $\log(C(x+y)+xy) = \log(CD(x,y) + D(x,y)^2/4)$.
However, in contrast to StableSwap, Curve considers a \emph{functional} parameter $\C(x,y)$; by dimensional analysis $\C(x,y)$ must have units in number of coins (so that $\C(x,y)(x+y)$ and $xy$ are both in number of coins squared).  Furthermore, with the notion that Curve desires a balanced pool, $\C(x,y)$ is constructed to be proportional to both the number of coins for the balanced pool ($D(x,y)$) and to a unitless measure of pool balance, i.e., $\C(x,y) \propto D(x,y) \times \left(\frac{xy}{D(x,y)^2/4}\right)$ where $\frac{xy}{D(x,y)^2/4} \in [0,1]$ provides a measure of pool balance.  With this functional parameter $\C(x,y) := \frac{C xy}{D(x,y)}$ with constant $C$, the balanced pool size $D(x,y)$ satisfies the equation
\begin{align}
\nonumber \log&\left(\C(x,y)(x+y) + xy\right) = \log\left(\C(x,y)D(x,y) + \frac{D(x,y)^2}{4}\right) \\ 
\label{eq:curve-D} &\Leftrightarrow \; D(x,y)^3 + 4(C-1)xyD(x,y) - 4C(x+y)xy = 0.
\end{align}
It is this balanced pool size $D(x,y)$ which defines the utility function for Curve.  
Specifically, \[u(x,y) = \log(D(x,y))\] for $x,y \geq 0$ where $D(x,y)$ is implicitly defined as the \emph{unique} nonnegative root of~\eqref{eq:curve-D} for $C \geq 1$.  Uniqueness of $D(x,y)$ follows from an application of Descartes' rule of signs.
With this utility function, the pricing oracle and price impact oracle are defined as 
\begin{align}
&P(x,y) = \frac{y[C(2x+y)-(C-1)D]}{x[C(x+2y)-(C-1)D]},\\
&\Imp(x,y) = \frac{y \left(C(x+y)-(C-1)D\right)\left(((C-1)D)^2 - 3C(C-1)(x+y)D + 3C^2(x^2+xy+y^2)\right)}{x^2 (C (x+2y)- (C-1)D)^3},\\
\end{align}
for $x,y \geq 0$ where, for simplicity, we set $D = D(x,y)$.  Despite the complex, implicit, structure of this pricing oracle, Curve permits pooling at the current ratio of reserves because it satisfies~\ref{scale}. 
Due to the implicit construction of the Curve AMM, (some of) the axioms presented within this paper can only be verified numerically.  Even so, as highlighted within Table~\ref{table:amm}, all axioms presented are satisfied for Curve either analytically or (for~\ref{convex},~\ref{deriv}, and~\eqref{eq:P-cond}) numerically.
Similar to L.StableSwap above, in order to have both a stable price near the ``balanced'' pool but provide liquidity throughout the price curve, Curve loses~\eqref{eq:P-cond-quad} and \eqref{eq:P-cond-quad-X}.

\subsection{Dodo}\label{sec:dodo}
In contrast to all other AMMs presented herein, Dodo is constructed based on an \emph{exogenous} pricing oracle (e.g., a centralized exchange) and does not provide its own pricing oracle.  In doing so, Dodo permits pooling in any combination of assets rather than guaranteeing the pricing oracle is kept constant.  To guarantee that the pool has the requisite liquidity, withdrawal fees may be assessed; these withdrawal fees take the place of the divergence loss (see Section~\ref{sec:divergence}) of other AMMs.
Mathematically, Dodo takes the form \[u(x,y) = \log(P\alpha(x,y)+\beta(x,y))\] for exogenous pricing signal $P$ and such that $\alpha,\beta: \R^2_+ \to \R_+$ satisfy price matching and equilibrium pooling, i.e.,
\begin{itemize}
\item \emph{price matching}: the value of $\alpha(x,y)$ is equal to $\beta(x,y)$, i.e., $P\alpha(x,y) = \beta(x,y)$; 
\item \emph{equilibrium pooling}: when an endogenized price $P f(x,y)$ (to account for the actual pool reserves) is used, the value of the portfolio $(x,y)$ should be equivalent to $(\alpha(x,y),\beta(x,y))$, i.e., $P f(x,y)(\alpha(x,y)-x) + (\beta(x,y) - y) = 0$.  Within the current construction of Dodo, the price modifier function $f: \R^2_+ \to \R_+$ is defined as \[f(x,y) := \begin{cases} 1 + C\left(\frac{\alpha(x,y)}{x}-1\right) &\text{if } Px \leq y, \\ \left[1 + C\left(\frac{\beta(x,y)}{y}-1\right)\right]^{-1} &\text{if } Px > y, \end{cases}\] with $C \in [0,1]$ for any $x,y \geq 0$.
\end{itemize}
Note that the construction of Dodo has a parameter $C \in [0,1]$ for the appropriate notion of equilibrium pooling.  If $C = 0$ then $u(x,y) = \log(Px + y)$ is equivalent to the mStable AMM; if $C = 1$ then $u(x,y) = \log(2\sqrt{Pxy}) \rightsquigarrow \log(x) + \log(y)$ is equivalent to the Uniswap V2 AMM.  In fact, a closed form for the Dodo construction can be provided for $C \in [0,1]$ such that
\begin{align*}
u(x,y) &= \begin{cases} \log\left(2\left[\frac{-(1-C)Px + \sqrt{(1-C)^2P^2x^2 + CPx((1-C)Px+y)}}{C}\right]\right) &\text{if } Px \leq y, \\ \log\left(2\left[\frac{-(1-C)y + \sqrt{(1-C)^2y^2 + Cy(Px + (1-C)y)}}{C}\right]\right) &\text{if } Px > y. \end{cases}
\end{align*}
As highlighted within Table~\ref{table:amm}, Dodo satisfies all \emph{relevant} axioms proposed within this work provided $C > 0$;~\eqref{eq:P-cond}, \eqref{eq:P-cond-quad}, and \eqref{eq:P-cond-quad-X} are not studied for Dodo due to the use of the exogenous pricing oracle.

\subsection{Visualization of Automated Market Makers}\label{sec:visualization}
\begin{figure}[h!]
\centering
\begin{subfigure}[t]{0.4\textwidth}
\centering
\includegraphics[width=\textwidth]{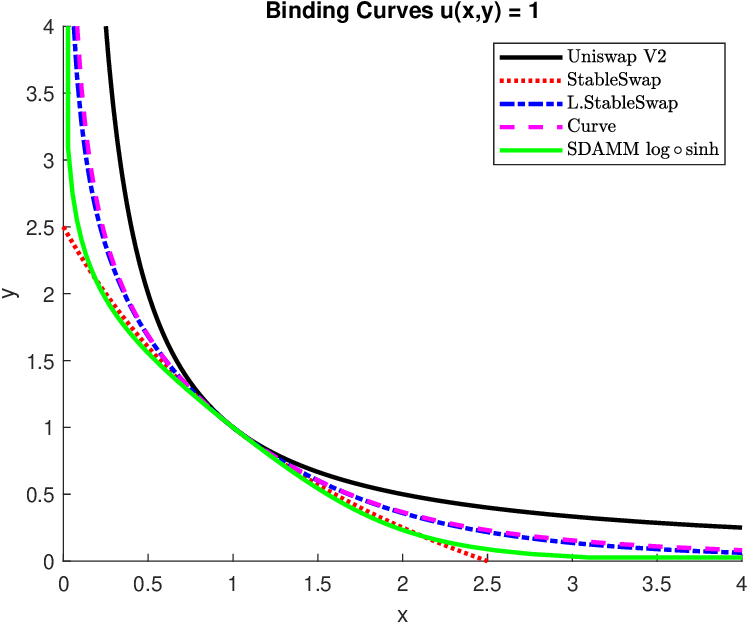}
\caption{Binding curve $u(x,y) = 1$ for various AMMs.}
\end{subfigure}
~~
\begin{subfigure}[t]{0.4\textwidth}
\centering
\includegraphics[width=\textwidth]{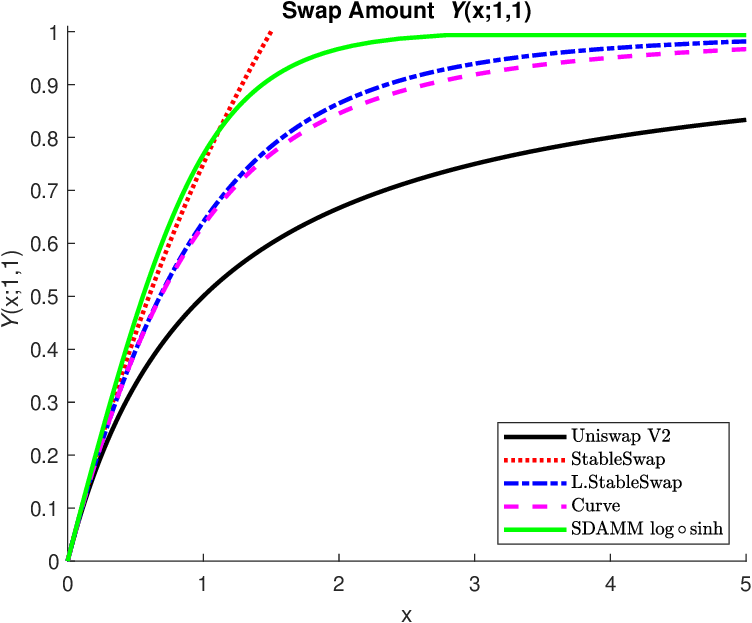}
\caption{Swap function $\Y(x;1,1)$ for various AMMs.}
\end{subfigure}
\\[10pt]
\begin{subfigure}[t]{0.4\textwidth}
\centering
\includegraphics[width=\textwidth]{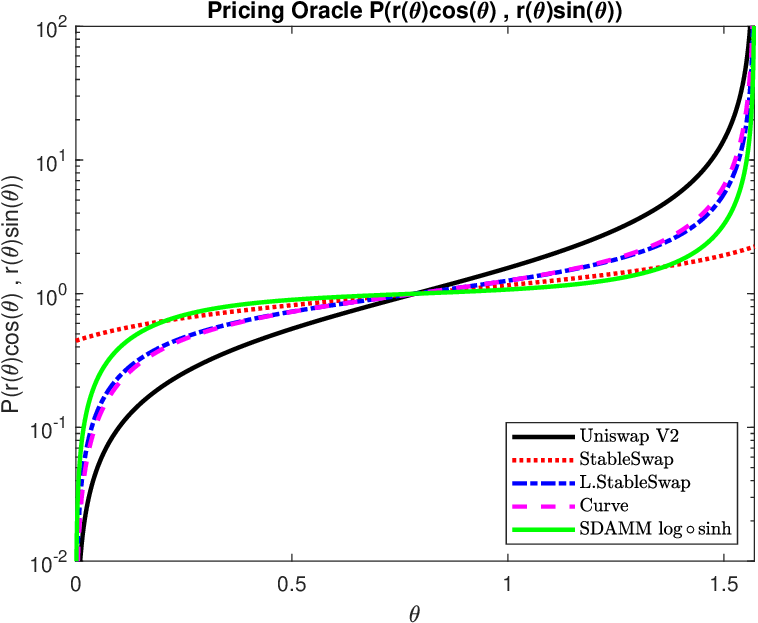}
\caption{Angled sweep of the pricing oracle $P(r(\theta)\cos(\theta),r(\theta)\sin(\theta))$ where $r(\theta)$ is such that $u(r(\theta),\cos(\theta),r(\theta)\sin(\theta)) = 1$ for various AMMs. The flatter the curve, the more stable the AMM prices.}
\end{subfigure}
~~
\begin{subfigure}[t]{0.4\textwidth}
\centering
\includegraphics[width=\textwidth]{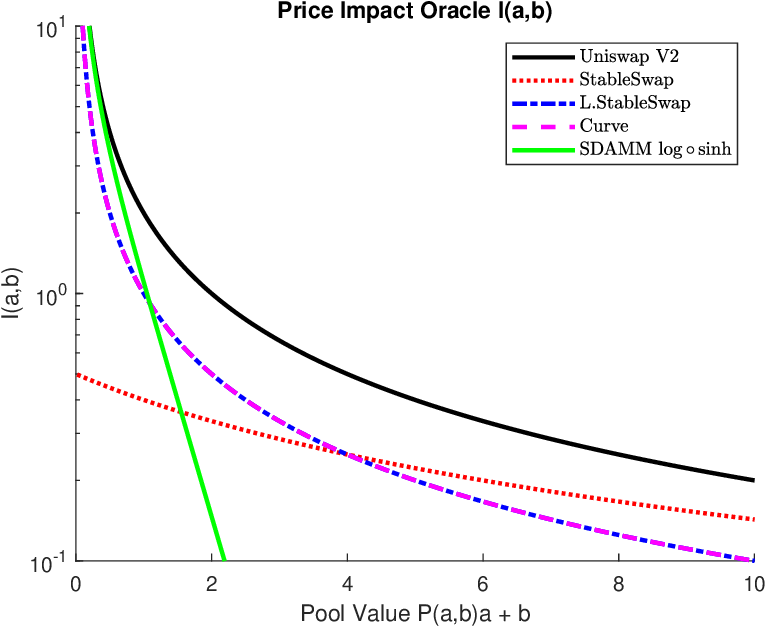}
\caption{Price impacts $\Imp(a,b)$ where $P(a,b) = 1$ (i.e., $a = b$) for various AMMs. The steeper the decline the faster the price impact decreases with liquidity.}
\label{fig:comparison-I}
\end{subfigure}
\caption{Comparison of Uniswap V2, StableSwap, L.StableSwap, Curve, and the hyperbolic sine SDAMM. }
\label{fig:comparison}
\end{figure}

We conclude this discussion of AMMs used in practice by comparing some of these constructions graphically. Demonstrations of Uniswap V2, StableSwap, L.StableSwap, Curve, and the hyperbolic sine SDAMM (see Example~\ref{ex:sinh}) are provided within Figure~\ref{fig:comparison}. In particular, we compare these AMMs in 4 dimensions: (i) the binding curve $u(z) = 1$; (ii) the swap function $\Y(x;1,1)$; (iii) the pricing oracle $P$; and (iv) the price impact oracle $\Imp$. These plots make clear that Uniswap V2 has the largest price impact, L.StableSwap and Curve have extremely similar behaviors, and, generally, that there are tradeoffs between price impacts and tail behavior (or nonexistence of possible trades in the case of StableSwap). 
In particular, in Figure~\ref{fig:comparison-I}, we observe the price impact oracle $\Imp(z)$ for a balanced pool $P(z) = 1$ as the value of the reserves $z$ vary. Here we observe that the price impacts are monotonic in pool value but to varying degrees with StableSwap having the most stable (but low) price impacts and the hyperbolic sine SDAMM having the greatest change in price impacts. Notably, and as expected, Uniswap V2 has the greatest price impacts for this pool as the 4 other AMM designs are all meant to have a stable price at $P(a,b) = 1$.

\section{Proofs}\label{sec:proofs}
\subsection{Proof of Lemma~\ref{lemma:amm-liq}}
\begin{proof}
\begin{enumerate}
\item Let $C(\bar z) := \{z \in \R^2_+ \; | \; u(z) \geq u(\bar z)\}$ denote the set of positions $z \in \R^2_+$ that exceed $\bar z \in \R^2_+$ in utility.  By (upper semi)continuity, $C(\bar z)$ is closed for every $\bar z \in \R^2_+$.  Therefore $\Y(x) = \sup\{y \; | \; y \in [0,b], \; (a+x,b-y) \in C(a,b)\}$ is the supremum over a compact set and thus is attained.
\item Assume $\Y(x) < b$.  Then, by construction of $\Y$, $u(a+x,b-[\Y(x)+\epsilon]) < u(a,b)$ for any $\epsilon \in (0,b-\Y(x)]$.  By (lower semi)continuity, $u(a+x,b-\Y(x)) \leq \liminf_{\epsilon \searrow 0} u(a+x,b-[\Y(x)+\epsilon]) \leq u(a,b)$.  
\item Assume $\Y(x) = b$ then $u(a+x,b-\Y(x)) = u(a+x,0) \geq u(a,b)$ contradicts~\ref{normalized}.
\end{enumerate}
\end{proof}
\subsection{Proof of Lemma~\ref{lemma:amm-monotone}}
\begin{proof}
\begin{enumerate}
\item By~\ref{strict_monotonic}, $u(a,b) > u(a,b-y)$ for any $y \in (0,b]$, which immediately results in $\Y(0) = 0$.
\item It trivially follows from \ref{strict_monotonic} that $\Y(x) \le \Y(x+\Delta)$ for any $\Delta \geq 0$. To prove the strict monotonicity, by contradiction, let $\Delta > 0$ and assume $\Y(x) = \Y(x+\Delta)$. Therefore, by \ref{strict_monotonic}, 
\[u(a+x+\Delta,b-\Y(x+\Delta)) = u(a+x+\Delta,b-\Y(x)) > u(a+x,b-\Y(x)) \geq u(a,b).\]  
From Lemma~\ref{lemma:amm-liq}\eqref{thm:amm-liquidity}, we can already conclude that $\Y(x) , \Y(x+\Delta)<b$. As such, we now obtain a contradiction to item Lemma~\ref{lemma:amm-liq}\eqref{thm:amm-util-2}. Therefore, it must follow that $\Y(x+\Delta) > \Y(x)$.
\item By the first claim in item~\eqref{thm:amm-monotone} within this theorem and monotone convergence, $\Y(x) \nearrow \Y^*$ for some $\Y^* \leq b$.  Assume $\Y^* < b$ (and therefore $\Y(x) \neq b$ for every $x > 0$), then
    \[u(a,b) = u(a+x,b-\Y(x)) \geq u(a+x,b-\Y^*) \quad \forall x > 0.\]
    Taking the limit as $x$ tends to infinity implies $\infty > u(a,b) \geq \lim_{\bar a \to \infty} u(\bar a,b-\Y^*) = \infty$ which forms a contradiction, i.e., $\Y^* = b$.
\end{enumerate}
\end{proof}
\subsection{Proof of Lemma~\ref{lemma:amm-concave}}
\begin{proof}
\begin{enumerate}
\item Note that $\bar C(x) := \{y \in [0,b] \; | \; u(a+x,b-y) \geq u(a,b)\}$ is upper continuous by the closed graph theorem \cite[Theorem 17.11]{AB07}.  Thus by \cite[Lemma 17.30]{AB07}, $\Y(x) = \sup\{y \; | \; y \in \bar C(x)\}$ is upper semicontinuous.
\item First, under axioms \ref{cont} and \ref{convex}, the hypograph 
\[\operatorname{hypo}\Y = \{(x,y) \in \R_+ \times \R \; | \; \Y(x) \geq y\} = \{(x,y) \in \R_+ \times \R \; | \; y \leq b, \; u(a+x,b-y) \geq u(a,b)\}\] is trivially convex.
    Now, instead, assume~\ref{deriv} holds.  By implicit differentiation, and with all derivatives of the utility $u$ being taken at $(a+x,b-\Y(x))$,
 \begin{align*}
 &\Y''(x) = \frac{u_B[u_B u_{AA} - u_A u_{AB}] + u_A[u_A u_{BB} - u_B u_{AB}]}{u_B^3} \leq 0.
 \end{align*}
\item By Lemma~\ref{lemma:amm-liq}\eqref{thm:amm-util}, $u(a+x,b-\Y(x;a,b)) \geq u(a,b)$.  An application of~\ref{scale} implies $u(t[a+x],t[b-\Y(x;a,b)]) \geq u(ta,tb)$ for any $t > 0$, i.e., $\Y(tx;ta,tb) \geq t\Y(x;a,b)$ by construction.  The reverse inequality ($\Y(tx;ta,tb) \leq t\Y(x;a,b)$) follows trivially by applying this result to $(tx;ta,tb)$ with scaling factor $t^{-1}$. 
\item By Lemma~\ref{lemma:amm-monotone}\eqref{thm:amm-zero}, $\Y(0) = 0$.  Further, by item~\eqref{thm:amm-concave}, $\Y$ is concave.  Combining these properties, $\Y(x_1) \geq \frac{x_1}{x_1+x_2} \Y(x_1+x_2)$ and $\Y(x_2) \geq \frac{x_2}{x_1+x_2} \Y(x_1+x_2)$.  Therefore, $\Y(x_1) + \Y(x_2) \geq \Y(x_1+x_2)$, i.e., $\Y$ is subadditive.
\end{enumerate}
\end{proof}
\subsection{Proof of Lemma~\ref{lemma:arbitrage}}
\begin{proof}
\begin{enumerate}
\item By Lemma~\ref{lemma:amm-liq}\eqref{thm:amm-util-2} and~\eqref{thm:amm-liquidity}, it immediately follows that $u(a+x,b-\Y(x;a,b)) = u(a,b)$ for any $x \geq 0$ and $a,b > 0$.  By~\ref{strict_monotonic}, $u(a+x,b-\Y(x;a,b)-\epsilon) < u(a,b)$ for any $x \geq 0$, $a,b > 0$, and $\epsilon \in (0,b-\Y(x;a,b)]$.  Therefore, if $u(a+x,b-y) = u(a,b)$ it must follow that $\Y(x;a,b) = y$.  Using this result we recover the desired property because:
\[u(a+x_1+x_2 , b - \Y(x_1;a,b) - \Y(x_2;a+x_1,b-\Y(x_1;a,b))) = u(a+x_1,b-\Y(x_1;a,b)) = u(a,b).\]
\item We will only prove the first (in)equality, the second follows comparably.  Denote $\Y := \Y(x;a,b)$ and $\X := \X(\Y;a+x,b-\Y)$.  Then, by Lemma~\ref{lemma:amm-liq}\eqref{thm:amm-util},
    \[u(a+x-\X,b) = u(a+x-\X,b-\Y+\Y) \geq u(a+x,b-\Y) \geq u(a,b).\]
    By~\ref{strict_monotonic}, this implies $a+x-\X \geq a$, i.e., $\X \leq x$.  If, additionally,~\ref{normalized} then the inequalities above hold as equalities and the result follows comparably.
\end{enumerate}
\end{proof}

\subsection{Proof of Theorem~\ref{thm:monotone-reserve}}
\begin{proof}
\begin{enumerate}
\item By implicit differentiation (and noting $u_A,u_B > 0$ by~\ref{deriv}):
    \begin{align*}
    \Y_a(x;a,b) &= \frac{u_A(a+x,b-\Y(x;a,b)) - u_A(a,b)}{u_B(a+x,b-\Y(x;a,b))} \\
    \Y_b(x;a,b) &= \frac{u_B(a+x,b-\Y(x;a,b)) - u_B(a,b)}{u_B(a+x,b-\Y(x;a,b))} < 1.
    \end{align*}
Therefore, the results hold if $\frac{\d}{\d x} u_A(a+x,b-\Y(x;a,b)) \leq 0$ and $\frac{\d}{\d x} u_B(a+x,b-\Y(x;a,b)) \geq 0$.  Consider these derivatives:
    \begin{align*}
    \frac{\d}{\d x} &u_A(a+x,b-\Y(x;a,b)) = u_{AA}(a+x,b-\Y(x;a,b)) - \Y'(x;a,b) u_{AB}(a+x,b-\Y(x;a,b))\\ 
        &= u_{AA}(a+x,b-\Y(x;a,b)) - \frac{u_A(a+x,b-\Y(x;a,b))}{u_B(a+x,b-\Y(x;a,b))}u_{AB}(a+x,b-\Y(x;a,b)) \leq 0, \\
    \frac{\d}{\d x} &u_B(a+x,b-\Y(x;a,b)) = u_{AB}(a+x,b-\Y(x;a,b)) - \Y'(x;a,b) u_{BB}(a+x,b-\Y(x;a,b)) \\
        &= u_{AB}(a+x,b-\Y(x;a,b)) - \frac{u_A(a+x,b-\Y(x;a,b))}{u_B(a+x,b-\Y(x;a,b))}u_{BB}(a+x,b-\Y(x;a,b)) \geq 0
    \end{align*}
    taking advantage of $\Y$ strictly increasing and differentiable.
\item \begin{enumerate}
    \item By item~\eqref{thm:mr-monotone} within this theorem and monotone convergence, $\Y(x;\bar a,b) \nearrow \Y^*$ for some $\Y^* \leq b$ as $\bar a \searrow 0$.  Assume $\Y^* < b$, then
        \[u(\bar a,b) = u(\bar a+x,b-\Y(x;\bar a,b) \geq u(\bar a + x,b-\Y^*) \quad \forall \bar a > 0.\]
        Taking the limit as $\bar a$ tends to 0 leads to $-\infty = \lim_{\bar a \searrow 0} u(\bar a,b) \geq \lim_{\bar a \searrow 0} u(\bar a + x,b-\Y^*) = u(x,b-\Y^*) > -\infty$ which forms a contradiction, i.e., $\Y^* = b$.
    \item For every $\epsilon > 0$, we have that $|\Y(x;a,\hat b)| < \hat b \leq \epsilon$ for any $\hat b \in (0,\epsilon]$ by Lemma~\ref{lemma:amm-liq}\eqref{thm:amm-liquidity}.
    \end{enumerate}
\item \begin{enumerate}
    \item By item~\eqref{thm:mr-monotone} within this theorem and monotone convergence, $\Y(x;\bar a,b) \searrow \Y^*$ for some $\Y^* \geq 0$ as $\bar a \nearrow \infty$.  Assume $\Y^* > 0$, then
        \[u(\bar a,b) = u(\bar a+x,b-\Y(x;\bar a,b)) \leq u(\bar a+x,b-\Y^*) \quad \forall \bar a > 0.\]
        By quasiconcavity 
        this inequality implies $x u_A(\bar a, b) \geq \Y^* u_B(\bar a,b)$ for every $\bar a > 0$; in particular, this implies $0 = x \lim_{\bar a \to \infty} u_A(\bar a,b) \geq \Y^* \lim_{\bar a \to \infty} u_B(\bar a,b) > 0$ which forms a contradiction due to~\ref{inada}, i.e., $\Y^* = 0$.
    \item By item~\eqref{thm:mr-monotone} within this theorem and monotone convergence, $\Y(x;a,\bar b) \nearrow \Y^*$ for some $\Y^* \leq \infty$ (possibly infinite) as $\bar b \nearrow \infty$.  Assume $\Y^* < \infty$, then
        \[u(a,\bar b) = u(a+x,\bar b-\Y(x;a,\bar b)) \geq u(a+x,\bar b-\Y^*) \quad \forall \bar b > \Y^*.\]
        By quasiconcavity 
        this inequality implies $\Y^* u_B(a + x,\bar b - \Y^*) \geq x u_A(a + x , \bar b - \Y^*)$; in particular, this implies $0 = \Y^* \lim_{\bar b \to \infty} u_B(a + x,\bar b - \Y^*) \geq x \lim_{\bar b \to \infty} u_A(a + x,\bar b - \Y^*) > 0$ which forms a contradiction due to~\ref{inada}, i.e., $\Y^* = \infty$.
    \end{enumerate}
\end{enumerate}
\end{proof}

\subsection{Proof of Proposition~\ref{prop:price}}
\begin{proof}
First, we will prove that $P$ is differentiable.  By explicitly providing the derivatives, monotonicity of the pricing oracle can be proven directly. With this result, we will prove surjectivity of $P$ by demonstrating that $\lim_{\bar a \searrow 0} P(\bar a,b) = \infty$, $\lim_{\bar b \searrow 0} P(a,\bar b) = 0$, $\lim_{\bar a \nearrow \infty} P(\bar a,b) = 0$, and $\lim_{\bar b \nearrow \infty} P(a,\bar b) = \infty$ for any $a,b > 0$.  

As mentioned above, we will prove differentiability and monotonicity by explicitly providing the partial derivatives of $P$:
\begin{align*}
\frac{\d}{\d a} P(a,b) &= \frac{u_B(a,b) u_{AA}(a,b) - u_A(a,b) u_{AB}(a,b)}{u_B(a,b)^2} \leq 0, \\
\frac{\d}{\d b} P(a,b) &= \frac{u_B(a,b) u_{AB}(a,b) - u_A(a,b) u_{BB}(a,b)}{u_B(a,b)^2} \geq 0,
\end{align*}
using~\ref{deriv} and the fact that $P(a,b) = \Y'(0;a,b) = u_A(a,b)/u_B(a,b)$. 

Consider now the limits:
\begin{itemize}
\item Consider $\lim_{\bar a \searrow 0} P(\bar a,b)$: By concavity of $\Y$, $P(a,b) = \Y'(0;a,b) \geq \frac{\Y(x;a,b)}{x}$ for any $x > 0$.  Therefore, by Theorem~\ref{thm:monotone-reserve}\eqref{thm:mr-limit-0}, $\lim_{\bar a \searrow 0} P(\bar a,b) \geq \lim_{\bar a \searrow 0} \Y(x;\bar a,b)/x = b/x$ for any $x > 0$.  As this inequality holds for any $x > 0$, it must follow that $\lim_{\bar a \searrow 0} P(\bar a,b) = \infty$.
\item Consider $\lim_{\bar b \searrow 0} P(a,\bar b)$: 
Recall that $P(a,b) = \frac{1}{\X'(0;a,b)}$ by the provided assumptions. Therefore $\lim_{\bar b \searrow 0} P(a,\bar b) = 0$ if $\lim_{\bar b \searrow 0} \X'(0;a,\bar b) = \infty$.  By symmetry of the assets, this is equivalent to $\lim_{\bar a \searrow 0} \Y'(0;\bar a,b) = \infty$ which is proven in the prior case.
\item Consider $\lim_{\bar a \nearrow \infty} P(\bar a,b)$: As in the prior case, recall that $P(a,b) = \frac{1}{\X'(0;a,b)}$ by the provided assumptions.  Therefore $\lim_{\bar a \nearrow \infty} P(\bar a,b) = 0$ if $\lim_{\bar a \nearrow \infty} \X'(0;\bar a,b) = \infty$.  By symmetry of the assets, this is equivalent to $\lim_{\bar b \nearrow \infty} \Y'(0;a,\bar b) = \infty$ which is proven in the next case.
\item Consider $\lim_{\bar b \nearrow \infty} P(a,\bar b)$: As in the first case, by concavity of $\Y$, $P(a,b) = \Y'(0;a,b) \geq \frac{\Y(x;a,b)}{x}$ for any $x > 0$.  Therefore, by Theorem~\ref{thm:monotone-reserve}\eqref{thm:mr-limit-inf}, $\lim_{\bar b \nearrow \infty} P(a,\bar b) \geq \lim_{\bar b \nearrow \infty} \Y(x;a,\bar b)/x = \infty$ for any $x > 0$.
\end{itemize}

Finally, the scale invariance of $P$ follows directly from the positive homogeneity of $\Y$ (see Lemma~\ref{lemma:amm-concave}\eqref{thm:amm-ph}).  Specifically, let $t > 0$ then
\[P(ta,tb) = \lim_{\epsilon \searrow 0} \frac{\Y(\epsilon;ta,tb)}{\epsilon} = \lim_{\epsilon \searrow 0} \frac{\Y(t\epsilon;ta,tb)}{t\epsilon} = \lim_{\epsilon \searrow 0} \frac{t\Y(\epsilon;a,b)}{t\epsilon} = \lim_{\epsilon \searrow 0} \frac{\Y(\epsilon;a,b)}{\epsilon} = P(a,b).\]
\end{proof}

\subsection{Proof of Theorem~\ref{thm:pooling}}\label{sec:proofs-pooling}
\begin{proof}
Fix $(a,b) \in \R^2_{++}$. Due to the symmetry of the AMM utility function, we will prove this result for the swap function $\Y$ only.
\begin{enumerate}
\item By Proposition~\ref{prop:price}\eqref{prop:price-si}, pooling is accomplished in proportion to the pool size, i.e., $(\alpha,\beta) = t(a,b)$ for some $t > 0$. Following Lemma~\ref{lemma:amm-concave}\eqref{thm:amm-ph}, \eqref{eq:pool-Y} trivially holds as
\[\Y(x;a+\alpha,b+\beta) = \Y(x;(1+t)a,(1+t)b) = (1+t)\Y(\frac{1}{1+t}x;a,b) \geq \Y(x;a,b)\]
where the final inequality holds since $\Y$ is concave (Lemma~\ref{lemma:amm-concave}\eqref{thm:amm-concave}) and $\Y(0) = 0$ (Lemma~\ref{lemma:amm-monotone}\eqref{thm:amm-zero}).
\item 
Define $\beta: [-a,\infty) \to [-b,\infty)$ such that $P(a,b) = P(a+\delta,b+\beta(\delta))$ for any $\delta \in (-a,\infty)$ which is guaranteed to exist by the surjective property of the pricing oracle $P$ as provided in Proposition~\ref{prop:price} (and with $\beta(-a) = -b$).  This proof could comparably be defined w.r.t.\ $\alpha: [-b,\infty) \to [-a,\infty)$ constructed as with $\beta$ but on the first asset.\footnote{
Though \ref{inada} was assumed for Theorem~\ref{thm:pooling}, it is not necessary for that result.  Specifically, \ref{inada} is only used to guarantee the existence of $\beta(\delta)$ for any $\delta \in [-a,\infty)$ through the surjectivity of $b \mapsto P(a,b)$ as provided in Proposition~\ref{prop:price}.
However, if $\beta(\delta)$ does not exist, then pooling would need to be done primarily in $A$ instead using the, similarly defined, function $\alpha: [-b,\infty) \to [-a,\infty)$. (At least one of $\alpha$ or $\beta$ can be appropriately defined as, from the proof of Proposition~\ref{prop:price}, $\lim_{\bar a \searrow 0} P(\bar a,b) = \infty$ and $\lim_{\bar b \searrow 0} P(a,\bar b) = 0$ for any $a,b > 0$ \emph{without} requiring~\ref{inada}.)
}

Recall $P(\bar a,\bar b) := \Y'(0;\bar a,\bar b) = u_A(\bar a,\bar b)/u_B(\bar a,\bar b)$ for any reserves $(\bar a,\bar b) \in \R^2_{++}$.  Therefore, by implicit differentiation, 
\[\beta'(\delta) = -\frac{P_A(\hat a,\hat b)}{P_B(\hat a,\hat b)},\]
where $(\hat a,\hat b) := (a+\delta,b+\beta(\delta))$ for any $\delta \in (-a,\infty)$.

We will prove this result only in the case of $\Y$; the monotonicity of $\X$ follows similarly.
For shorthand, define $\bar\Y(\delta) := \Y(x;a+\delta,b+\beta(\delta))$ for fixed $x > 0$.
Then $u(a+\delta+x,b+\beta(\delta)-\bar\Y(\delta)) = u(a+\delta,b+\beta(\delta))$ by construction. 
Using the same construction of $(\hat a,\hat b) := (a+\delta,b+\beta(\delta))$ as above, by implicit differentiation
\[\bar\Y'(\delta) = \frac{\left[u_A(\hat a+x,\hat b-\bar\Y(\delta))+\beta'(\delta)u_B(\hat a+x,\hat b-\bar\Y(\delta))\right] - \left[u_A(\hat a,\hat b) + \beta'(\delta)u_B(\hat a,\hat b)\right]}{u_B(\hat a+x,\hat b-\bar\Y(\delta))}.\]
Therefore $\bar\Y'(\delta) \geq 0$ if 
\[\underbrace{\frac{\d}{\d x}\left[u_A(\hat a+x,\hat b-\Y(x;\hat a,\hat b))+\beta'(\delta)u_B(\hat a+x,\hat b-\Y(x;\hat a,\hat b))\right]}_{(\ast)} \geq 0.\]
Explicitly computing the derivative $(\ast)$, we recover the equivalent condition:
\[\beta'(\delta) = -\frac{P_A(\hat a,\hat b)}{P_B(\hat a,\hat b)} \geq -\frac{P_A(\hat a+x,\hat b-\Y(x;\hat a,\hat b))}{P_B(\hat a+x,\hat b-\Y(x;\hat a,\hat b))}.\] 
In particular, this holds if
\[\underbrace{\frac{\d}{\d x}\left(\frac{P_A(\hat a+x,\hat b-\Y(x;\hat a,\hat b))}{P_B(\hat a+x,\hat b-\Y(x;\hat a,\hat b))}\right)}_{(\ast\ast)} \geq 0.\]
Explicitly computing the derivative $(\ast\ast)$, we recover the desired monotonicity:
\[(\ast\ast) := \frac{P_B(\hat z)P_{AA}(\hat z) - \left[P(\hat z) P_B(\hat z) + P_A(\hat z)\right]P_{AB}(\hat z) + P(\hat z) P_A(\hat z) P_{BB}(\hat z)}{P_B(\hat z)^2} \geq 0\]
by assumption where $\hat z := (\hat a + x , \hat b - \Y(x;\hat a,\hat b))$.
\end{enumerate}
\end{proof}

\subsection{Proof of Corollary~\ref{cor:impacts}}
\begin{proof}
The decrease of the price impact from swapping inequalities, i.e.\ the inequalities in \eqref{eq:impct1} follow immediately from the definition of $\Imp_\Y,~\Imp_\X$ and \eqref{eq:pool-Y} shown in  
Theorem~\ref{thm:pooling}.
\end{proof}

\subsection{Proof of Proposition~\ref{prop:quadratic-approx}}
\begin{proof}
To show the approximating power of the price impact oracle, we will only investigate the bounds for $\Y$; similar arguments can be made for the bounds on $\X$.
Fix $(a,b) \in \R^2_{++}$ and let $z(x) := (a+x,b-\Y(x;a,b))$.
By simple differentiation, we find that
\begin{align}
\Y'(x) &= P(z(x)), \label{eq:ineq-proof1}\\
\Y''(x) &= P_A(z(x)) - P(z(x))P_B(z(x)), \\
\Y'''(x) &= P_{AA}(z(x)) - 2 P(z(x)) P_{AB}(z(x)) + P(z(x))^2 P_{BB}(z(x))\\
&\quad + (P(z(x)) P_B(z(x)) - P_A(z(x))) P_B(z(x)) \label{eq:ineq-proof3}
\end{align}
for any $x \geq 0$. 
Therefore, since $\Y$ is thrice continuously differentiable at $x = 0$, we have the Taylor expansion of $\Y$ around zero as
\begin{align}
\Y(x) &= \Y(0) + a\Y'(0)\frac{x}{a} +\frac{a^2}{2}\Y''(0)\left(\frac{x}{a}\right)^2 + O\left(\left(\frac{x}{a}\right)^3\right)\\
 &= P(a,b)x  - \frac{1}{2} (P(a,b) P_B(a,b) - P_A(a,b))x^2 + O\left(\left(\frac{x}{a}\right)^3\right). 
\label{eq:Taylor-I_Y}
\end{align}

For the second part of the proposition, first let the conditions of Theorem~\ref{thm:pooling} hold.
Let $\beta: [-a,\infty) \to [-b,\infty)$ be defined as in the proof of Theorem~\ref{thm:pooling}.\footnote{As noted within the proof of Theorem~\ref{thm:pooling}, we can take $\alpha: [-b,\infty) \to [-a,\infty)$ instead if $\beta$ is not well-defined for every reserve level.}
To simplify notation, let $(\hat{a},\hat{b}) := (a+\delta,b+\beta(\delta))$ for $\delta \in (-a,\infty)$.  The result follows by taking the derivative of the price impact oracle w.r.t.\ the added liquidity $\delta$ (noting that $P(a,b) = P(\hat{a},\hat{b})$):
\begin{align*}
\frac{\partial}{\partial\delta}& \Imp(\hat{a},\hat{b}) = \frac{\partial}{\partial\delta} [P(a,b) P_B(\hat{a},\hat{b}) - P_A(\hat{a},\hat{b})] \\
&= -\frac{P_B(\hat{a},\hat{b})P_{AA}(\hat{a},\hat{b}) - (P(a,b) P_B(\hat{a},\hat{b}) + P_A(\hat{a},\hat{b}))P_{AB}(\hat{a},\hat{b}) + P(a,b) P_A(\hat{a},\hat{b}) P_{BB}(\hat{a},\hat{b})}{P_B(\hat{a},\hat{b})} \\
&= -\frac{P_B(\hat{a},\hat{b})P_{AA}(\hat{a},\hat{b}) - (P(\hat{a},\hat{b}) P_B(\hat{a},\hat{b}) + P_A(\hat{a},\hat{b}))P_{AB}(\hat{a},\hat{b}) + P(\hat{a},\hat{b}) P_A(\hat{a},\hat{b}) P_{BB}(\hat{a},\hat{b})}{P_B(\hat{a},\hat{b})} \leq 0.
\end{align*}

Now assume \ref{deriv} and \ref{scale}. Note that, as in Remark~\ref{rem:pooling}, $\beta(\delta) = \delta b/a$ for any $\delta \in (-a,\infty)$. Therefore, the modified liquidity $(\hat{a},\hat{b}) = (ta,tb)$ for some $t > 0$ and, as such, we can prove the desired monotonicity in liquidity by demonstrating that the price impact oracle is positive homogeneous of degree $-1$. Fix $t > 0$ then, 
by the positive homogeneity of $\Y$ (see Lemma~\ref{lemma:amm-concave}\eqref{thm:amm-ph}) and the scale invariance of $P$ (see Proposition~\ref{prop:price}\eqref{prop:price-si}),
\begin{align*}
\Imp(ta,tb) &= -\frac{1}{2}\Y''(0;ta,tb) = -\frac{1}{2} \lim_{\epsilon \searrow 0} \frac{\Y'(\epsilon;ta,tb) - \Y'(0;ta,tb)}{\epsilon} \\
    &= -\frac{1}{2} \lim_{\epsilon \searrow 0} \frac{\Y'(t\epsilon;ta,tb) - \Y'(0;ta,tb)}{t\epsilon} = -\frac{1}{2t} \lim_{\epsilon \searrow 0} \frac{P(t[a+\epsilon],t[b-\Y(\epsilon;a,b)]) - P(ta,tb)}{\epsilon} \\
    &= -\frac{1}{2t} \lim_{\epsilon \searrow 0} \frac{P(a+\epsilon,b-\Y(\epsilon;a,b)) - P(a,b)}{\epsilon} = \frac{1}{t} \Imp(a,b).
\end{align*}
\end{proof}

\subsection{Proof of Corollary~\ref{cor:quadratic-approx}}
\begin{proof}
In the same setting as in the proof of Proposition \ref{prop:quadratic-approx}, the upper and the lower bounds trivially follow from the Taylor expansion of $\Y$ in \eqref{eq:Taylor-I_Y}, and the assumptions of the lemma that the derivatives in \eqref{eq:ineq-proof1}--\eqref{eq:ineq-proof3} satisfy $\Y'(x) > 0$, $\Y''(x) \leq 0$, and $\Y'''(x) \geq 0$ for any $x \geq 0$. The bounds for $\Imp_\X$ can similarly be proven. 
Finally, by Proposition~\ref{prop:price}\eqref{prop:price-monotone}, $\Imp(a,b) \geq 0$  for any $(a,b) \in \R^2_{++}$.
\end{proof}

\subsection{Proof of Proposition~\ref{prop:sdamm}}
\begin{enumerate}
\item Under the assumptions, trivially $u(x,0) = U(x) + U(0) = -\infty$, $u(0,y) = U(0) + U(y) = -\infty$ and $u(z) = U(z_1) + U(z_2) > -\infty$ for any $x,y \geq 0$ and $z \in \R^2_{++}$.
\item Under the assumptions, trivially $\lim_{\bar x \to \infty} u(\bar x,y) = \lim_{\bar x \to \infty} U(\bar x) + U(y) = \infty$ and $\lim_{\bar y \to \infty} u(x,\bar y) = U(x) + \lim_{\bar y \to \infty} U(\bar y) = \infty$ for any $x,y > 0$.
\item Let $z - \bar z \in \R^2_+ \backslash \{0\}$ for $z,\bar z \in \R^2_{++}$, i.e., $z_1 \geq \bar z_1$, $z_2 \geq \bar z_2$ and $z \neq \bar z$. Therefore at least one of $z_1 > \bar z_1$ or $z_2 > \bar z_2$. Under the assumptions, the desired strict monotonicity follows: $u(z) = U(z_1) + U(z_2) > U(\bar z_1) + U(\bar z_2) = u(\bar z)$.
\item First we note that $u(x,y) = U(x) + U(y)$ is concave as the sum of two concave functions, i.e., \ref{convex} holds trivially. Now let us consider the properties of \ref{deriv} under the added assumption that $U'(z) > 0$ for every $z > 0$: Fix $z \in \R^2_{++}$.
    \begin{itemize}
    \item $u_A(z) = U'(z_1) > 0$ and $u_B(z) = U'(z_2) > 0$.
    \item $u_B(z)u_{AA}(z) = U'(z_2) U''(z_1) \leq 0 = u_A(z) u_{AB}(z)$ and $u_A(z)u_{BB}(z) = U'(z_1) U''(z_2) \leq 0 = u_B(z) u_{AB}(z)$ as $u_{AB} \equiv 0$.
    \end{itemize}
\item To prove \ref{inada}, fix $x,y > 0$: 
    \begin{itemize}
    \item $\lim_{\bar x \to \infty} u_A(\bar x,y) = \lim_{\bar x \to \infty} U'(\bar x) = 0$ and $\lim_{\bar y \to \infty} u_B(x,\bar y) = \lim_{\bar y \to \infty} U'(\bar y) = 0$. 
    \item $\lim_{\bar x \to \infty} u_B(\bar x,y) = U'(y) \in (0,\infty)$ and $\lim_{\bar y \to \infty} u_A(x,\bar y) = U'(x) \in (0,\infty)$. 
    \item $\lim_{\bar x \to 0} u_A(\bar x,y) = \lim_{\bar x \to 0} U'(\bar x) = \infty$ and $\lim_{\bar y \to 0} u_B(x,\bar y) = \lim_{\bar y \to 0} U'(\bar y) = \infty$. 
    \item $\lim_{\bar x \to 0} u_B(\bar x,y) = U'(y) \in (0,\infty)$ and $\lim_{\bar y \to 0} u_A(x,\bar y) = U'(x) \in (0,\infty)$. 
    \end{itemize}
\item First we recall that $P(z) = U'(z_1)/U'(z_2)$ for any $z \in \R^2_{++}$. Note that, by assumption, $U'(z) U'''(z) \geq U''(z)^2$ for any $z > 0$ by assumption. Fix $z \in \R^2_{++}$ then:
    \begin{align*}
    &P_B(z) P_{AA}(z) - (P(z)P_B(z) + P_A(z))P_{AB}(z) + P(z)P_A(z)P_{BB}(z) \\
    &\quad = -\frac{U'(z_1)U''(z_2)}{U'(z_2)^2}\frac{U'''(z_1)}{U'(z_2)} + \left(-\frac{U'(z_1)}{U'(z_2)}\frac{U'(z_1)U''(z_2)}{U'(z_2)^2} + \frac{U''(z_1)}{U'(z_2)}\right)\frac{U''(z_1)U''(z_2)}{U'(z_2)^2} \\
    &\qquad + \frac{U'(z_1)}{U'(z_2)}\frac{U''(z_1)}{U'(z_2)}\frac{U'(z_1)\left[2 U''(z_2)^2 - U'(z_2) U'''(z_2)\right]}{U'(z_2)^3} \\
    &\quad = \frac{U'(z_1)^2 U''(z_1) [U''(z_2)^2 - U'(z_2)U'''(z_2)] + U'(z_2)^2 U''(z_2) [U''(z_1)^2 - U'(z_1)U'''(z_1)]}{U'(z_2)^5} \geq 0.
    \end{align*}
\item Recall that $P(z) = U'(z_1)/U'(z_2)$ for any $z \in \R^2_{++}$. Therefore, by construction:
    \begin{align*}
    \Psi(z) &= P_{AA}(z) - 2 P(z) P_{AB}(z) + P(z)^2 P_{BB}(z) + (P(z) P_B(z) - P_A(z)) P_B(z)\\
        &= \frac{U'''(z_1)}{U'(z_2)} + 2\frac{U'(z_1)}{U'(z_2)} \frac{U''(z_1)U''(z_2)}{U'(z_2)^2} + \frac{U'(z_1)^2}{U'(z_2)^2}\frac{U'(z_1)\left[2U''(z_2)^2 - U'(z_2)U'''(z_2)\right]}{U'(z_2)^3}\\
        &\qquad + \left(\frac{U'(z_1)}{U'(z_2)} \frac{U'(z_1)U''(z_2)}{U'(z_2)^2} + \frac{U''(z_1)}{U'(z_2)}\right) \frac{U'(z_1)U''(z_2)}{U'(z_2)^2}\\
        &= \frac{U'(z_1)^3 [3 U''(z_2)^2 - U'(z_2) U'''(z_2)] + U'(z_2)^2 [U'(z_2)^2 U'''(z_1) + 3 U'(z_1) U''(z_1) U''(z_2)]}{U'(z_2)^5}
    \end{align*}
for any $z \in \R^2_{++}$.
    Trivially, by the assumptions, $\Psi(z) \geq 0$ for any $z \in \R^2_{++}$, i.e., \eqref{eq:P-cond-quad} is satisfied.
    Consider, now, \eqref{eq:P-cond-quad-X} and fix $z \in \R^2_{++}$
    \begin{align*}
    &-P(z)\Psi(z) + 3 (P(z) P_B(z) - P_A(z))^2 \\
    &\quad = \frac{U'(z_2)^3 [3 U''(z_1)^2 - U'(z_1) U'''(z_1)] + U'(z_1)^2 [U'(z_1)^2 U'''(z_2) + 3 U'(z_2) U''(z_1) U''(z_2)]}{U'(z_2)^5}.
    \end{align*}
    By the same logic as above, \eqref{eq:P-cond-quad-X} is satisfied.
\end{enumerate}

\subsection{Proof of Lemma~\ref{lemma:divergence}}
\begin{proof}
We will prove these results for $p < P(a,b)$ only; the case for $p > P(a,b)$ follows similarly. To simplify notation, let $\Y_{\gamma}(p) := \Y_{\gamma}(x_p;(1+\delta)a,(1+\delta)b)$. Furthermore, recall that $p = P((1+\delta)a+x_p,(1+\delta)b-\Y_{\gamma}(p)) = \frac{1}{1-\gamma}\Y_{\gamma}'(x_p;(1+\delta)a,(1+\delta)b)$.
First, we wish to demonstrate that we recover the simplified version of the divergence loss:
\begin{align*}
\Delta(p) &= \delta[pa+b] - \frac{\delta}{1+\delta}\left[p((1+\delta)a+x_p) + ((1+\delta)b-\Y_{\gamma}(p)\right] \\
&= \frac{\delta}{1+\delta}\left[(1+\delta)(pa+b) - p((1+\delta)a+x_p) - ((1+\delta)b-\Y_{\gamma}(p))\right] \\
&= \frac{\delta}{1+\delta}[\Y_{\gamma}(p) - px_p].
\end{align*}
Therefore the sign of the divergence loss $\Delta(p)$ is completely characterized by the sign of $\Y_{\gamma}(p) - px_p$.
As there is a one-to-one relation between $\Delta$ and $\bar\Delta$, we will consider the question of the sign of $\Y_{\gamma}(x;(1+\delta)a,(1+\delta)b) - P((1+\delta)a+x,(1+\delta)b-\Y_{\gamma}(x;(1+\delta)a,(1+\delta)b)) x$ as $x \geq 0$ varies (corresponding to $p \leq P(a,b)$).  
To simplify notation, as before we will drop the arguments of these functions where the meaning is clear.
Note that $[\Y_{\gamma} - Px]_{x = 0} = 0$ by construction and $\frac{\d}{\d x}[\Y_{\gamma} - Px] = -\gamma P + x((1-\gamma)P P_B - P_A)$.  Recall from Proposition~\ref{prop:price} that $P_A \leq 0$ and $P_B \geq 0$. 
Therefore, at $\gamma = 0$, $\Y_{\gamma} - Px \geq 0$ for any $x > 0$ with equality if and only if $P((1+\delta)a+x,(1+\delta)b-\Y_{\gamma}(x;(1+\delta)a,(1+\delta)b)) = P(a,b)$, i.e., where $p = P(a,b)$. 
If $\gamma \in (0,1)$ then, for $x$ small enough, $\frac{\d}{\d x}[\Y_{\gamma} - Px] < 0$; thus there exists some $x_* \in (0,\infty]$ such that $\Y_{\gamma} - Px < 0$ for every $x \in (0,x^*)$ and $[\Y_{\gamma}-Px]_{x = x^*} = 0$. 
By the relation between $x$ and $p$, we can define $p_* := P((1+\delta)a + x_* , (1+\delta)b - \Y_{\gamma}(x_*;(1+\delta)a,(1+\delta)b))$.
\end{proof}

\subsection{Proof of Corollary~\ref{cor:divergence}}
\begin{proof}{Proof}
Let $\delta := \frac{P(a,b)\alpha + \beta}{P(a,b)a + b}$. For the purposes of this proof, we will consider the case with $p < P(a,b)$ generated by a swap of $x > 0$ for $\Y(x;a+\alpha,b+\beta) > 0$.
Note that $\bar\Delta(0) = 0$. Fix $z := (a + \alpha + x , b + \beta - \Y(x;a+\alpha,b+\beta))$ then
\begin{align*}
\bar\Delta'(x) &= (P_A(z) - P(z) P_B(z))\alpha - \frac{\delta}{1+\delta}[P(z) + (P_A(z) - P(z) P_B(z))(a + \alpha + x) - P(z)]\\
    &= \frac{P_A(z) - P(z) P_B(z)}{1+\delta}(\alpha - \delta(a+x)).
\end{align*}
Recall, from Proposition~\ref{prop:price}, that $P_A(z) \leq 0$ and $P_B(z) \geq 0$; in fact, without loss of generality, we can take $P_A(z) - P(z) P_B(z) < 0$ as otherwise there exists some $x^* > x$ such that $P(z) = P(a+\alpha+x^*,b+\beta-\Y(x^*;a+\alpha,b+\beta))$ which can be taken instead.
Therefore, for $x > 0$ small enough, $\bar\Delta'(x) < 0$ if and only if $\alpha - \delta a > 0$, i.e., $\alpha < \delta a$ and $\beta > \delta b$, i.e., $\alpha/\beta < a/b$.
Similarly, considering a swap $y > 0$ for $\X(y;a+\alpha,b+\beta)$ results in $\bar\Delta'(-y) < 0$ if and only if $\beta - \delta b > 0$.
\end{proof}

\section{Properties of fees on the marginal price}\label{sec:fee-price-details}
Herein we wish to provide the properties that our novel fee structure (Definition~\ref{defn:fees}) satisfies.
Recall that the intuition behind the constructions \eqref{eq:Y-gamma} and \eqref{eq:X-gamma} is that the AMM should be indifferent to the size of transactions; a sequence of unidirectional small trades should have the same result for the pool as a single large trade (assuming nothing happens in between the transactions). 
In other words, a trade of $\Y_\gamma(x;a,b)$ and then another (infinitesimal) trade $\Y_{\gamma}(dx;a+x,b-\Y_{\gamma}(x;a,b))$ should be equivalent to a single trade of $\Y_\gamma(x+dx;a,b)$ from the perspective of the pool. 
Charging $\gamma$ proportion in fees, and using the fact that $\Y_{\gamma}(dx;a+x,b-\Y_{\gamma}(x;a,b)) \approx \Y_{\gamma}'(0;a+x,b-\Y_{\gamma}(x;a,b))dx = (1-\gamma) P(a+x,b-\Y_{\gamma}(x;a,b))dx$, we arrive at an equivalent ODE formulation for \eqref{eq:Y-gamma}:
\begin{equation}\label{eq:fee-diffeq}
\Y_{\gamma}'(x) = (1-\gamma)P(a+x,b-\Y_{\gamma}(x)) =: (1-\gamma) g(x,\Y_{\gamma}(x)) \quad \forall x \geq 0
\end{equation}
with initial value $\Y_{\gamma}(0) = 0$. (An ODE can similarly be provided for $\X_{\gamma}$). 

Before continuing with the discussion of these fees, we will prove that the AMM with fees is well-defined in that there exists unique swap functions $\Y_{\gamma},\X_{\gamma}$ given by Definition~\ref{defn:fees}. 
\begin{lemma}\label{lemma:exist}
Let $u: \R^2_+ \to \R \cup\{-\infty\}$ be an AMM satisfying Assumption~\ref{ass:fees}.  There exists unique swap functions $\Y_{\gamma},\X_{\gamma}$ for any $\gamma \in [0,1]$.
\end{lemma}
\begin{proof}
We will prove the result for $\Y_{\gamma}$ only, the proof for $\X_{\gamma}$ follows comparably.  
Consider the ODE representation~\eqref{eq:fee-diffeq} and note the domain $\dom g := \R_+ \times [0,b)$. 
By Proposition~\ref{prop:price}, $g$ and $\frac{\d}{\d x}g, \frac{\d}{\d y}g$ are continuous, and thus bounded, on this domain.  Therefore, by the Picard-Lindel\"of Theorem and extension theorem (e.g., \cite[Theorem II.1.1 and Theorem II.3.1]{Hartman}), there exists a unique solution $\Y_{\gamma}(x)$ on some maximal domain $x \in [0,x^*)$ for some $x^* > 0$. 
Furthermore, by the extension theorem, if $x^* < \infty$ then $\lim_{x \to x^*} \Y_{\gamma}(x) \in \{0,b\}$.  Trivially, $\Y_{\gamma}(x) > 0$ for $x > 0$ by $g(x,y) > 0$ for any $y \in [0,b)$. 
Assume now that $x^* < \infty$ and $\lim_{x \to x^*} \Y_{\gamma}(x) = b$. 
To complete this proof, we will demonstrate that $\Y_{\gamma}(x) \leq \Y(x)$ for any $x \in [0,x^*)$,  from which it will follow that $\lim_{x \to x^*} \Y_{\gamma}(x) \leq \lim_{x \to x^*} \Y(x) = \Y(x^*) < b$ (by Lemma~\ref{lemma:amm-liq}\eqref{thm:amm-liquidity}) to reach a contradiction. 
To show that  $\Y_{\gamma}(x) \leq \Y(x)$ for any $x \in [0,x^*)$, assume by contradiction that this inequality is false and take $x^\dagger := \inf\{x \in [0,x^*) \; | \; \Y_{\gamma}(x) > \Y(x)\} < x^*$.  Note that $\Y_{\gamma}(x^\dagger) = \Y(x^\dagger)$ by a simple continuity argument, and $\Y_{\gamma}(x^\dagger+\epsilon) > \Y(x^\dagger+\epsilon)$ for every $\epsilon \in (0,\delta)$ for some $\delta > 0$.  However, this implies $\Y_{\gamma}'(x^\dagger) = (1-\gamma)g(x^\dagger,\Y_{\gamma}(x^\dagger)) \leq g(x^\dagger,\Y(x^\dagger)) = \Y'(x^\dagger)$ which results in a simple contradiction.
\end{proof}

Before studying the properties of the modified swap functions and pricing oracle, we wish to provide an explicit example of the swap function $\Y_{\gamma}$ under a real-world AMM construction.
\begin{example}\label{ex:uniswapv2}
Consider the Uniswap V2 utility function $u(x,y) = \log(x) + \log(y)$ as discussed in Example~\ref{ex:uniswap}.  For this AMM, the pricing oracle $P(x,y) = y/x$ is the ratio of the reserves as noted in Example~\ref{ex:uniswap-P}.  The swap for $x \geq 0$ with fee level $\gamma \in [0,1]$ can be found to be 
\[\Y_{\gamma}(x;a,b) = b\left(1 - \frac{a^{1-\gamma}}{(a+x)^{1-\gamma}}\right)\]
by solving the differential equation~\eqref{eq:fee-diffeq}. 
\end{example}

We now consider our first property of the AMM with fees.  Specifically, as expected, as the fees $\gamma$ increase then the pool will pay out less in a swap. 
\begin{proposition}\label{prop:amm-fee-bound}
Let $u: \R^2_+ \to \R \cup\{-\infty\}$ be an AMM satisfying Assumption~\ref{ass:fees}.  For any $0 \leq \gamma_1 < \gamma_2 \leq 1$, 
\begin{align}
&0 \leq \Y_{\gamma_2}(x;a,b) < \Y_{\gamma_1}(x;a,b) \leq \Y(x;a,b),\label{eq:Y-ineq}\\
&0 \leq \X_{\gamma_2}(y;a,b) < \X_{\gamma_1}(y;a,b) \leq \X(y;a,b),\label{eq:X-ineq}
\end{align}
for every $x,y > 0$ and $a,b > 0$.  
\end{proposition}
\begin{proof}
As in prior proofs, we will provide the proof of this result for $\Y_{\gamma}$ only as the proof for $\X_{\gamma}$ follows comparably.
First, we wish to note that the non-strict monotonicity encoded in \eqref{eq:Y-ineq} follows as a simple application of \cite[Theorem III.4.1]{Hartman}. 
We now wish to prove the strict monotonicity of \eqref{eq:Y-ineq}.
Consider the ODE satisfied by $\frac{\d}{\d \gamma} \Y_{\gamma}(x;a,b)$, i.e.,
\begin{align}
&\frac{\d}{\d \gamma} \Y_{\gamma}'(x;a,b) =-P(a+x,b-\Y_{\gamma}(x;a,b)) -  (1-\gamma)\frac{\d}{\d \gamma} \Y_{\gamma}(x;a,b)P_B(a+x,b-\Y_{\gamma}(x;a,b)) \quad
\label{eq:ODE-Y-gamma-dgamma}
\end{align}
where, by~\ref{deriv}, we have that $\frac{\d}{\d \gamma} \Y_{\gamma}' = \Big(\frac{\d}{\d \gamma} \Y_{\gamma}\Big)'$.
Furthermore, note that we can view $P(a+x,b-\Y_{\gamma}(x;a,b))$ and $P_B(a+x,b-\Y_{\gamma}(x;a,b))$ as functions of $x$ (with fixed reserves $a,b > 0$). 
Therefore, together with the initial condition $\frac{\d}{\d \gamma} \Y_{\gamma}(0;a,b) = 0$, we can explicitly solve~\eqref{eq:ODE-Y-gamma-dgamma} for $\frac{\d}{\d\gamma}\Y_{\gamma}(x;a,b)$, i.e., for any $x \geq 0$
\begin{align*}
\frac{\d}{\d \gamma} \Y_{\gamma}(x;a,b) = -\int_0^x \e{-(1-\gamma) \int_t^x  P_B(a+u,b-\Y_{\gamma}(u;a,b)) du } P(a+t,b-\Y_{\gamma}(t;a,b)) dt.
\end{align*}
As this is strictly negative for any $x > 0$, \eqref{eq:Y-ineq} follows.
\end{proof}

We now turn our attention to the fundamental properties of AMMs that, without fees, were studied previously in Section~\ref{sec:swap}.  Within the following corollary, we find that all relevant properties are satisfied by $\Y_{\gamma}$ with strictly positive fees. For this result we restrict ourselves to fee levels $\gamma \in (0,1)$; if $\gamma = 0$ then we recover the original swap values (as discussed in Remark~\ref{rem:zero-fee}) and if $\gamma = 1$ then $\Y_1 \equiv 0$ by construction.
\begin{corollary}\label{cor:amm-fee}
Consider an AMM $u: \R^2_+ \to \R \cup \{-\infty\}$ satisfying Assumption~\ref{ass:fees}.  Fix the fee level $\gamma \in (0,1)$, initial pool reserves $a,b > 0$, and swap amounts $x,x_1,x_2 \geq 0$. 
\begin{enumerate}
\item $u(a+x,b-\Y_{\gamma}(x)) \geq u(a,b)$ with strict inequality if $x > 0$, i.e., market utility never drops.
\item \textbf{\nameref{SSI} and \nameref{SCC}:} $\Y_{\gamma}$ is strictly increasing, concave, and subadditive in $x$.
\item \textbf{\nameref{SNW}:} If, additionally, \ref{inf}, then $\lim\limits_{x\to\infty} \Y_\gamma(x) =b$.
\item If, additionally,~\ref{scale}, then $\Y_\gamma$ is positive homogeneous in $(x;a,b)$, i.e., $\Y_{\gamma}(tx;ta,tb) = t\Y_{\gamma}(x;a,b)$ for any $t > 0$.
\item \textbf{\nameref{SPI}:} $\Y_{\gamma}(x_1+x_2;a,b) = \Y_{\gamma}(x_1;a,b) + \Y_{\gamma}(x_2;a+x_1,b-\Y_{\gamma}(x_1;a,b))$.
\end{enumerate}
\end{corollary}
\begin{proof}
\begin{enumerate}
\item Let $x > 0$ as the case of equality when $x = 0$ is trivial.  By Lemma~\ref{lemma:amm-liq}\eqref{thm:amm-util-2} and Proposition~\ref{prop:amm-fee-bound}, and recalling from Remark~\ref{rem:zero-fee} that $\Y_0 \equiv \Y$, $u(a+x,b-\Y_{\gamma}(x)) > u(a+x,b-\Y(x)) = u(a,b)$.
\item \begin{enumerate}
    \item Strict monotonicity of $\Y_{\gamma}$ in $x$ follows trivially from its integral representation as the pricing oracle is strictly positive on positive pool sizes.
    \item By implicit differentiation of~\eqref{eq:fee-diffeq} and the monotonicity of the pricing oracle as provided in Proposition~\ref{prop:price}, $\Y_{\gamma}''(x) = (1-\gamma)[P_A(a+x,b-\Y_{\gamma}(x)) - (1-\gamma)P(a+x,b-\Y_{\gamma}(x)) P_B(a+x,b-\Y_{\gamma}(x))] \leq 0$.
    \item Subadditivity follows from concavity and $\Y_{\gamma}(0) = 0$ as demonstrated in the proof of Lemma~\ref{lemma:amm-concave}\eqref{thm:amm-subadditive}.
    \end{enumerate}
\item Note that $c_B(b) := \sup_{\bar a > 0} u_B(\bar a,b) < \infty$ for every $b > 0,$ because $u_B$ is continuous with limiting behavior $\lim\limits_{\bar a \to 0} u_B(\bar a,b),\lim\limits_{\bar a \to \infty} u_B(\bar a,b) < \infty$. 
Assume by contradiction that $\lim\limits_{x\to\infty} \Y_{\gamma}(x) = \Y_{\gamma}^* < b.$ Recall $\Y_{\gamma}$ is strictly monotonic, therefore $\Y_{\gamma}(x) < \Y_{\gamma}^*$ for every $x$. Therefore,
\begin{align}
\lim\limits_{x\to\infty}  \Y_{\gamma}(x) &= \lim_{x \to \infty} (1-\gamma) \int_0^{x} P(a+y,b-\Y_{\gamma}(y)) dy\geq (1-\gamma) \int_0^{\infty} P(a+y,b-\Y_{\gamma}^*) dy\\
&= (1-\gamma) \int_0^{\infty} \frac{u_A(a+y,b-\Y_{\gamma}^*)}{u_B(a+y,b-\Y_{\gamma}^*)} dy\geq \frac{(1-\gamma) \int_0^{\infty} u_A(a+y,b-\Y_{\gamma}^*) dy}{c_B(b-\Y_{\gamma}^*)}\\
&= \frac{1-\gamma}{c_B(b-\Y_{\gamma}^*)} \Big(\lim_{x \to \infty} u(a+x,b-\Y_{\gamma}^*) - u(a,b-\Y_{\gamma}^*)\Big) = \infty.
\end{align}
\item Let $t > 0$.  From \eqref{eq:fee-diffeq} it follows that
\begin{align*}
\frac{\d}{\d x} \frac{1}{t}\Y_{\gamma}(tx;ta,tb) &= (1-\gamma)P(ta+tx,tb-\Y_{\gamma}(tx;ta,tb)) = (1-\gamma)P(a+x,b-\frac{1}{t}\Y_{\gamma}(tx;ta,tb)).
\end{align*}
Since for $x = 0$, we have that both $\frac{1}{t}\Y_{\gamma}(0;ta,tb) = 0 = \Y_{\gamma}(0;a,b)$, we can appeal to the existence and uniqueness result of Lemma \ref{lemma:exist} to recover $\frac{1}{t}\Y_{\gamma}(tx;ta,tb) = \Y_{\gamma}(x;a,b)$, i.e., positive homogeneity.
\item From \eqref{eq:fee-diffeq} it follows that
\begin{align}
\Y_{\gamma}'(x_1+x_2;a,b) =(1-\gamma)P(a+x_1+x_2,b-\Y_{\gamma}(x_1;a,b) - [\Y_{\gamma}(x_1+x_2;a,b) -\Y_{\gamma}(x_1;a,b)]).
\label{eq:tmp-Y'1}
\end{align}
Note also that 
\begin{align}
\Y_{\gamma}'(x_1+x_2;a,b) = \frac{\d}{\d x_2} \left[ \Y_{\gamma}(x_1+x_2;a,b) -\Y_{\gamma}(x_1;a,b)\right].
\label{eq:tmp-Y'2}
\end{align}
Combining \eqref{eq:tmp-Y'1} and \eqref{eq:tmp-Y'2}, it follows that
\begin{align}
\frac{\d}{\d x_2} &\left[\Y_{\gamma}(x_1+x_2;a,b) -\Y_{\gamma}(x_1;a,b)\right]\\
&= (1-\gamma)P(a+x_1+x_2,b-\Y_{\gamma}(x_1;a,b) - [\Y_{\gamma}(x_1+x_2;a,b) -\Y_{\gamma}(x_1)]).
\end{align}
We also have, by construction of~\eqref{eq:fee-diffeq}, that
\begin{align}
\Y_{\gamma}'&(x_2;a+x_1,b-\Y_{\gamma}(x_1;a,b)) \\
&= (1-\gamma) P(a+x_1+x_2,b-\Y_{\gamma}(x_1;a,b) -\Y_{\gamma}(x_2;a+x_1,b-\Y_{\gamma}(x_1;a,b)))
\end{align}
Since for $x_2=0$, we have that both $\Y_{\gamma}(x_1+x_2) -\Y_{\gamma}(x_1) =0=\Y_{\gamma}(x_2;a+x_1,b-\Y_{\gamma}(x_1;a,b))$, we can appeal to the existence and uniqueness result of Lemma \ref{lemma:exist} to conclude
\begin{align}
\Y_{\gamma}(x_1+x_2;a,b) -\Y_{\gamma}(x_1;a,b) = \Y_{\gamma}(x_2;a+x_1,b-\Y_{\gamma}(x_1;a,b))
\end{align}
for any $x_1,x_2 \geq 0$.
\end{enumerate}
\end{proof}

\begin{remark}\label{rem:bid-ask}
In contrast to the fee-less construction provided within Section~\ref{sec:amm} (and as expected), introducing a fee $\gamma \in (0,1)$ immediately introduces a bid-ask spread.  Specifically, the bid price is given by $\Y_{\gamma}'(0) = (1-\gamma)P(a,b)$ and the ask price is given by $\X_{\gamma}'(0)^{-1} = (1-\gamma)^{-1}P(a,b)$ for any pool reserves $(a,b) \in \R^2_{++}$.  Note that the pricing oracle $P(a,b)$ is between the bid and ask prices but is \emph{not} the mid-price.
\end{remark}

As a direct consequence of the bid-ask spread discussed in Remark~\ref{rem:bid-ask}, the introduction of fees guarantees a strict no-arbitrage condition.
\begin{corollary}\label{cor:no-arb}
\textbf{\nameref{NA}:} Consider an AMM $u: \R^2_+ \to \R \cup \{-\infty\}$ satisfying Assumption~\ref{ass:fees}.  If $\gamma \in (0,1]$ then $\X_{\gamma}(\Y_{\gamma}(x;a,b);a+x,b-\Y_{\gamma}(x;a,b)) < x$ and $\Y_{\gamma}(\X_{\gamma}(y;a,b);a-\X_{\gamma}(y;a,b),b+y) < y$ for any $x,y > 0$ and $a,b > 0$.
\end{corollary}
\begin{proof}
We will only prove the first inequality, the second follows similarly.  
By Lemma~\ref{lemma:amm-monotone}\eqref{thm:amm-monotone}, Theorem~\ref{thm:monotone-reserve}\eqref{thm:mr-monotone}, and Proposition~\ref{prop:amm-fee-bound},
\begin{align*}
\X_{\gamma}(\Y_{\gamma}(x;a,b);a+x,b-\Y_{\gamma}(x;a,b)) &< \X(\Y_{\gamma}(x;a,b);a+x,b-\Y_{\gamma}(x;a,b)) \\
&\leq \X(\Y_{\gamma}(x;a,b);a+x,b-\Y(x;a,b)) \\
&< \X(\Y(x;a,b);a+x,b-\Y(x;a,b)) = x. 
\end{align*}
\end{proof}

We wish to conclude the discussion of properties of the swap amounts with fees by studying the dependence of $\Y_{\gamma}$ on the reserves $(a,b)$ as well as the fees $\gamma$.
\begin{corollary}\label{cor:monotone-reserve}
\textbf{\nameref{ML}:} Consider an AMM $u: \R^2_+ \to \R \cup \{-\infty\}$ satisfying Assumption~\ref{ass:fees}.  Fix the fee level $\gamma \in (0,1]$. Then
$\frac{\d}{\d a} \Y_{\gamma}(x;a,b) \leq 0$ and $\frac{\d}{\d b}\Y_{\gamma}(x;a,b) \in [0,1)$ for any $x \geq 0$ and $a,b > 0$.
\end{corollary}
\begin{proof}
We will prove this result for $\frac{\d}{\d a}\Y_{\gamma}(x;a,b) \leq 0$ only; the proof for $\frac{\d}{\d b}\Y_{\gamma}(x;a,b) \in (0,1]$ follows similarly.
Following the same strategy as in the proof of Proposition~\ref{prop:amm-fee-bound}, we can construct the ODE
\begin{align*}
\frac{\d}{\d a} \Y_{\gamma}'(x;a,b) = (1-\gamma)[P_A(a+x,b-\Y_{\gamma}(x;a,b)) - \frac{\d}{\d a} \Y_{\gamma}(x;a,b)P_B(a+x,b-\Y_{\gamma}(x;a,b))]
\end{align*}
with initial condition $\frac{\d}{\d a} \Y_{\gamma}(0;a,b) = 0$ for every $a,b > 0$.
By solving this ODE, we recover
\begin{align}
\frac{\d}{\d a} \Y_{\gamma}(x;a,b) = (1-\gamma)\int_0^x \e{-(1-\gamma) \int_t^x  P_B(a+u,b-\Y_{\gamma}(u;a,b)) du } P_A(a+t,b-\Y_{\gamma}(t;a,b)) dt,~\forall x\ge0. 
\end{align}
From Proposition \ref{prop:price} it now follows that $\frac{\d}{\d a} \Y_{\gamma}(x;a,b)\le0$.
\end{proof}

\begin{remark}\label{rem:fee-structure}
The fee structure introduced within Section~\ref{sec:fees} is not the only approach that can be used for an AMM.  We wish to highlight two alternate structures which can be viewed as assessing the fees on the assets sold or bought by the investor respectively.  With both of these structures we wish to briefly comment on how an investor may want to optimally split a transaction when transacting with the pool; this lack of indifference to trade splitting leads to strange implications for the pool (i.e., the liquidity provider should only care about the final state of the pool rather than how trades are executed). 
\begin{enumerate}
\item Consider the fees collected on the asset being sold to the pool.  That is, a fraction $\gamma \in [0,1]$ of $x$ is taken by the pool to compensate it for acting as a liquidity provider.  Mathematically, this is encoded by 
\begin{equation*}
\bar\Y_{\gamma}(x;a,b) := \Y((1-\gamma)x;a,b).
\end{equation*}  
That is, the collected fees are $\gamma x$ of asset $A$ so that the realized pool size, after the swap is completed, is still of the form $(a+x,b-\bar\Y_{\gamma}(x;a,b))$.
We wish to note that this is the fee structure considered in, e.g.,~\cite{angeris2020improved,lipton2021automated}.

Though a simple structure to implement, this fee structure provides a discount to buying in bulk, i.e., $\bar\Y_{\gamma}(x_1+x_2;a,b) \geq \bar\Y_{\gamma}(x_1;a,b) + \bar\Y_{\gamma}(x_2;a+x_1,b-\bar\Y_{\gamma}(x_1;a,b))$, and aside from extreme cases, leading to strict inequality violating \nameref{SPI}.  By imposing costs to an investor who splits her transaction, the pool is subsidizing large traders.
\item Consider the fees collected on the asset being bought by the trader.  That is, a fraction $\gamma \in [0,1]$ of $\Y$ is taken by the pool to compensate it for acting as a liquidity provider.  Mathematically, this is encoded by 
\begin{equation*}
\bar\Y^{\gamma}(x;a,b) := (1-\gamma)\Y(x;a,b).
\end{equation*}  
That is, the collected fees are $\gamma \Y(x;a,b)$ of asset $B$ so that the realized pool size, after the swap is completed, is still of the form $(a+x,b-\bar\Y^{\gamma}(x;a,b))$. 

In contrast to the fees on the sold asset, imposing fees on the bought asset provides a benefit to an investor who splits her trade over time, i.e., $\bar\Y^\gamma(x_1+x_2;a,b) \leq \bar\Y^\gamma(x_1;a,b) + \bar\Y^\gamma(x_2;a+x_1,b-\bar\Y^\gamma(x_1;a,b))$, and aside from extreme cases, leading to strict inequality violating \nameref{SPI}.  By encouraging these small transactions, the pool incentivizes traders to make a series of infinitesimally small transactions; that is, an intelligent trader will, in fact, implement the integral strategy $\Y_{\gamma}$ we propose in Definition~\ref{defn:fees}. 
\end{enumerate}
\end{remark}

\section{Accounting Profits and Losses}\label{sec:pnl}
As highlighted by, e.g., Figure~\ref{fig:pnl}, it is possible for both the liquidity provider and an arbitrageur (taking advantage of the pool's stale price) to have accounting profits. Within this section, we wish to detail the strategies of both types of market participants.

\subsection{Liquidity Providers}\label{sec:pnl-LP}
Recall from Remark~\ref{rem:pnl} that the accounting profits and losses of the liquidity providing position can be provided by:
\begin{align*}
\Pi_L(p) &:= \begin{cases} \frac{P(a,b)\alpha + \beta}{P(a,b)[a+\alpha] + [b+\beta]}\left[p(a+\alpha+x_p)+(b+\beta-\Y_\gamma(x_p;a+\alpha,b+\beta)\right] - (P(a,b)\alpha+\beta) &\text{if } p < P(a,b) \\
    \frac{P(a,b)\alpha + \beta}{P(a,b)[a+\alpha] + [b+\beta]}\left[p(a+\alpha-\X_\gamma(y_p;a+\alpha,b+\beta)) + (b+\beta+y_p)\right] - (P(a,b)\alpha+\beta) &\text{if } p > P(a,b) \\
    0 &\text{if } p = P(a,b) \end{cases}
\end{align*}
where $x_p,y_p$ are as in Definition~\ref{defn:divergence}. 
Though it is possible that with high enough fees $\gamma \in (0,1)$, there is also an accounting profit for the liquidity provide even in case of price decrease $p< P(a,b)$, we will consider the more intuitive case when the price increases $p > P(a,b)$ and the accounting profits are always positive.
For simplicity of constructions, herein we consider $\bar\Pi_L$ as a function of the swap amounts rather than prices (as taken with $\bar\Delta$ in Section~\ref{sec:divergence}). Noting that $\bar\Pi_L(0) = 0$, we wish to consider the change in the profitability as the swap size grows, i.e.,
\begin{align*}
\bar\Pi_L'(y) &= \frac{P(a,b)\alpha+\beta}{P(a,b)[a+\alpha]+[b+\beta]}\\
\times&\left[\gamma + \left(-\frac{(1-\gamma)P_A(a+\alpha-\X_\gamma(y),b+\beta+y)}{P(a+\alpha-\X_\gamma(y),b+\beta+y)} + P_B(a+\alpha-\X_\gamma(y),b+\beta+y)\right)\left(a+\alpha-\X_\gamma(y)\right)\right] \\
&\geq 0
\end{align*}
    by $P_A \leq 0$ and $P_B \geq 0$. In particular, if $y > 0$ and $\gamma \in (0,1)$ then $\bar\Pi_L(y) > 0$. For Uniswap V2 (see Example~\ref{ex:uniswapv2}), these profits are given by $\Pi_L(p) = \frac{2\alpha y_p}{a+\alpha} > 0$. As highlighted in Figure~\ref{fig:comparison-delta}, though the liquidity providers would be guaranteed positive profits, the divergence loss may still exist.
The aforementioned properties indicate that the liquidity providers are purchasing a long position in $A$ (where $B$ is utilized as the num\'eraire asset). By symmetry, it becomes clear that the profitability of the pool depends crucially on the choice of num\'eraire.\footnote{This symmetry holds also for the buy-and-hold strategy. For example, letting $p_A,p_B > 0$ be the price of $A$ and $B$ respectively in some third asset (e.g., US dollars) then the value of this strategy in units of $B$ is monotonically increasing in $p_A/p_B$ while the value in units of $A$ would be monotonically decreasing.}

\subsection{Arbitrageurs}\label{sec:pnl-arb}
In contrast to the long position in $A$ taken by the liquidity providers, the arbitrageurs are directionally agnostic. As such, it is possible for both the liquidity provider and an arbitrageur (taking advantage of the pool's stale price) to have accounting profits. Consider the strategy of the arbitrageur: 
\begin{itemize}
\item If the arbitrageur is able to buy $A$ frictionlessly at a price of $p < (1-\gamma)P(a,b)$ on some external market, then she will purchase $x_{(1-\gamma)^{-1}p}$ units of $A$ externally (for a cost of $px_{(1-\gamma)^{-1}p}$) swap for $\Y_\gamma(x_{(1-\gamma)^{-1}p};a+\alpha,b+\beta)$ through the AMM (for a value of $\Y_\gamma(x_{(1-\gamma)^{-1}p};a+\alpha,b+\beta)$). This leads to a strictly positive profit of $\Pi_A(p) := \Y_\gamma(x_{(1-\gamma)^{-1}p};a+\alpha,b+\beta) - px_{(1-\gamma)^{-1}p} > 0$.
\item If the arbitrageur is able to sell $A$ frictionlessly at a price of $p > (1-\gamma)^{-1}P(a,b)$ on some external market, then she will borrow $y_{(1-\gamma)p}$ units of $B$ externally (at unit cost), swap for $\X_\gamma(y_{(1-\gamma)p};a+\alpha,b+\beta)$ through the AMM, and liquidate those $A$ assets (for a value of $p\X_\gamma(y_{(1-\gamma)p};a+\alpha,b+\beta)$). This leads to a strictly positive profit of $\Pi_A(p) := p\X_\gamma(y_{(1-\gamma)p};a+\alpha,b+\beta) - y_{(1-\gamma)p} > 0$.
\end{itemize}
Note that, due to the fee structure, the arbitrageur would not profit (and, therefore, not operate) if the external market quotes a price within the bid-ask spread of the pool $(1-\gamma)P(a,b) < (1-\gamma)^{-1}P(a,b)$.\footnote{Notably, the necessary condition for profitability of the liquidity provider coupled with the behavior of the arbitrageur, guarantees that the liquidity provider cannot profit off of an arbitrageur for any price $p < P(a,b)$.} The profitability of the arbitrageur follows from the construction of $\Y_\gamma,\X_\gamma$ as provided in Definition~\ref{defn:fees}.

\section{Comparison to Prior Axioms and Definitions}\label{sec:comparison}

Within this section, we wish to highlight four other contemporary papers which, independently, provide generalized definitions of AMMs. Within this section we highlight how the axioms proposed herein relate to the definitions used within~\cite{capponi2021adoption,ferreira2022credible,schlegel2022axioms,angeris2023geometry}. For ease of reference, these results are summarized within Table~\ref{table:comparison}.

First, \cite{capponi2021adoption} introduces four properties for the utility function $u: \R^2_+ \to \R \cup \{-\infty\}$ of an AMM within Assumption 1 of that work.  Briefly, that paper assumes that:
\begin{enumerate}
\item {\bf Positive derivatives: $u_A(z),u_B(z) > 0$ for every $z \in \R^2_+$} is a stronger version of \ref{strict_monotonic} though, we note, we also assume this property within \ref{deriv};
\item {\bf Convexity: $u_{AA}(z),u_{BB}(z) < 0$, $u_{AB}(z) > 0$ for every $z \in \R^2_+$} was discussed in more details within Remark~\ref{rem:deriv} and (along with the prior property) implies \ref{deriv}; 
\item {\bf Homogenous of degree $l$: $\exists l > 0:~\forall c \geq 0,~c^l u(z) = u(cz)$ for every $z \in \R^2_+$} is strictly stonger than \ref{scale}; and
\item {\bf Surjective in price: $\lim_{x \to 0} P(x,y) = \infty,~\lim_{x \to \infty} P(x,y) = 0,$\newline$\lim_{y \to 0} P(x,y) = 0,~\lim_{y \to \infty} P(x,y) = \infty$} is strongly related to \nameref{SIL} as highlighted in Remark~\ref{rem:price-sil}; in particular, we find this property is implied by \ref{normalized}, \ref{convex}, \ref{inada}, and \ref{deriv} within Proposition~\ref{prop:price}\eqref{prop:price-b} and \eqref{prop:price-a}.
\end{enumerate}

Second, \cite{ferreira2022credible} independently considered utility functions $u: \R^2_+ \to \R \cup \{-\infty\}$ that were {\bf strictly increasing} (i.e., \ref{strict_monotonic}) and {\bf quasiconcave} (i.e., \ref{convex}). Notably, that work makes no further assumptions upon the AMMs under consideration. We highlight in Table~\ref{table:properties} how these two axioms alone are insufficient to guarantee most desired properties for the markets made by an AMM.

Third, \cite{schlegel2022axioms} introduces a number of properties for the utility functions $u: \R^2_+ \to \R \cup \{-\infty\}$.\footnote{We wish to note that \cite{schlegel2022axioms} proposes all of these axioms within a multi-asset market. For ease of comparison, we consider only the two asset case herein.} Briefly, that paper assumes that:
\begin{enumerate}
\item {\bf Existence of marginal prices: $u$ is differentiable everywhere}  implies~\ref{cont} and, in turn, is implied by~\ref{deriv} where we assume twice-differentiability;
\item {\bf Aversion to permament loss: $\{z \in \R^2_+ \; | \; u(z) \geq u(\bar z)\}$ is convex for every $\bar z \in \R^2_+$} is equivalent to \ref{convex};
\item {\bf Sufficient funds: $u(z) = u(\bar z)$ with $\bar z \in \R^2_{++}$ implies $z \in \R^2_{++}$} provides the same implications as \ref{normalized} used herein;
\item {\bf Scale invariance: $u(z) = u(\bar z)$ implies $u(t z) = u(t \bar z)$ for any $z,\bar z \in \R^2_+$ and $t > 0$} is equivalent to \ref{scale}; 
\item {\bf Homogenity in liquidity: $u(t z) = t u(z)$ for any $z \in \R^2_+$ and $t > 0$} is strictly stronger than \ref{scale}; 
\item {\bf Translation invariance: $u(z) = u(\bar z)$ implies $u(z + t \vec{1}) = u(\bar z + t \vec{1})$ for any feasible $t \in \R$} is only applicable for proving the equivalence to scoring rules for prediction markets (as in~\cite{chen2012utility});
\item {\bf One invariance: $u(z+t \vec{1}) = u(z) + t$} is a stronger translation invariance property; and
\item {\bf Symmetry: $u(a,b) = u(b,a)$} is not explicitly considered herein though we note that all examples in Section~\ref{sec:examples} satisfy this property.
\end{enumerate}

Finally, \cite{angeris2023geometry} considers a geometric representation for AMMs rather than the functional approach taken elsewhere. Specifically, in that work, the authors consider the reachable set $S_{z} = \{\bar z \in \R^2 \; | \; u(\bar z) \geq u(z)\}$ as the set of positions that can be attained through trading with the pool (the equivalence to the utility formulation provided is the functional equivalent).\footnote{We wish to note that \cite{angeris2023geometry} proposes all of these axioms within a multi-asset market. For ease of comparison, we consider only the two asset case herein.} Briefly, that paper assumes that:
\begin{enumerate}
\item {\bf Non-empty and non-negative reserves: $\emptyset \neq S_{z} \subseteq \R^2_+$ for any $z \in \R^2_{++}$} is equivalent to $\dom u \subseteq \R^2_+$;
\item {\bf Closed: $S_{z}$ is closed for any $z \in \R^2_{++}$} is equivalent to the upper semicontinuity of $u$ (and therefore implied by, e.g.,~\ref{cont});
\item {\bf Convex: $S_{z}$ is convex for any $z \in \R^2_{++}$} is equivalent to \ref{convex}; and
\item {\bf Upward closed: $S_{z} + \R^2_+ = S_{z}$ for any $z \in \R^2_{++}$} is implied by \ref{strict_monotonic} (and is equivalent to the non-strict monotonicity).
\end{enumerate}

\begin{landscape}
\begin{table}
\centering
{\footnotesize
\begin{tabular}{|c|p{2.8in}||c|c|c|c|c|c|c|c|c|c|c|}
\hline
{\bf Reference} & \multicolumn{1}{c||}{\bf Property} & \ref{normalized} & \ref{inf} & \ref{strict_monotonic} & \ref{cont} & \ref{convex} & \ref{scale} & \ref{inada} & \ref{deriv} & \eqref{eq:P-cond} & \eqref{eq:P-cond-quad} & \eqref{eq:P-cond-quad-X} \\ \hline\hline
\multirow{4}{*}{\cite{capponi2021adoption}} & Positive derivatives \newline $u_A,u_B > 0$ &  &  & $\Rightarrow$ & $\Rightarrow$ &  &  &  & $\Leftarrow$ &  &  & \\ \cline{2-13}
& Convexity \newline $u_{AA},u_{BB} < 0,~u_{AB} > 0$ &  &  &  &  &  &  &  & $\Rightarrow^{\ast}$ &  &  & \\ \cline{2-13}
& Homogenous of degree $l$ \newline $\exists l > 0: \; c^l u(z) = u(cz)$ &  &  &  &  &  & $\Rightarrow$ &  &  &  &  & \\ \cline{2-13}
& Surjective in price \newline $\lim_{x \to 0} P(x,y) = \infty,~\lim_{x \to \infty} P(x,y) = 0,$\newline$\lim_{y \to 0} P(x,y) = 0,~\lim_{y \to \infty} P(x,y) = \infty$ & $\Leftarrow$ &  &  &  & $\Leftarrow$ &  & $\Leftarrow$ & $\Leftarrow$ &  &  & \\ \hline\hline
\multirow{2}{*}{\cite{ferreira2022credible}} & Strictly increasing &  &  & X &  &  &  &  &  &  &  & \\ \cline{2-13}
& Quasiconcave &  &  &  &  & X &  &  &  &  &  & \\ \hline\hline
\multirow{8}{*}{\cite{schlegel2022axioms}} & Existence of marginal prices \newline $u$ is differentiable &  &  &  & $\Rightarrow$ &  &  &  & $\Leftarrow$ &  &  & \\ \cline{2-13}
& Aversion to permanent loss \newline $\{z \; | \; u(z) \geq u(\bar z)\}$ convex &  &  &  &  & X &  &  &  &  &  & \\ \cline{2-13}
& Sufficient funds \newline $u(z) = u(\bar z)$ then $z \in \R^2_{++} \Rightarrow \bar{z} \in \R^2_{++}$ & X &  &  &  &  &  &  &  &  &  & \\ \cline{2-13}
& Scale invariance \newline $u(z) = u(\bar z)$ implies $u(t z) = u(t\bar z)$ &  &  &  &  &  & X &  &  &  &  & \\ \cline{2-13}
& Homogeneity in liquidity \newline $u(t z) = t u(z)$ &  &  &  &  &  & $\Rightarrow$ &  &  &  &  & \\ \cline{2-13}
& Translation invariance \newline $u(z) = u(\bar{z}) \Rightarrow u(z+t\vec{1}) = u(\bar{z} + t\vec{1})$ & \multicolumn{11}{c|}{Only appropriate for prediction markets} \\ \cline{2-13}
& One invariance \newline $u(z+t\vec{1}) = u(z) + t$ & \multicolumn{11}{c|}{Only appropriate for prediction markets} \\ \cline{2-13}
& Symmetry \newline $u(a,b) = u(b,a)$ &  &  &  &  &  &  &  &  &  &  & \\ \hline\hline
\multirow{4}{*}{\cite{angeris2023geometry}} & Non-empty and non-negative reserves \newline $\emptyset \neq S_{z} := \{\bar z \; | \; u(\bar z) \geq u(z)\} \subseteq \R^2_+$ & \multicolumn{11}{c|}{By construction of the reachable set} \\ \cline{2-13}
& Closed \newline $S_z$ is closed &  &  &  & $\Leftarrow$ &  &  &  &  &  &  & \\ \cline{2-13}
& Convex \newline $S_z$ is convex &  &  &  &  & X &  &  &  &  &  & \\ \cline{2-13}
& Upward closed \newline $S_z + \R^2_+ = S_z$  &  &  & X$^\dagger$ &  &  &  &  &  &  &  & \\ \hline
\end{tabular}
}
\caption{Summary of properties defined within~\cite{capponi2021adoption,ferreira2022credible,schlegel2022axioms} and the relation to the axioms provided herein.\\ 
X: Equivalence of axioms.\\
$\Rightarrow$: The property from the external paper implies our axiom.\\
$\Leftarrow$: The collection of our axioms imply the property from the external paper.\\
$\ast$: Convexity ($u_{AA},u_{BB} < 0,~u_{AB} > 0$) implies \ref{deriv} if positive derivatives ($u_A,u_B > 0$) is also assumed.\\
$\dagger$: Upward closed ($S_z + \R^2_+ = S_z$) is equivalent to nondecreasing without requiring the strict monotonicity.}
\label{table:comparison}
\end{table}
\end{landscape}

\end{document}